\documentclass[10pt]{article}
\pdfoutput=1

\usepackage[utf8]{inputenc}
\usepackage[T2A]{fontenc}

\input{epsf}
\usepackage{epsfig}
\usepackage{amssymb}
\usepackage{amsfonts}
\usepackage{amsbsy}
\usepackage[cmtip,all]{xy}
\usepackage{amsmath}
\usepackage{esint}
\usepackage{collectbox}
\usepackage{amscd}
\usepackage{mathrsfs}
\usepackage{amsmath,amsthm}
\usepackage{enumitem}
\usepackage{dsfont}
\usepackage{soul}
\usepackage{comment}
\usepackage{cite} 
\usepackage[most]{tcolorbox}
\usepackage{stmaryrd}
\usepackage{xcolor}
\definecolor{burgundy}{rgb}{0.5, 0.0, 0.13}
\definecolor{olive}{rgb}{0.50, 0.50, 0.0}
\usepackage[linktocpage=true,colorlinks=true,linkcolor=burgundy,citecolor=black!20!blue,urlcolor=violet]{hyperref}
\usepackage{accents}
\usepackage{xfrac}

\usepackage{tikz}
\usepackage{tikz-cd}
\usetikzlibrary{arrows}
\usetikzlibrary{arrows.meta}
\usetikzlibrary{positioning}
\usetikzlibrary{shapes,snakes}
\usetikzlibrary{fit}
\usetikzlibrary{decorations.pathmorphing,decorations.pathreplacing,decorations.markings}
\usetikzlibrary{calc}

\usepackage[font=footnotesize,labelfont=bf]{caption}

\theoremstyle{definition}

\DeclareMathAlphabet{\mathpzc}{OT1}{pzc}{m}{it}

\def\exp{{\rm exp}}

\def\I{{\rm i}}

\def\log{{\rm log}}

\def\Tr{{\rm Tr}}

\def\p{\partial}

\def\CC {{\cal C}}
\def\CD {{\cal D}}

\def\CH {{\cal H}}

\def\CK {{\cal K}}
\def\CL {{\cal L}}

\def\CN {{\cal N}}
\def\CO {{\cal O}}

\def\CV {{\cal V}}

\def\CO {{\cal O}}

\def\CH {{\cal H}}

\def\IC{\mathbb{C}}

\def\IP{\mathbb{P}}
\def\IR{{\mathbb{R}}}

\def\fg{\mathfrak{g}}

\def\fl{\mathfrak{l}}

\def\fs{\mathfrak{s}}

\def\fs{\mathfrak{s}}

\def\fQ{\mathfrak{Q}}

\def\bphi{{\boldsymbol{\phi}}}

\def\lm{\limits}

\hoffset= - 1.0in

\numberwithin{equation}{section}

\tcbset{boxrule=0.6pt,colback=white!97!green}

\DeclareSymbolFont{bbsymbol}{U}{bbold}{m}{n}
\DeclareMathSymbol{\bbzero}{\mathbin}{bbsymbol}{"30}
\DeclareMathSymbol{\bbone}{\mathbin}{bbsymbol}{"31}
\DeclareMathSymbol{\bbtwo}{\mathbin}{bbsymbol}{"32}
\DeclareMathSymbol{\bbthree}{\mathbin}{bbsymbol}{"33}
\DeclareMathSymbol{\bbfour}{\mathbin}{bbsymbol}{"34}
\DeclareMathSymbol{\bbfive}{\mathbin}{bbsymbol}{"35}
\DeclareMathSymbol{\bbsix}{\mathbin}{bbsymbol}{"36}
\DeclareMathSymbol{\bbseven}{\mathbin}{bbsymbol}{"37}
\DeclareMathSymbol{\bbeight}{\mathbin}{bbsymbol}{"38}
\DeclareMathSymbol{\bbnine}{\mathbin}{bbsymbol}{"39}

\def\myblue{white!40!blue}

\newcommand\sqbox[1]{{
	\setbox0=\hbox{\mbox{$\Box$}}
	\setbox1=\hbox{\mbox{\raisebox{0.35ex}{\tiny #1}}}
	\mbox{\raisebox{-0.2ex}{\rlap{\hbox to \wd0{\hss{\box1}\hss}}\box0}}
}}

\begin{document}

\hfill MIPT/TH-05/25

\hfill ITEP/TH-05/25

\hfill IITP/TH-05/25

\vskip 1.5in
\begin{center}
	
    {\bf\Large Tunnels Under Geometries \\ \vskip 0.2in (or Instantons Know Their Algebras)}
	
	\vskip 0.2in
	\renewcommand{\thefootnote}{\fnsymbol{footnote}}
	{Dmitry Galakhov
		\footnote[2]{e-mail: d.galakhov.pion@gmail.com, galakhov@itep.ru} and  Alexei Morozov
		\footnote[3]{e-mail: morozov@itep.ru}}\\
	
	\vskip 0.2in
	\renewcommand{\thefootnote}{\roman{footnote}}
	{\small{
			\textit{
				MIPT, 141701, Dolgoprudny, Russia
				}
			\vskip 0 cm
			\textit{
				NRC “Kurchatov Institute”, 123182, Moscow, Russia
				}
			\vskip 0 cm
			\textit{
				IITP RAS, 127051, Moscow, Russia}
			\vskip 0 cm
			\textit{
				ITEP, Moscow, Russia}
	}}
\end{center}

\vskip 0.2in
\baselineskip 16pt

\centerline{ABSTRACT}

\bigskip

{\footnotesize
    In the tight binding model with multiple degenerate vacua we might treat wave function overlaps as instanton tunnelings between
    different wells (vacua).
    An amplitude for such a tunneling process might be constructed as $\mathsf{T}_{i\to j}\sim e^{-S_{\mathrm{ inst}}}{\mathbf{v}}_j^{+}{\mathbf{v}}_i^{-}$, where there is canonical instanton action suppression, and $\mathbf{v}_i^{-}$ annihilates a particle in the $i^{\mathrm{th}}$ vacuum, whereas $\mathbf{v}_j^{+}$ creates a particle in the $j^{\mathrm{th}}$ vacuum.
    Adiabatic change of the wells leads to a Berry-phase evolution of the couplings, which is described by the zero-curvature
    Gauss-Manin connection, i.e. by a quantum $R$-matrix.
    Zero-curvature is actually a consequence of level repulsion or topological protection, and its implication is the Yang-Baxter relation for the $R$-matrices.
    In the simplest case the story is pure Abelian and not very exciting.
    But when the model becomes more involved, incorporates supersymmetry, gauge and other symmetries, such amplitudes obtain more intricate structures.
    Operators $\mathbf{v}_i^{-}$, $\mathbf{v}_j^{+}$ might also evolve from ordinary Heisenberg operators into a more sophisticated algebraic object  -- a ``tunneling algebra''.
    The result for the tunneling algebra would depend strongly on geometry of the QFT we started with, and, unfortunately, at the moment we are unable to solve the reverse engineering problem.
    In this note we revise few successful cases of the aforementioned correspondence: quantum algebras $U_q(\mathfrak{g})$ and affine Yangians $Y(\hat{\mathfrak{g}})$.
    For affine Yangians we demonstrate explicitly how instantons ``perform'' equivariant integrals over associated quiver moduli spaces appearing in the alternative geometric construction.

}

\bigskip

\bigskip

\tableofcontents

\bigskip

\section{Introduction}

Over recent years the attention in the mathematical physics community to {\it hidden symmetries} in various systems
was split into two essentially different directions.
They are not truly independent and will merge in the future, still may look somewhat unsimilar at the present stage.
One direction concerns all kinds of integrability in the non-perturbative domain \cite{Morozov:1992ea,Mironov:1993wi,Morozov:1994hh,Mironov:1994sf,Gorsky:1995zq,Mironov:2022fsr},
and we do not touch it in this paper.
Another -- the one of interest for us here --
has re-incarnated into the popularity of ``categorical'' and ``non-invertible'' symmetries \cite{Bah:2022wot,Cordova:2022ruw}.
As the name suggests, these symmetries are no longer restricted to
conventional Lie (super-)groups and (super-)algebras or discrete symmetry groups.
Rather the non-perturbative and anomalous effects  deform these naive symmetries and promote them
into what one might call {\it algebraic objects} --
satisfying weaker axioms,  like groupoids,
or demonstrate new parameters responsible for deviations from the usual Lie brackets etc., like quantum algebras.
Yet, though formally deformed, these algebraic objects remember their origins and
continue to reveal themselves in just the same physical contexts: Ward identities and more relations between amplitudes,
conservation laws etc.

This paper suggests
a program to extract algebraic objects from simple physics,
e.g. the structure of quantum and Yangian algebras from a simple model of particles in the multi-well potential.
The internal structure of wells is described by representations of the algebra,
the adiabatic evolution of wells leads to the change of transition amplitudes (instantons),
which is described in terms of the Gauss-Manin (Berry) connection.
Its $P$-exponential is an $R$-matrix.
The zero-curvature condition has its origin in level repulsion, and implies Yang-Baxter relations
for the $R$-matrix.
Thus the quantum and Yangian algebras follow from the underlying {\it tunneling algebra},
which is fully controlled by the multi-particle Shr\"odinger equation in elementary quantum mechanics.

Surely, our framework has a huge linage of cousins in the modern literature: algebras of supersymmetric interfaces \cite{Gaiotto:2015aoa,Dedushenko:2021mds,Dedushenko:2023qjq,Haouzi:2023doo,Khan:2024yiy,Cordova:2024iti}, monopole algebras of 3d theories \cite{Bullimore:2016hdc,Braverman:2016wma,Bullimore:2021rnr,Garner:2023izn,Chen:2025xoe}, vertex operator algebras of class S theories \cite{Beem:2013sza,Arakawa:2018egx,Coman:2023xcq,Elliot:2024hat} and many more others.
Here for our journey the major landmark is a ``physical naturalness'' of the construction, we would make an attempt to learn how physical objects such as tunneling amplitudes, instantons ``know'' algebraic and geometric peculiarities in such a variety of contexts.

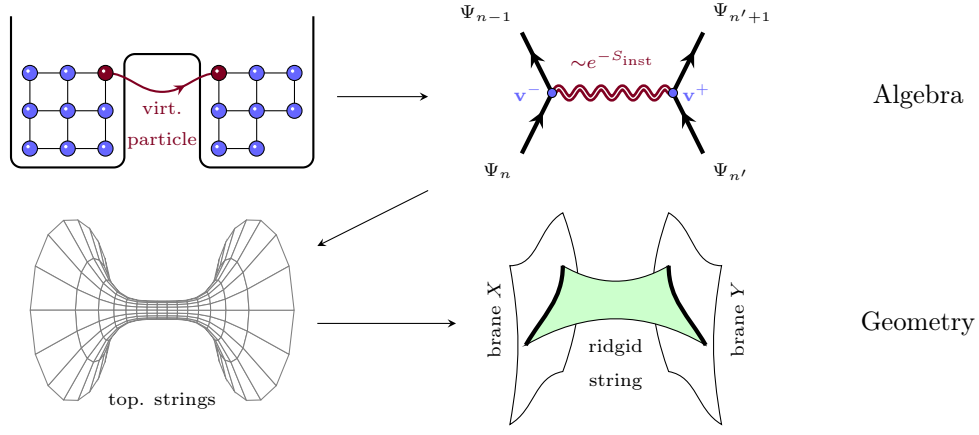
\begin{figure}[ht!]
	\centering
	\begin{tikzpicture}
		\node (E) at (0,0) {$\begin{array}{c}
				\begin{tikzpicture}[scale=0.5]
					\draw[thick, rounded corners=4] (-4,4) -- (-4,0) -- (-1,0) -- (-1,3) -- (1,3) -- (1,0) -- (4,0) -- (4,4);
					\draw[thick, burgundy, postaction={decorate}, decoration={markings, mark= at position 0.7 with {\arrow{stealth}}}] (-1.5,2.5) to[out=0,in=180] (0,2) to[out=0,in=180] (1.5,2.5);
					\begin{scope}[shift={(-3.5,0.5)}]
						\draw (0,0) -- (2,0) (0,1) -- (2,1) (0,2) -- (2,2) (0,0) -- (0,2) (1,0) -- (1,2) (2,0) -- (2,2);
						\foreach \x/\y in {0/0, 0/1, 0/2, 1/0, 1/1, 1/2, 2/0, 2/1}
						{
							\draw[fill=\myblue] (\x,\y) circle (0.2);
							\draw[fill=white, draw=\myblue] (\x-0.05,\y+0.05) circle (0.06);
						}
						\foreach \x/\y in {2/2}
						{
							\draw[fill=burgundy] (\x,\y) circle (0.2);
							\draw[fill=white, draw=burgundy] (\x-0.05,\y+0.05) circle (0.06);
						}
					\end{scope}
					\begin{scope}[shift={(1.5,0.5)}]
						\draw (0,0) -- (1,0) (0,1) -- (2,1) (0,2) -- (2,2) (0,0) -- (0,2) (1,0) -- (1,2) (2,1) -- (2,2);
						\foreach \x/\y in {0/0, 0/1, 1/0, 1/1, 1/2, 2/1, 2/2}
						{
							\draw[fill=\myblue] (\x,\y) circle (0.2);
							\draw[fill=white, draw=\myblue] (\x-0.05,\y+0.05) circle (0.06);
						}
						\foreach \x/\y in {0/2}
						{
							\draw[fill=burgundy] (\x,\y) circle (0.2);
							\draw[fill=white, draw=burgundy] (\x-0.05,\y+0.05) circle (0.06);
						}
					\end{scope}
					\node(A) [burgundy, below] at (0,2) {\scriptsize virt.};
					\node[burgundy, below] at (A.south) {\scriptsize particle};
				\end{tikzpicture}
			\end{array}$};
		\node (B) at (6,0) {$\begin{array}{c}
				\begin{tikzpicture}[scale=0.4]
					\draw[ultra thick, postaction={decorate}, decoration={markings, mark= at position 0.7 with {\arrow{stealth}}}] (-3,-2) -- (-2,0);
					\draw[ultra thick, postaction={decorate}, decoration={markings, mark= at position 0.7 with {\arrow{stealth}}}] (-2,0) -- (-3,2);
					\draw[ultra thick, postaction={decorate}, decoration={markings, mark= at position 0.7 with {\arrow{stealth}}}] (3,-2) -- (2,0);
					\draw[ultra thick, postaction={decorate}, decoration={markings, mark= at position 0.7 with {\arrow{stealth}}}] (2,0) -- (3,2);
					\draw[burgundy, line width = 2.5,decorate,decoration={snake,amplitude=2,segment length=8}] (-2,0) -- (2,0);
					\draw[white, line width = 0.5,decorate,decoration={snake,amplitude=2,segment length=8}] (-2,0) -- (2,0);
					\draw[fill=\myblue] (-2,0) circle (0.15) (2,0) circle (0.15);
					\node[\myblue, left] at (-2,0) {$\scriptstyle {\bf v}^-$};
					\node[\myblue, right] at (2,0) {$\scriptstyle {\bf v}^+$};
					\node[below left] at (-3,-2) {$\scriptstyle \Psi_{n}$};
					\node[above left] at (-3,2) {$\scriptstyle \Psi_{n-1}$};
					\node[below right] at (3,-2) {$\scriptstyle \Psi_{n'}$};
					\node[above right] at (3,2) {$\scriptstyle \Psi_{n'+1}$};
					\node[above, burgundy] at (0,0.5) {$\scriptstyle \sim e^{-S_{\rm inst}}$};
				\end{tikzpicture}
			\end{array}$};
		\node at (10,0) {Algebra};
		\node at (10,-3) {Geometry};
		\node (D) at (6,-3) {$\begin{array}{c}
				\begin{tikzpicture}[scale=0.7]
					\draw (-2,1) to[out=0,in=210] (-1,2) to[out=290,in=70] (-1,-1) to[out=190,in=30] (-2,-2) to[out=80,in=280] cycle;
					\begin{scope}[xscale=-1]
						\draw (-2,1) to[out=0,in=210] (-1,2) to[out=290,in=70] (-1,-1) to[out=190,in=30] (-2,-2) to[out=80,in=280] cycle;
					\end{scope}
					\draw[fill=white!80!green] (-1.7,-0.5) to[out=60,in=270] (-1,1) to[out=330,in=210] (1,1) to[out=270,in=120] (1.7,-0.5) to[out=150,in=30] cycle;
					\draw[ultra thick] (-1.7,-0.5) to[out=60,in=270] (-1,1) (1,1) to[out=270,in=120] (1.7,-0.5);
					\node[rotate=90] at (-2.3,0) {\scriptsize brane $X$};
					\node[rotate=90] at (2.3,0) {\scriptsize brane $Y$};
					\node(A)[below] at (0,-0.2) {\scriptsize ridgid};
					\node[below] at (A.south) {\scriptsize string};
				\end{tikzpicture}
			\end{array}$};
		\node (C) at (0,-3) {$\begin{array}{c}
				\begin{tikzpicture}[scale=0.8]
					\foreach \a/\b/\c/\d/\e/\f/\g/\h in {-0.948989/0.863717/-0.894983/0.575765/-0.850354/0.397458/-0.879728/0.581728,-0.879728/-0.581728/-0.850354/-0.397458/-0.894983/-0.575765/-0.948989/-0.863717,0.850354/-0.397458/0.879728/-0.581728/0.948989/-0.863717/0.894983/-0.575765,0.894983/0.575765/0.948989/0.863717/0.879728/0.581728/0.850354/0.397458,-1.0535/1.12718/-0.958361/0.724784/-0.894983/0.575765/-0.948989/0.863717,-0.948989/-0.863717/-0.894983/-0.575765/-0.958361/-0.724784/-1.0535/-1.12718,0.894983/-0.575765/0.948989/-0.863717/1.0535/-1.12718/0.958361/-0.724784,0.958361/0.724784/1.0535/1.12718/0.948989/0.863717/0.894983/0.575765,-1.2/1.35/-1.04/0.827001/-0.958361/0.724784/-1.0535/1.12718,-1.0535/-1.12718/-0.958361/-0.724784/-1.04/-0.827001/-1.2/-1.35,0.958361/-0.724784/1.0535/-1.12718/1.2/-1.35/1.04/-0.827001,1.04/0.827001/1.2/1.35/1.0535/1.12718/0.958361/0.724784,-1.04/0.827001/-0.88/0.497046/-0.83714/0.449695/-0.958361/0.724784,-0.958361/-0.724784/-0.83714/-0.449695/-0.88/-0.497046/-1.04/-0.827001,0.83714/-0.449695/0.958361/-0.724784/1.04/-0.827001/0.88/-0.497046,0.88/0.497046/1.04/0.827001/0.958361/0.724784/0.83714/0.449695,-0.88/0.497046/-0.72/0.30552/-0.698033/0.281702/-0.83714/0.449695,-0.83714/-0.449695/-0.698033/-0.281702/-0.72/-0.30552/-0.88/-0.497046,0.698033/-0.281702/0.83714/-0.449695/0.88/-0.497046/0.72/-0.30552,0.72/0.30552/0.88/0.497046/0.83714/0.449695/0.698033/0.281702,-0.0812557/0.144921/0.0812557/0.144921/0.08/0.150024/-0.08/0.150024,-0.08/-0.150024/0.08/-0.150024/0.0812557/-0.144921/-0.0812557/-0.144921,-0.243815/0.146781/-0.0812557/0.144921/-0.08/0.150024/-0.24/0.15192,-0.24/-0.15192/-0.08/-0.150024/-0.0812557/-0.144921/-0.243815/-0.146781,0.08/-0.150024/0.24/-0.15192/0.243815/-0.146781/0.0812557/-0.144921,0.0812557/0.144921/0.243815/0.146781/0.24/0.15192/0.08/0.150024,-0.406908/0.159455/-0.243815/0.146781/-0.24/0.15192/-0.4/0.164815,-0.4/-0.164815/-0.24/-0.15192/-0.243815/-0.146781/-0.406908/-0.159455,0.24/-0.15192/0.4/-0.164815/0.406908/-0.159455/0.243815/-0.146781,0.243815/0.146781/0.406908/0.159455/0.4/0.164815/0.24/0.15192,-0.572195/0.201071/-0.406908/0.159455/-0.4/0.164815/-0.56/0.206913,-0.56/-0.206913/-0.4/-0.164815/-0.406908/-0.159455/-0.572195/-0.201071,0.4/-0.164815/0.56/-0.206913/0.572195/-0.201071/0.406908/-0.159455,0.406908/0.159455/0.572195/0.201071/0.56/0.206913/0.4/0.164815,-0.743395/0.300008/-0.572195/0.201071/-0.56/0.206913/-0.72/0.30552,-0.72/-0.30552/-0.56/-0.206913/-0.572195/-0.201071/-0.743395/-0.300008,0.56/-0.206913/0.72/-0.30552/0.743395/-0.300008/0.572195/-0.201071,0.572195/0.201071/0.743395/0.300008/0.72/0.30552/0.56/0.206913,-0.927486/0.498227/-0.743395/0.300008/-0.72/0.30552/-0.88/0.497046,-0.88/-0.497046/-0.72/-0.30552/-0.743395/-0.300008/-0.927486/-0.498227,0.72/-0.30552/0.88/-0.497046/0.927486/-0.498227/0.743395/-0.300008,0.743395/0.300008/0.927486/0.498227/0.88/0.497046/0.72/0.30552,-0.0824227/0.125047/0.0824227/0.125047/0.0812557/0.144921/-0.0812557/0.144921,-0.0812557/-0.144921/0.0812557/-0.144921/0.0824227/-0.125047/-0.0824227/-0.125047,-0.247363/0.126676/-0.0824227/0.125047/-0.0812557/0.144921/-0.243815/0.146781,-0.243815/-0.146781/-0.0812557/-0.144921/-0.0824227/-0.125047/-0.247363/-0.126676,0.0812557/-0.144921/0.243815/-0.146781/0.247363/-0.126676/0.0824227/-0.125047,0.0824227/0.125047/0.247363/0.126676/0.243815/0.146781/0.0812557/0.144921,-0.413348/0.137787/-0.247363/0.126676/-0.243815/0.146781/-0.406908/0.159455,-0.406908/-0.159455/-0.243815/-0.146781/-0.247363/-0.126676/-0.413348/-0.137787,0.243815/-0.146781/0.406908/-0.159455/0.413348/-0.137787/0.247363/-0.126676,0.247363/0.126676/0.413348/0.137787/0.406908/0.159455/0.243815/0.146781,-0.583662/0.174469/-0.413348/0.137787/-0.406908/0.159455/-0.572195/0.201071,-0.572195/-0.201071/-0.406908/-0.159455/-0.413348/-0.137787/-0.583662/-0.174469,0.406908/-0.159455/0.572195/-0.201071/0.583662/-0.174469/0.413348/-0.137787,0.413348/0.137787/0.583662/0.174469/0.572195/0.201071/0.406908/0.159455,-1.13684/0.859765/-0.927486/0.498227/-0.88/0.497046/-1.04/0.827001,-1.04/-0.827001/-0.88/-0.497046/-0.927486/-0.498227/-1.13684/-0.859765,0.88/-0.497046/1.04/-0.827001/1.13684/-0.859765/0.927486/-0.498227,0.927486/0.498227/1.13684/0.859765/1.04/0.827001/0.88/0.497046,-0.083373/0.0918997/0.083373/0.0918997/0.0824227/0.125047/-0.0824227/0.125047,-0.0824227/-0.125047/0.0824227/-0.125047/0.083373/-0.0918997/-0.083373/-0.0918997,-0.250252/0.093111/-0.083373/0.0918997/-0.0824227/0.125047/-0.247363/0.126676,-0.247363/-0.126676/-0.0824227/-0.125047/-0.083373/-0.0918997/-0.250252/-0.093111,0.0824227/-0.125047/0.247363/-0.126676/0.250252/-0.093111/0.083373/-0.0918997,0.083373/0.0918997/0.250252/0.093111/0.247363/0.126676/0.0824227/0.125047,-0.418605/0.101382/-0.250252/0.093111/-0.247363/0.126676/-0.413348/0.137787,-0.413348/-0.137787/-0.247363/-0.126676/-0.250252/-0.093111/-0.418605/-0.101382,0.247363/-0.126676/0.413348/-0.137787/0.418605/-0.101382/0.250252/-0.093111,0.250252/0.093111/0.418605/0.101382/0.413348/0.137787/0.247363/0.126676,-0.765843/0.262909/-0.583662/0.174469/-0.572195/0.201071/-0.743395/0.300008,-0.743395/-0.300008/-0.572195/-0.201071/-0.583662/-0.174469/-0.765843/-0.262909,0.572195/-0.201071/0.743395/-0.300008/0.765843/-0.262909/0.583662/-0.174469,0.583662/0.174469/0.765843/0.262909/0.743395/0.300008/0.572195/0.201071,-0.593094/0.128807/-0.418605/0.101382/-0.413348/0.137787/-0.583662/0.174469,-0.583662/-0.174469/-0.413348/-0.137787/-0.418605/-0.101382/-0.593094/-0.128807,0.413348/-0.137787/0.583662/-0.174469/0.593094/-0.128807/0.418605/-0.101382,0.418605/0.101382/0.593094/0.128807/0.583662/0.174469/0.413348/0.137787,-0.0839948/0.0486749/0.0839948/0.0486749/0.083373/0.0918997/-0.083373/0.0918997,-0.083373/-0.0918997/0.083373/-0.0918997/0.0839948/-0.0486749/-0.0839948/-0.0486749,-0.252144/0.0493212/-0.0839948/0.0486749/-0.083373/0.0918997/-0.250252/0.093111,-0.250252/-0.093111/-0.083373/-0.0918997/-0.0839948/-0.0486749/-0.252144/-0.0493212,0.083373/-0.0918997/0.250252/-0.093111/0.252144/-0.0493212/0.0839948/-0.0486749,0.0839948/0.0486749/0.252144/0.0493212/0.250252/0.093111/0.083373/0.0918997,-0.422052/0.0537384/-0.252144/0.0493212/-0.250252/0.093111/-0.418605/0.101382,-0.418605/-0.101382/-0.250252/-0.093111/-0.252144/-0.0493212/-0.422052/-0.0537384,0.250252/-0.093111/0.418605/-0.101382/0.422052/-0.0537384/0.252144/-0.0493212,0.252144/0.0493212/0.422052/0.0537384/0.418605/0.101382/0.250252/0.093111,-0.0842112/0./0.0842112/0./0.0839948/0.0486749/-0.0839948/0.0486749,-0.0839948/-0.0486749/0.0839948/-0.0486749/0.0842112/0./-0.0842112/0.,-0.252802/0./-0.0842112/0./-0.0839948/0.0486749/-0.252144/0.0493212,-0.252144/-0.0493212/-0.0839948/-0.0486749/-0.0842112/0./-0.252802/0.,0.0839948/-0.0486749/0.252144/-0.0493212/0.252802/0./0.0842112/0.,0.0842112/0./0.252802/0./0.252144/0.0493212/0.0839948/0.0486749,-0.423253/0./-0.252802/0./-0.252144/0.0493212/-0.422052/0.0537384,-0.422052/-0.0537384/-0.252144/-0.0493212/-0.252802/0./-0.423253/0.,0.252144/-0.0493212/0.422052/-0.0537384/0.423253/0./0.252802/0.,0.252802/0./0.423253/0./0.422052/0.0537384/0.252144/0.0493212,-0.599312/0.0684281/-0.422052/0.0537384/-0.418605/0.101382/-0.593094/0.128807,-0.593094/-0.128807/-0.418605/-0.101382/-0.422052/-0.0537384/-0.599312/-0.0684281,0.418605/-0.101382/0.593094/-0.128807/0.599312/-0.0684281/0.422052/-0.0537384,0.422052/0.0537384/0.599312/0.0684281/0.593094/0.128807/0.418605/0.101382,-1.39382/1.4913/-1.13684/0.859765/-1.04/0.827001/-1.2/1.35,-1.2/-1.35/-1.04/-0.827001/-1.13684/-0.859765/-1.39382/-1.4913,1.04/-0.827001/1.2/-1.35/1.39382/-1.4913/1.13684/-0.859765,1.13684/0.859765/1.39382/1.4913/1.2/1.35/1.04/0.827001,-0.784647/0.195704/-0.593094/0.128807/-0.583662/0.174469/-0.765843/0.262909,-0.765843/-0.262909/-0.583662/-0.174469/-0.593094/-0.128807/-0.784647/-0.195704,0.583662/-0.174469/0.765843/-0.262909/0.784647/-0.195704/0.593094/-0.128807,0.593094/0.128807/0.784647/0.195704/0.765843/0.262909/0.583662/0.174469,-0.974945/0.445504/-0.765843/0.262909/-0.743395/0.300008/-0.927486/0.498227,-0.927486/-0.498227/-0.743395/-0.300008/-0.765843/-0.262909/-0.974945/-0.445504,0.743395/-0.300008/0.927486/-0.498227/0.974945/-0.445504/0.765843/-0.262909,0.765843/0.262909/0.974945/0.445504/0.927486/0.498227/0.743395/0.300008,-0.601485/0./-0.423253/0./-0.422052/0.0537384/-0.599312/0.0684281,-0.599312/-0.0684281/-0.422052/-0.0537384/-0.423253/0./-0.601485/0.,0.422052/-0.0537384/0.599312/-0.0684281/0.601485/0./0.423253/0.,0.423253/0./0.601485/0./0.599312/0.0684281/0.422052/0.0537384,-0.797215/0.104536/-0.599312/0.0684281/-0.593094/0.128807/-0.784647/0.195704,-0.784647/-0.195704/-0.593094/-0.128807/-0.599312/-0.0684281/-0.797215/-0.104536,0.593094/-0.128807/0.784647/-0.195704/0.797215/-0.104536/0.599312/-0.0684281,0.599312/0.0684281/0.797215/0.104536/0.784647/0.195704/0.593094/0.128807,-0.801639/0./-0.601485/0./-0.599312/0.0684281/-0.797215/0.104536,-0.797215/-0.104536/-0.599312/-0.0684281/-0.601485/0./-0.801639/0.,0.599312/-0.0684281/0.797215/-0.104536/0.801639/0./0.601485/0.,0.601485/0./0.801639/0./0.797215/0.104536/0.599312/0.0684281,-1.01621/0.337378/-0.784647/0.195704/-0.765843/0.262909/-0.974945/0.445504,-0.974945/-0.445504/-0.765843/-0.262909/-0.784647/-0.195704/-1.01621/-0.337378,0.765843/-0.262909/0.974945/-0.445504/1.01621/-0.337378/0.784647/-0.195704,0.784647/0.195704/1.01621/0.337378/0.974945/0.445504/0.765843/0.262909,-1.2411/0.79843/-0.974945/0.445504/-0.927486/0.498227/-1.13684/0.859765,-1.13684/-0.859765/-0.927486/-0.498227/-0.974945/-0.445504/-1.2411/-0.79843,0.927486/-0.498227/1.13684/-0.859765/1.2411/-0.79843/0.974945/-0.445504,0.974945/0.445504/1.2411/0.79843/1.13684/0.859765/0.927486/0.498227,-1.0446/0.182325/-0.797215/0.104536/-0.784647/0.195704/-1.01621/0.337378,-1.01621/-0.337378/-0.784647/-0.195704/-0.797215/-0.104536/-1.0446/-0.182325,0.784647/-0.195704/1.01621/-0.337378/1.0446/-0.182325/0.797215/-0.104536,0.797215/0.104536/1.0446/0.182325/1.01621/0.337378/0.784647/0.195704,-1.05475/0./-0.801639/0./-0.797215/0.104536/-1.0446/0.182325,-1.0446/-0.182325/-0.797215/-0.104536/-0.801639/0./-1.05475/0.,0.797215/-0.104536/1.0446/-0.182325/1.05475/0./0.801639/0.,0.801639/0./1.05475/0./1.0446/0.182325/0.797215/0.104536,-1.33851/0.625626/-1.01621/0.337378/-0.974945/0.445504/-1.2411/0.79843,-1.2411/-0.79843/-0.974945/-0.445504/-1.01621/-0.337378/-1.33851/-0.625626,0.974945/-0.445504/1.2411/-0.79843/1.33851/-0.625626/1.01621/-0.337378,1.01621/0.337378/1.33851/0.625626/1.2411/0.79843/0.974945/0.445504,-1.63155/1.48495/-1.2411/0.79843/-1.13684/0.859765/-1.39382/1.4913,-1.39382/-1.4913/-1.13684/-0.859765/-1.2411/-0.79843/-1.63155/-1.48495,1.13684/-0.859765/1.39382/-1.4913/1.63155/-1.48495/1.2411/-0.79843,1.2411/0.79843/1.63155/1.48495/1.39382/1.4913/1.13684/0.859765,-1.40955/0.346366/-1.0446/0.182325/-1.01621/0.337378/-1.33851/0.625626,-1.33851/-0.625626/-1.01621/-0.337378/-1.0446/-0.182325/-1.40955/-0.346366,1.01621/-0.337378/1.33851/-0.625626/1.40955/-0.346366/1.0446/-0.182325,1.0446/0.182325/1.40955/0.346366/1.33851/0.625626/1.01621/0.337378,-1.4358/0./-1.05475/0./-1.0446/0.182325/-1.40955/0.346366,-1.40955/-0.346366/-1.0446/-0.182325/-1.05475/0./-1.4358/0.,1.0446/-0.182325/1.40955/-0.346366/1.4358/0./1.05475/0.,1.05475/0./1.4358/0./1.40955/0.346366/1.0446/0.182325,-1.88696/1.24777/-1.33851/0.625626/-1.2411/0.79843/-1.63155/1.48495,-1.63155/-1.48495/-1.2411/-0.79843/-1.33851/-0.625626/-1.88696/-1.24777,1.2411/-0.79843/1.63155/-1.48495/1.88696/-1.24777/1.33851/-0.625626,1.33851/0.625626/1.88696/1.24777/1.63155/1.48495/1.2411/0.79843,-2.09781/0.729292/-1.40955/0.346366/-1.33851/0.625626/-1.88696/1.24777,-1.88696/-1.24777/-1.33851/-0.625626/-1.40955/-0.346366/-2.09781/-0.729292,1.33851/-0.625626/1.88696/-1.24777/2.09781/-0.729292/1.40955/-0.346366,1.40955/0.346366/2.09781/0.729292/1.88696/1.24777/1.33851/0.625626,-2.18182/0./-1.4358/0./-1.40955/0.346366/-2.09781/0.729292,-2.09781/-0.729292/-1.40955/-0.346366/-1.4358/0./-2.18182/0.,1.40955/-0.346366/2.09781/-0.729292/2.18182/0./1.4358/0.,1.4358/0./2.18182/0./2.09781/0.729292/1.40955/0.346366}
					{
						\draw[thin, rounded corners=0.05, draw=gray, fill=white] (\a,\b) -- (\c,\d) -- (\e,\f) -- (\g,\h) -- cycle;
					}
					\node at (0,-1.5) {\scriptsize top. strings};
				\end{tikzpicture}
			\end{array}$};
		\path (E) edge[-stealth] (B) (B) edge[-stealth] (C) (C) edge[-stealth] (D);
	\end{tikzpicture}
	\caption{
		Tight binding model for the tunneling algebra and geometry probes.
		We expect that the short-range interaction prevails in our picture, and when a particle leaves the potential well, it propagates under the barrier virtually, staying almost free, experiencing only the barrier background field.
		Therefore one might think of this intrinsically  non-perturbative process as a ``perturbative'' t-channel Feynman diagram with a \emph{thick} propagator $D_0(x,x')\sim e^{-S_{\rm int}}$.
		The thick propagator might be thought of as a heat kernel $(-\p_x^2+V(x)-E)D_E(x,x')=\delta(x-x')$.
		The type of the tunneling algebra is defined eventually by D-brane geometry.
		For example, to distinguish between rational, trigonometric and elliptic Yangians one should consider different compactifications (see \cite[eq.(1.1)]{Galakhov:2021vbo}).
	}\label{fig:main_story}
\end{figure}

In this program we need thorough investigation of the following points:
\begin{itemize}
 \item{} A tight binding model with multiple degenerate vacua, described by the instanton technique, or simply by the wave function overlaps.
 We choose the tight binding model as, in principle, its potential wells in the nodes may contain any physical (even strongly interacting) subsystem.
 If particle tunneling is suppressed by the potential barrier height, the tunneling amplitude introduces a new small parameter, so the perturbative methods in some incarnation are applicable to this system.

 \item{} Its algebraic description through the t-channel tunneling (instanton) amplitudes  $\mathsf{T}_{i\to j}\sim e^{-S_{\rm inst}}{\bf v}_j^{+}{\bf v}_i^{-}$,
 which generate the {\it tunneling algebra}.
 This structure of the tunneling amplitudes is expected naturally (see Fig.\ref{fig:main_story}).
 The tunneling particle propagates under the barrier, so it is virtual, its amplitude is suppressed by the Euclidean action.
 Operators ${\bf v}_{i,j}^{\pm}$ annihilate (decrease the number of) particles or create (increase the number of) particles in the corresponding potential wells.

 \item{}  Supersymmetrization, needed to separate anti-instantons from instantons (up to complete elimination of the former).
 It will also imply the  applicability of localization techniques.
 In general, we can not justify the absolute necessity of supersymmetry for this model to work.
 Yet the localization techniques ensure stability of the result, preserve it from the influence of the renormalization group.
 Surely, it would be interesting to study the latter effects separately.

 \item{}  Adiabatic variations of parameters the wells (positions in space, depths, etc.) and its description in terms of the  Berry phase and the Gauss-Manin connection \cite{manin1958algebraic,Moore:2017byz,Dedushenko:2022pem,Ferrari:2023hza}.
 The Berry connection is nearly flat in many cases due to level repulsion, and the Gauss-Manin connection (we will also refer to as a projected Berry connection in the case of supersymmetric systems) is exactly flat.
 If the parameters evolve in a such a way that in the result they are permuted on the real
 axis or intertwined in the complex plane we will call the respective evolution operator,
 given by a path ordered exponential of the Gauss-Manin/Berry connection, an R-matrix.

 \item{}  Yang-Baxter relation for $R$-matrices, provided by the $P$-exponentials of the zero-curvature Gauss-Manin connection.

 \item{}  Integration over instanton moduli spaces provides coefficients in the representation theory.
 Moduli spaces can be complex and symplectic.
 This sentence is provided by an heuristic picture depicted in Fig.\ref{fig:main_story}, where the t-channel exchange is represented by a string exchange between D-branes.
 This picture could be justified more for topological string models \cite{Aganagic:2003qj,Neitzke:2004ni}.
 For topological strings scales sometimes do not matter, and we could harden the string to its maximum rigidity, so that the path integral over its fluctuation modes probes geometry of D-branes instead.

 \item{}  Localization technique reduces integrals to particular points, what can be considered as reduction from
 fluctuating string wondering along whole $D$-brane world-volumes to a rather tight ``stick'' between the closest points.

 \item{}  Examples for various algebras emerging as tunneling algebras.
 Here for simplicity  we concentrate in detail solely on examples of quantum algebra $U_q(\fg)$ and affine Yangians $Y(\hat{\fg})$.
 Undoubtedly, this list could be continued far beyond.
\end{itemize}
This paper is just a sketch of such investigation,
much work remains to be done.

This paper is organized as follows.
In Sec.\ref{sec:tunnel} we define the tunneling algebras and consider a few simple examples.
Since we are planning to incorporate supersymmetry and Berry transport in our quantum mechanical model in Sec.\ref{sec:dynSUSY} we review a simple realization of the model with non-stationary supercharges.
In Sec.\ref{sec:QA} we consider supersymmetric Landau-Ginzburg quantum mechanics and confirm that for a specific model whose superpotential is related to WZWN conformal blocks the tunneling algebra is quantum algebra $U_q(\fg)$.
In Sec.\ref{sec:Yangians} we consider supersymmetric quantum mechanics with a quiver target space (sigma-model), and in the particular case when the quiver variety is a Hilbert scheme of points in $\IC^2$ we extract the tunneling algebra to be affine Yangian $Y(\widehat{\fg\fl}_1)$.


\section{Tunneling algebra}\label{sec:tunnel}

\subsection{Motivation}

The basic object of our interest in this note is an algebra we call a \emph{tunneling algebra}, and a possibility to associate to it an R-matrix solution to \emph{Yang-Baxter equations}.

We are motivated to introduce such an object by an attempt to extend the usual concepts of weakly coupled field theories such as the Fock space and a system of Heisenberg creation-annihilation operators to strongly coupled systems.
Surely, in truly strongly coupled systems neither the concept of a particle, nor operations of creating and annihilating free particles, do not make sense.
Yet it is not hard to number families of strongly coupled systems having quantized charge operators at which one might look as at colored particle numbers.
Respectively, it seems to be a natural desire to introduce operators acting on wave functions and shifting eigen values of those charges.

As a model example of such a system one might have in mind some matter in its crystal phase.
The role of particles is played by atoms in the crystal lattice, they are manifestly indivisible elementary objects, however one can not call such a system weakly coupled, as the mutual interaction of atoms forces them to occupy distinguished positions in the lattice nodes.
Neither, there is no well-defined Fock space where we might have united wave functions for crystals consisting of any number of atoms.
Nevertheless if this crystal is put in contact with liquid in the melting point we could easily imagine as an analog of ``creation/annihilation'' operators is played by processes of gluing an atom from the liquid bulk to the crystal body and, vice versa, melting an atom away.

We would like to divide the motivation for our construction in the following scheme:
\begin{enumerate}
	\item The first obvious issue in this route is that the strongly coupled Hamiltonian is naively diagonal in the particle number, so there is no ``natural'' way to organize a lack or excess of particles.
	\item However if there are at least two potential wells and tunneling between them, the Hamiltonian is no longer diagonal in the particle number for a single well.
	The off-diagonal corrections are small and are controlled by the tunneling amplitude as a small parameter.
	\item Concentrating on one of the wells we will introduce operators ${\bf v}^+$ and ${\bf v}^-$ respectively adding a particle that tunneled from another well and subtracting a particle that tunneled away.
	These are our new ``creation/annihilation'' operators and the algebra they form (probably with a mixture of other local or non-local operators) we call a \emph{tunneling algebra}.
	\item Even in simple examples we will see that the resulting algebra is not necessarily a clone of the (super) Heisenberg algebra.
	This raises natural questions about properties of the tunneling algebras, their classification and a relation to the properties, symmetries of the original system.
	However in this note we are not even trying to chase an attempt to describe all the variety of the tunneling algebras in general, rather we would try to describe just a few examples.
	\item Now having constructed the tunneling algebra one might try to rewrite the Hamiltonian in these terms.
	The Hamiltonian and the evolution operator would contain tunneling terms as $U\sim\bbone+{\bf v}_i^+{\bf v}_j^-+\ldots$ that correspond to a particle hopping from well $j$ to well $i$.
	This form is quite reminiscent of the R-matrix solutions for other algebras to Yang-Baxter equations.
	This observation raises the following questions:
	\begin{enumerate}
		\item Are evolution operators in our construction \emph{R-matrices} in some new or studied integrable system?
		\item If so, does this construction of an R-matrix make it \emph{automatically} compatible with the tunneling algebra?
		Does the tunneling algebra have a universal Hopf algebra structure and an associated R-matrix?
	\end{enumerate}
	\item It seems to be natural to expect that if some mechanism of connection of the tunneling algebras and R-matrices exists it originates in the Berry connection.
	Characteristic elements we usually assign to a system of Yang-Baxter equations seem to be already implemented:
	\begin{enumerate}
		\item A structure of the tensor product: we might treat the natural wave function product rule for weakly interacting systems (separated by a potential wall in our case) as a natural tensor product.
		\item R-matrices mixing, intertwining vectors from various tensor product components: tunneling processes are supposed to do the job.
		\item Yang-Baxter (YB) equation: the Berry connection is mostly flat (or exactly flat if supersymmetry is involved) implying that parallel transports of the system along two different yet homotopic paths are equivalent.
		This constraint imposes relations on evolution operators that we might treat as YB equations eventually.
	\end{enumerate}
\end{enumerate}

Yet, unfortunately, one should not expect from this construction a universal way to construct the tunneling algebra and associated R-matrices solving the YB equations for it.
The presence of quantum tunnels allows a particle to just spread its position between wells with some probabilistic measure.
And the adiabatic parameter variation necessary for the Berry connection to be valid implies that non-degenerate Hamiltonian eigen functions remain at their energy levels, that do not intersect due to level repulsion in quantum mechanics.
One of resolution for this situation is to incorporate supersymmetry.
It is well-known that supersymmetric theories are related to integrability\footnote{This is an extremely broad subject in the modern literature.
Here we solely indicate some directions \cite{Gorsky:1995zq, Nekrasov:2009ui, Gaiotto:2013bwa, Awata:2017lqa, Yamazaki:2018xbx, Costello:2017dso,Koroteev:2021lvp}.} and there was substantial success in studying projected Berry connection to supercharge cohomologies \cite{Moore:2017byz,Dedushenko:2022pem,Ferrari:2023hza} including applications to integrability.
In this note the majority of our attention would be devoted to the latter case.
However even without supersymmetry we managed to construct a simple model where adiabatic evolution produces non-trivial R-matrices (see Sec.\ref{sec:TowYBeq} for details).

\subsection{Tunneling algebra construction}\label{sec:TAC}

We assume that our system is represented by $n$ mutually interacting particles put in an external potential, we could schematize a Hamiltonian of such a system in the following way:
\begin{equation}
    H=-\sum\lm_{i=1}^n\vec{\nabla}_i^2+\sum\lm_{1\leq i<j\leq n} u\left(\vec{x}_i-\vec{x}_j\right)+\sum\lm_{i=1}^n V(\vec{x}_i)\,.
\end{equation}
Such a Hamiltonian acts on $n$-particle Hilbert space $\mathscr{H}_n$ represented by functions of $n$ vector variables.
Furthermore we assume that the external potential $V$ consists of $N$ deep potential wells.
In general, we would like to consider families of theories with different $N$ and well positions.

The wells are assumed to be deep enough so that the tunneling processes are suppressed, then one would be able to approximate wave functions and respective Hilbert spaces in the $N$-well potential as being decomposed over Hilbert spaces in the 1-well potential.
Say, for two wells we call $A$ and $B$ we have the following decomposition:\footnote{Zero-particle Hamiltonian $H=0$, and the Hilbert space is 1-dimensional $\mathscr{H}_0=\IC$.}
\begin{equation}\label{pwf}
    \mathscr{H}_n(A,B)\approx\bigoplus\lm_{k=0}^n \mathscr{H}_k(A)\otimes \mathscr{H}_{n-k}(B)\,.
\end{equation}

As an ultimate goal for this section we would like to construct an algebra of raising/lowering (``creation/annihilation'') operators acting on 1-well Hilbert spaces and shifting $n$:
\begin{equation}\label{tunneling}
    {\bf v}^+:\;\;\mathscr{H}_n(A)\longrightarrow \mathscr{H}_{n+1}(A),\quad {\bf v}^-:\;\;\mathscr{H}_n(A)\longrightarrow \mathscr{H}_{n-1}(A)\,.
\end{equation}

The construction of the algebra goes as follows:
\begin{enumerate}
    \item Consider well $A$ of our interest, and another probe well $P$.
	Let us denote the wave functions from $\mathscr{H}_0(P)$ and $\mathscr{H}_1(P)$ when there are no particles and there is a particle in well $P$ as $|0\rangle$ and $|1\rangle$ respectively.
    \item Consider Euclidean evolution operator $e^{-\tau H}$.
	Our product wave functions \eqref{pwf} are not eigen in general for Hamiltonian $H$, and we expect that the evolution operator contains operators performing tunneling processes mapping $\mathscr{H}_0(P)\otimes\mathscr{H}_{n+1}(A)$ into $\mathscr{H}_1(P)\otimes\mathscr{H}_{n}(A)$.
    \item Similarly to how one defines partial trace of the density matrix during entanglement entropy calculation we construct our operators  by taking partial correlations:
    \begin{tcolorbox}
    \begin{equation}\label{algebra}
	{\bf v}^+={}_P\langle 0|\;e^{-\tau H}\;|1\rangle_P,\quad {\bf v}^-={}_P\langle 1|\;e^{-\tau H}\;|0\rangle_P\,.
    \end{equation}
    \end{tcolorbox}
\end{enumerate}

A few comments are in order:
\begin{itemize}
    \item We did not have a particular reason to select the Euclidean evolution operator among other operators, simply in examples we will be usually interested in the IR dynamics of the systems in questions, and the Euclidean evolution suppresses high energies naturally.
    \item Naturally, we will be interested in inserting into correlators other operators and assigning to particles internal degrees of freedom like color.
	This would modify operators ${\bf v}^{\pm}$ to have extra indices.
    \item For a system admitting free particles flying to spacial infinity we could have considered deep inelastic scattering process of a light particle $P$ on a heavy system $A$ rather than tunneling.
    The resulting algebra would be defined similarly to so called BPS algebras \cite{Harvey:1995fq} via a residue of the S-matrix at the bound state pole.
    However the latter method seems inapplicable in the cases when there are no free particles or the S-matrix is hard to handle for some reason.
    Our tunneling proposal could be considered as a scattering alternative in those cases.
    \item A rather famous method to estimate the tunneling amplitudes is the instanton calculus \cite{Vainshtein:1981wh,Coleman:1978ae}.
	These methods are conventionally referred to as non-perturbative methods, which makes our algebra a close cousin of other fusion algebras for non-perturbative defects appearing in considerations of non-invertible symmetries (see a review of this subject in \cite{Cordova:2022ruw}) and those algebras arising from topological objects in supersymmetric theories (this subject is vastly broad, see some directions in \cite{Gaiotto:2015aoa, Bullimore:2016hdc, Dedushenko:2021mds, Dedushenko:2023qjq}, see also a comparison of Yangian algebras we discuss in this note and monopole algebras in \cite{Chen:2025xoe}).
    \item Our algebra setup \eqref{algebra} is rather generic and, therefore, rather far from a definition including various peculiarities.
    As we mentioned before a related structures of BPS algebras in solely string theory \cite{Harrison:2021gnp} incorporates countless possibilities, and there is no reason to expect from the tunneling algebra to be more robust in this sense.
    As well we could not estimate completeness for construction \eqref{algebra}.
    Depending on the structure of the theory in question one might need to complete this structure with other operators.
    For example, these operators might be just local operators, cf. fusion of monopole and anti-monopole operators in supersymmetric 3d theories (see e.g.\cite{Seiberg:1994pq}).
    As well particles in our picture might start clustering, so that tunneling of a multi-particle cluster is not decomposed as tunneling of cluster constituents separately.
    In this case one would be forced to introduce higher tunneling operators.
\end{itemize}

Before discussing examples of the tunneling algebra use let us revise the attempt of constructing Fock spaces in this context.
We could simply define:
\begin{equation}
    \mathscr{F}(A):=\bigoplus\lm_n \mathscr{H}_n(A)\,.
\end{equation}
The action of ${\bf v}^{\pm}$ extends to $\mathscr{F}$ naturally.
If the wells are deep enough the tunneling  between well $A$ and well $B$ is suppressed by the Euclidean (instanton) action $S_{AB}$ of a tunneling particle in the inverse potential.
We could construct tunneling corrections to the Hamiltonian in a complete analogy with the tight binding model in the case of multiple wells:
\begin{equation}\label{instantons}
    H=H_0+\sum\lm_{A\neq B}{\bf v}^{+}_A \Delta_{AB} e^{-S_{AB}} {\bf v}_B^-+\ldots\,,
\end{equation}
where $H_0$ is a sum of effective Hamiltonians for one-well systems.
Then substituting this expression back into \eqref{algebra} we arrive to a self-consistency up to unimportant numerical multipliers:
\begin{equation}
\begin{aligned}
    & {}_P\langle 0|\;e^{-\tau H}\;|1\rangle_P=\left[\tau\,e^{-\tau E_0-S_{AP}}\Delta_{AP}\times\langle 0|{\bf v}_P^-|1\rangle\right]\times{\bf v}_A^+\,,\\
    & {}_P\langle 1|\;e^{-\tau H}\;|0\rangle_P=\left[\tau\,e^{-\tau E_0-S_{AP}}\Delta_{AP}\times\langle 1|{\bf v}_P^+|0\rangle\right]\times{\bf v}_A^-\,,
\end{aligned}
\end{equation}
where $E_0$ is an unperturbed eigen value of $H_0$.
This setting allows us to treat the tunneling algebra as a ``square root'' of the algebra of tunneling operators.

Let us also indicate quickly another handy formula that allows one to define matrix coefficients of the tunneling algebra via instanton amplitudes.
An instanton is a trajectory supporting the addle point contribution to the path integral representation of the Euclidean evolution operator $e^{-\tau H}$.
Sandwiching this operator with states in a two-well problem and expanding the evolution operator up to the first order we derive:
\begin{tcolorbox}
\begin{equation}\label{sandwitch}
    \langle \tilde\Psi|e^{-\tau H}|\Psi\rangle=\langle \tilde\psi_A|{\bf v}_A^+|\psi_A\rangle\times\langle \tilde\psi_B|{\bf v}_B^-|\psi_B\rangle\times\left(\tau\Delta_{AB}\right)e^{-S_{AB}-E_0\tau}\,.
\end{equation}
\end{tcolorbox}
We will apply this formula to determine matrix coefficients of the tunneling algebra operators in the cases when it is easier to calculate the instanton amplitude in the l.h.s. with the help of various tricks like supersymmetric localization.

\subsection{Warm-up examples}
In the rest of this section we consider just a few simple examples illustrating simple constructions of the tunneling algebra for various setups: Heisenberg algebra, $\fs\fl_2$ and $\fs\fl_3$.

\subsubsection{Heisenberg algebra}

The Heisenberg algebra emerges in the form of the tunneling algebra in a natural way as an algebra of creation/annihilation operators in the second quantization formalism applied to a system of particles where the mutual interaction is negligible \cite{abrikosov2012methods}.

Let us start with a two-well potential in 1d and a single particle.
We will not pick among particular models, let us solely assume that the wells are symmetric with respect to $x=0$.
In this situation it is well known that two lowest energy levels correspond to symmetric wave function $\psi_+(x)$ without zeroes and to anti-symmetric wave function $\psi_-(x)$ with a single zero respectively:
\begin{equation}
    H\psi_+=(\epsilon_0-\delta)\psi_+,\quad H\psi_-=(\epsilon_0+\delta)\psi_-,\quad \epsilon_0>\delta>0\,.
\end{equation}
If the wells are rather deep, level energy splitting $\delta$ is rather small and can be estimated via, say, instanton calculus \cite{Vainshtein:1981wh} or uniform WKB methods \cite{Dunne:2014bca}.
Out of these two functions one could construct other wave functions representing a particle concentrated in either the left or the right well.
These functions are no longer eigen function of the Hamiltonian:
\begin{equation}
    \varphi_1(x):=\frac{\psi_+(x)+\psi_-(x)}{\sqrt{2}},\quad \varphi_2:=\frac{\psi_+(x)-\psi_-(x)}{\sqrt{2}},\quad H\left(\begin{array}{c}\varphi_1\\ \varphi_2\end{array}\right)=\left(\begin{array}{cc} \epsilon_0 & -\delta\\ -\delta &\epsilon_0\end{array}\right)\left(\begin{array}{c}\varphi_1\\ \varphi_2\end{array}\right)\,.
\end{equation}

Now consider $n+1$ particles put in these two wells and assume that the first well is well $A$ of our interest and the second well is the probe well $P$.
We could explicitly describe the situations when all the particles are put in well $A$ and one particle migrated to well $P$ by one-dimensional Hilbert spaces with respectively symmetrized wave functions:
\begin{equation}
\begin{aligned}
    &\Psi_{(n+1,0)}=\varphi_A(x_1)\varphi_A(x_2)\cdots\varphi_A(x_{n+1})\,,\\
    &\Psi_{(n,1)}=\frac{1}{\sqrt{n+1}}\left(\varphi_P(x_1)\varphi_A(x_2)\cdots\varphi_A(x_{n+1})+\varphi_A(x_1)\varphi_P(x_2)\cdots\varphi_A(x_{n+1})+\ldots+\varphi_A(x_1)\cdots\varphi_P(x_{n+1})\right)\,.
\end{aligned}
\end{equation}

We calculate the Euclidean evolution operator up to the first order in $\delta$:
\begin{equation}
    e^{-\tau H}\left(\begin{array}{c}
	\Psi_{(n+1,0)}\\ \Psi_{(n,0)}
	\end{array}\right)=\left(\begin{array}{cc}
	    e^{-\tau (n+1) \epsilon_0} & -e^{-\tau (n+1) \epsilon_0}\tau \delta \sqrt{n+1}\\
	    -e^{-\tau (n+1) \epsilon_0}\tau \delta \sqrt{n+1} & e^{-\tau (n+1) \epsilon_0}
    \end{array}\right)\left(\begin{array}{c}
	\Psi_{(n+1,0)}\\ \Psi_{(n,0)}
    \end{array}\right)\,.
\end{equation}
It is clear that up to insufficient numerical factors including average energy eigenvalue exponent and the tunneling coefficient the off diagonal elements of this matrix represent exactly operators shifting the filling numbers of 1d Hilbert spaces $\mathscr{H}_n(A)$.
If we denote the vector spanning 1d space $\mathscr{H}_n(A)$ as $|n\rangle$ then the matrix elements of our tunneling algebra generators are a representation of the Heisenberg algebra:
\begin{equation}
    {\bf v}^+|n\rangle=\sqrt{n+1}\;|n+1\rangle, \quad {\bf v}^-|n\rangle=\sqrt{n}\;|n-1\rangle,\quad \left[{\bf v}^-,{\bf v}^+\right]=1\,.
\end{equation}

\subsubsection{Algebra $\fs\fl_2$}

Let us revise the previous problem, however in the opposite regime when the mutual interaction is very strong and repulsive.
In this case multi-particle states in a single well have a very high energy.
If this energy level is much larger than $\epsilon_0$ a contribution of such states in the evolution operator is suppressed.
We could truncate effectively the Fock spaces to contain only 0-particle and 1-particle components:
\begin{equation}
    \mathscr{F}(A)=\mathscr{H}_0(A)\oplus \mathscr{H}_1(A)\,.
\end{equation}
In this case we find:
\begin{equation}
    {\bf v}^+|0\rangle=|1\rangle,\quad {\bf v}^+|1\rangle=0,\quad {\bf v}^-|0\rangle=0,\quad {\bf v}^-|1\rangle=|0\rangle\,.
\end{equation}
In this case the representation is not of the Heisenberg algebra rather of $\fs\fl_2$:
\begin{equation}
    {\bf v}^-=e=\left(\begin{array}{cc}
	0 & 1\\
	0 & 0\\
    \end{array}\right),\quad {\bf v}^+=f=\left(\begin{array}{cc}
	0 & 0\\
	1 & 0\\
	\end{array}\right),\quad h=\left[e,f\right]=\left(\begin{array}{cc} 1 & 0\\ 0 & -1\end{array}\right),\quad \left[h,e\right]=2e,\quad\left[h,f\right]=-2f\,.
\end{equation}
As we see in this example there is no guarantee that our procedure would produce all the generators at once (generator $h$ in this case).
One might need to extend the algebra of ${\bf v}^{\pm}$ by commutators and other operators not switching the particle numbers.

\subsubsection{Algebra $\fs\fl_3$}

Let us again modify the previous problem in the following way.
Introduce two particle colors: 1 and 2.
We leave the mutual interaction of particles of the same color large and repulsive, whereas the interaction of particles of opposite colors mild and attractive.
Wells also acquire colors, so that a particle ``sees'' a well of the same color.

Consider a well of color 1.
There could be no particles at all, let us denote the respective wave function as $|0\rangle$.
We could add there a single particle of color 1 via tunneling, let us denote this state as ${\bf v}_1^+|0\rangle$.
A particle of color 2 does not ``see'' potential well of color 1, therefore it can not tunnel to that position.
It implies that ${\bf v}_2^+$ annihilates $|0\rangle$, ${\bf v}_2^{+}|0\rangle=0$.
However if the well of color 1 contains a particle of color 1 it creates an effective attracting potential for particle of color 2.
This particle could tunnel to the new effective well, this process produces a new state ${\bf v}_2^+{\bf v}_1^+|0\rangle$.
Similarly, if the well contains particles of both colors, a particle of color 1 can not tunnel away from this well as this would destroy the bound state, therefore ${\bf v}_1^-{\bf v}_2^+{\bf v}_1^+|0\rangle=0$.

This reasoning indicates that in this case the representation of the tunneling algebra is isomorphic to the fundamental representation of $\fs\fl_3$:
\begin{equation}
    {\bf v}_1^-=e_1=\left(\begin{array}{ccc}
	0 & 1 & 0\\
	0 & 0 & 0\\
	0 & 0 & 0\\
    \end{array}\right),\;
    {\bf v}_2^-=e_2=\left(\begin{array}{ccc}
	0 & 0 & 0\\
	0 & 0 & 1\\
	0 & 0 & 0\\
    \end{array}\right),\;
    {\bf v}_1^+=f_1=\left(\begin{array}{ccc}
	0 & 0 & 0\\
	1 & 0 & 0\\
	0 & 0 & 0\\
    \end{array}\right),\;
    {\bf v}_2^+=f_2=\left(\begin{array}{ccc}
	0 & 0 & 0\\
	0 & 0 & 0\\
	0 & 1 & 0\\
    \end{array}\right).
\end{equation}
As in the previous case we should complete them with Cartan operators:
\begin{equation}
    h_1=\left[e_1,f_1\right]=\left(\begin{array}{ccc}
	1 & 0 & 0\\
	0 & -1 & 0\\
	0 & 0 & 0\\
    \end{array}\right),\quad h_2=\left[e_2,f_2\right]=\left(\begin{array}{ccc}
	0 & 0 & 0\\
	0 & 1 & 0\\
	0 & 0 & -1\\
    \end{array}\right)\,.
\end{equation}
If we had started with a well of color 2 we would arrive to the canonical form of the anti-fundamental representation of $\fs\fl_3$.

\subsection{Towards Yang-Baxter equation}\label{sec:TowYBeq}

The Yang-Baxter originates \cite{Jimbo:1989qm} in the statistical mechanics problems as a self-consistency of factorization for scattering matrices (R-matrices) in the Bethe ansatz technique.
Pictorially this relation is represented as a ``star-triangle'' relation, where we have two ways to resolve a triple intersection of three non-parallel lines in three pairwise intersections.
Each pairwise intersection is treated as a rank-4 tensor -- R-matrix, and the diagrams prescribe how their indices should be convoluted.
For the quantum algebras related to the knot theory R-matrices are promoted to a representation of the braid group, and the Young-Baxter equation is depicted now as a braid relation:
\begin{equation}
	\begin{array}{c}
		\begin{tikzpicture}
			\node (A) at (0,0) {$\begin{array}{c}\begin{tikzpicture}[scale=0.8]
						\draw[thick] (0,0.5) -- (2,0.5);
						\draw[thick] (0,0) to[out=60,in=180] (2,1);
						\draw[thick] (0,1) to[out=0,in=120] (2,0);
						\node[left] at (0,0) {$\scriptstyle h_1$};
						\node[left] at (0,0.5) {$\scriptstyle h_2$};
						\node[left] at (0,1) {$\scriptstyle h_3$};
						\draw[-stealth] (-0.7,-0.5) -- (2.5,-0.5);
						\draw[-stealth] (-0.7,-0.5) -- (-0.7,1.3);
						\node[right] at (2.5,-0.5) {$\scriptstyle t$};
						\node[above] at (-0.7,1.3) {$\scriptstyle h$};
				\end{tikzpicture}\end{array}$};
			\node (B) at (4,0) {$\begin{array}{c}\begin{tikzpicture}[scale=0.8]
						\draw[thick] (0,0.5) -- (2,0.5);
						\draw[thick] (0,0) to[out=0,in=240] (2,1);
						\draw[thick] (0,1) to[out=300,in=180] (2,0);
						\node[left] at (0,0) {$\scriptstyle h_1$};
						\node[left] at (0,0.5) {$\scriptstyle h_2$};
						\node[left] at (0,1) {$\scriptstyle h_3$};
						\draw[-stealth] (-0.7,-0.5) -- (2.5,-0.5);
						\draw[-stealth] (-0.7,-0.5) -- (-0.7,1.3);
						\node[right] at (2.5,-0.5) {$\scriptstyle t$};
						\node[above] at (-0.7,1.3) {$\scriptstyle h$};
				\end{tikzpicture}\end{array}$};
			\draw[stealth-stealth] (1.8,0) -- (2.2,0) node[pos=0.5,above] {\tiny homotopy};
		\end{tikzpicture}
	\end{array},\quad \begin{array}{c}
	\begin{tikzpicture}
		\node (A) at (0,0) {$\begin{array}{c}\begin{tikzpicture}[scale=0.8]
					\draw[thick] (0,0.5) -- (2,0.5);
					\draw[line width=1mm, white] (0,0) to[out=60,in=180] (2,1);
					\draw[thick] (0,0) to[out=60,in=180] (2,1);
					\draw[line width=1mm, white] (0,1) to[out=0,in=120] (2,0);
					\draw[thick] (0,1) to[out=0,in=120] (2,0);
					\node[left] at (0,0) {$\scriptstyle h_1$};
					\node[left] at (0,0.5) {$\scriptstyle h_2$};
					\node[left] at (0,1) {$\scriptstyle h_3$};
					\draw[-stealth] (-0.7,-0.5) -- (2.5,-0.5);
					\draw[-stealth] (-0.7,-0.5) -- (-0.7,1.3);
					\node[right] at (2.5,-0.5) {$\scriptstyle t$};
					\node[above] at (-0.7,1.3) {$\scriptstyle h$};
			\end{tikzpicture}\end{array}$};
		\node (B) at (4,0) {$\begin{array}{c}\begin{tikzpicture}[scale=0.8]
					\draw[thick] (0,0.5) -- (2,0.5);
					\draw[line width=1mm, white] (0,0) to[out=0,in=240] (2,1);
					\draw[thick] (0,0) to[out=0,in=240] (2,1);
					\draw[line width=1mm, white] (0,1) to[out=300,in=180] (2,0);
					\draw[thick] (0,1) to[out=300,in=180] (2,0);
					\node[left] at (0,0) {$\scriptstyle h_1$};
					\node[left] at (0,0.5) {$\scriptstyle h_2$};
					\node[left] at (0,1) {$\scriptstyle h_3$};
					\draw[-stealth] (-0.7,-0.5) -- (2.5,-0.5);
					\draw[-stealth] (-0.7,-0.5) -- (-0.7,1.3);
					\node[right] at (2.5,-0.5) {$\scriptstyle t$};
					\node[above] at (-0.7,1.3) {$\scriptstyle h$};
			\end{tikzpicture}\end{array}$};
		\draw[stealth-stealth] (1.8,0) -- (2.2,0) node[pos=0.5,above] {\tiny homotopy};
	\end{tikzpicture}
	\end{array}\,.
\end{equation}
We would like to treat both these pictures as homotopy equivalences of paths $h_i(t)$ in the parameter spaces, when $h_i(t)$ are real or complex correspondingly.
The Yang-Baxter equation emerges in our situation in the following way:
\begin{enumerate}
	\item The system is parallel transported along a trajectory in the parameter space.
	Moreover, the transport is flat.
	\item Due to a motion in the parameter space we might say that the system is out of equilibrium.
	So in addition to usual constant smearing of the wave function between vacua tunneling rate might depend on the position in the parameter space.
	Therefore even homotopic, yet distinct  paths for parallel transport prescribe different factorizations for the transport amplitudes.
	\item It is natural to expect the tunneling amplitudes grow due to some specific ``catalytic'' disposition of parameters that we depict as an intersection.
	Then factorization equivalence for two homotopic paths leads to the Yang-Baxter equation.
\end{enumerate}

Let us illustrate this principle in the following simple model.
Consider a situation of $n$ delta-function shaped wells.
The respective Schr\"odinger equation reads:
\begin{equation}
	\left[-\frac{d^2}{dx^2}-\sum\lm_{i=1}^n\kappa_i\delta(x-s_i)\right]\psi(x)=-p^2\psi(x)\,.
\end{equation}

We search for the wave function solution to this equation as a piece-wisely smooth function:
\begin{equation}
	\begin{array}{c}
		\begin{tikzpicture}
			\node[above] at (-5,0) {$\scriptstyle\psi(x)=c_i^+e^{px}+c_i^-e^{-px}$};
			\node[above] at (0,0) {$\scriptstyle\psi(x)=c_{i+1}^+e^{px}+c_{i+1}^-e^{-px}$};
			\node[above] at (5,0) {$\scriptstyle\psi(x)=c_{i+2}^+e^{px}+c_{i+2}^-e^{-px}$};
			\draw[-stealth] (-7,0) -- (7,0);
			\node[right] at (7,0) {$\scriptstyle x$};
			\draw (-2.5,-0.5) -- (-2.5,0.5) (2.5,-0.5) -- (2.5,0.5);
			\begin{scope}[shift={(-2.5,0)}]
				\draw[thick, \myblue] (-0.2,0) to[out=0,in=180] (0,-0.4) to[out=0,in=180] (0.2,0);
				\node[above left] {$\scriptstyle s_i$};
				\node[left, \myblue] at (-0.1,-0.3) {$\scriptstyle\kappa_i$};
			\end{scope}
			\begin{scope}[shift={(2.5,0)}]
				\draw[thick, \myblue] (-0.2,0) to[out=0,in=180] (0,-0.4) to[out=0,in=180] (0.2,0);
				\node[above left] {$\scriptstyle s_{i+1}$};
				\node[left, \myblue] at (-0.1,-0.3) {$\scriptstyle\kappa_{i+1}$};
			\end{scope}
		\end{tikzpicture}
	\end{array}
\end{equation}

Coefficients in gaps between wells are related with the following transfer matrix:
\begin{equation}\label{transfer}
	\left(\begin{array}{c}
		c_{i+1}^+ \\ c_{i+1}^-
	\end{array}\right)=
	\left(\begin{array}{cc}
		1-\frac{\kappa_i}{2p} & -\frac{\kappa_i}{2p}e^{-2 p s_i}\\
		\frac{\kappa_i}{2p}e^{2 p s_i} & 1+\frac{\kappa_i}{2p}
	\end{array}\right)
	\left(\begin{array}{c}
		c_{i}^+ \\ c_{i}^-
	\end{array}\right)\,.
\end{equation}
A requirement for $\psi(x)$ to belong to $L^2(\IR)$ imposes the following boundary conditions $c_1^-=c_{n+1}^+=0$.

Modulo higher exponential terms this set of equations have a solution if $p$ satisfies the following equation:
\begin{equation}
	(2p-\kappa_1)\cdot(2p-\kappa_2)\cdot\ldots\cdot(2p-\kappa_n)\approx 0\,,
\end{equation}
Apparently, a solution $p\approx\kappa_i/2$ when parameters $\kappa_i$ are separated far apart corresponds to a particle localized in well $i$.
Probability to find a particle in any other well is exponentially small.

Now we could consider a problem with time-dependent coupling constants $\kappa_i(t)$.
Clearly nothing interesting happens unless the levels overlap, i.e. unless there is a critical time moment $t_*$ when $\kappa_i(t_*)=\kappa_j(t_*)$ for some $i$ and $j$.
Although even in this case respective solution $p_i(t_*)$ never intersects $p_j(t_*)$ due to the canonical level repulsion \cite{Landau:1991wop}.
Let us concentrate for a moment on the two well case, when two delta-wells are located at $s_1=-\frac{s}{2}$, $s_2=\frac{s}{2}$ and have coupling constants $\kappa_1=\kappa-\Delta \kappa$, $\kappa_2=\kappa+\Delta\kappa$.

\begin{figure}[ht!]
	\centering
	\begin{tikzpicture}[scale=0.7]
		\draw[-stealth] (-9,0) -- (8,0);
		\draw[-stealth] (-9,0) -- (-9,3.2);
		\node[above] at (-9,3.2) {\footnotesize couplings, momenta};
		\node[right] at (8,0) {$\scriptstyle t$};
		\begin{scope}[shift={(-7,-2)}]
			\begin{scope}[scale=0.5]
				\draw [-stealth] (-3,0) -- (3,0);
				\draw [-stealth] (0,-1.5) -- (0,1.5);
				\draw [ultra thick] (-1,-0.1) -- (-1,0.1) (1,-0.1) -- (1,0.1);
				\node [above] at (0,1.5) {$\scriptstyle{\color{burgundy}\psi_-}$};
				\node [right] at (3,0) {$\scriptstyle x$};
				\draw[thick, burgundy] (-3,0) to[out=0,in=225] (-1,1) to[out=315,in=170] (1,-0.1) to[out=10,in=180] (3,0);
			\end{scope}
		\end{scope}
		\begin{scope}[shift={(-3.5,-2)}]
			\begin{scope}[scale=0.5]
				\draw [-stealth] (-3,0) -- (3,0);
				\draw [-stealth] (0,-1.5) -- (0,1.5);
				\draw [ultra thick] (-1,-0.1) -- (-1,0.1) (1,-0.1) -- (1,0.1);
				\node [above] at (0,1.5) {$\scriptstyle{\color{burgundy}\psi_-}$};
				\node [right] at (3,0) {$\scriptstyle x$};
				\draw[thick, burgundy] (-3,0) to[out=0,in=225] (-1,0.9) to[out=315,in=160] (1,-0.4) to[out=20,in=180] (3,0);
			\end{scope}
		\end{scope}
		\begin{scope}[shift={(0,-2)}]
			\begin{scope}[scale=0.5]
				\draw [-stealth] (-3,0) -- (3,0);
				\draw [-stealth] (0,-1.5) -- (0,1.5);
				\draw [ultra thick] (-1,-0.1) -- (-1,0.1) (1,-0.1) -- (1,0.1);
				\node [above] at (0,1.5) {$\scriptstyle{\color{burgundy}\psi_-}$};
				\node [right] at (3,0) {$\scriptstyle x$};
				\draw[thick, burgundy] (-3,0) to[out=0,in=225] (-1,0.8) to[out=315,in=145] (1,-0.8) to[out=45,in=180] (3,0);
			\end{scope}
		\end{scope}
		\begin{scope}[shift={(3.5,-2)}]
			\begin{scope}[scale=0.5]
				\draw [-stealth] (-3,0) -- (3,0);
				\draw [-stealth] (0,-1.5) -- (0,1.5);
				\draw [ultra thick] (-1,-0.1) -- (-1,0.1) (1,-0.1) -- (1,0.1);
				\node [above] at (0,1.5) {$\scriptstyle{\color{burgundy}\psi_-}$};
				\node [right] at (3,0) {$\scriptstyle x$};
				\begin{scope}[scale=-1]
					\draw[thick, burgundy] (-3,0) to[out=0,in=225] (-1,0.9) to[out=315,in=160] (1,-0.4) to[out=20,in=180] (3,0);
				\end{scope}
			\end{scope}
		\end{scope}
		\begin{scope}[shift={(7,-2)}]
			\begin{scope}[scale=0.5]
				\draw [-stealth] (-3,0) -- (3,0);
				\draw [-stealth] (0,-1.5) -- (0,1.5);
				\draw [ultra thick] (-1,-0.1) -- (-1,0.1) (1,-0.1) -- (1,0.1);
				\node [above] at (0,1.5) {$\scriptstyle{\color{burgundy}\psi_-}$};
				\node [right] at (3,0) {$\scriptstyle x$};
				\begin{scope}[scale=-1]
					\draw[thick, burgundy] (-3,0) to[out=0,in=225] (-1,1) to[out=315,in=170] (1,-0.1) to[out=10,in=180] (3,0);
				\end{scope}
			\end{scope}
		\end{scope}
		\begin{scope}[shift={(0,-2.2)}]
			\begin{scope}[shift={(-7,-2)}]
				\begin{scope}[scale=0.5]
					\draw [-stealth] (-3,0) -- (3,0);
					\draw [-stealth] (0,-1.5) -- (0,1.5);
					\draw [ultra thick] (-1,-0.1) -- (-1,0.1) (1,-0.1) -- (1,0.1);
					\node [above] at (0,1.5) {$\scriptstyle{\color{black!60!green}\psi_+}$};
					\node [right] at (3,0) {$\scriptstyle x$};
					\draw[thick, black!60!green] (-3,0) to[out=0,in=185] (-1,0.1) to[out=-5,in=250] (1,1.5) to[out=-70,in=180] (3,0);
				\end{scope}
			\end{scope}
			\begin{scope}[shift={(-3.5,-2)}]
				\begin{scope}[scale=0.5]
					\draw [-stealth] (-3,0) -- (3,0);
					\draw [-stealth] (0,-1.5) -- (0,1.5);
					\draw [ultra thick] (-1,-0.1) -- (-1,0.1) (1,-0.1) -- (1,0.1);
					\node [above] at (0,1.5) {$\scriptstyle{\color{black!60!green}\psi_+}$};
					\node [right] at (3,0) {$\scriptstyle x$};
					\draw[thick, black!60!green] (-3,0) to[out=0,in=200] (-1,0.4) to[out=-20,in=240] (1,1) to[out=-60,in=180] (3,0);
				\end{scope}
			\end{scope}
			\begin{scope}[shift={(0,-2)}]
				\begin{scope}[scale=0.5]
					\draw [-stealth] (-3,0) -- (3,0);
					\draw [-stealth] (0,-1.5) -- (0,1.5);
					\draw [ultra thick] (-1,-0.1) -- (-1,0.1) (1,-0.1) -- (1,0.1);
					\node [above] at (0,1.5) {$\scriptstyle{\color{black!60!green}\psi_+}$};
					\node [right] at (3,0) {$\scriptstyle x$};
					\draw[thick, black!60!green] (-3,0) to[out=0,in=240] (-1,0.8) to[out=-60,in=240] (1,0.8) to[out=-60,in=180] (3,0);
				\end{scope}
			\end{scope}
			\begin{scope}[shift={(3.5,-2)}]
				\begin{scope}[scale=0.5]
					\draw [-stealth] (-3,0) -- (3,0);
					\draw [-stealth] (0,-1.5) -- (0,1.5);
					\draw [ultra thick] (-1,-0.1) -- (-1,0.1) (1,-0.1) -- (1,0.1);
					\node [above] at (0,1.5) {$\scriptstyle{\color{black!60!green}\psi_+}$};
					\node [right] at (3,0) {$\scriptstyle x$};
					\begin{scope}[xscale=-1]
						\draw[thick, black!60!green] (-3,0) to[out=0,in=200] (-1,0.4) to[out=-20,in=240] (1,1) to[out=-60,in=180] (3,0);
					\end{scope}
				\end{scope}
			\end{scope}
			\begin{scope}[shift={(7,-2)}]
				\begin{scope}[scale=0.5]
					\draw [-stealth] (-3,0) -- (3,0);
					\draw [-stealth] (0,-1.5) -- (0,1.5);
					\draw [ultra thick] (-1,-0.1) -- (-1,0.1) (1,-0.1) -- (1,0.1);
					\node [above] at (0,1.5) {$\scriptstyle{\color{black!60!green}\psi_+}$};
					\node [right] at (3,0) {$\scriptstyle x$};
					\begin{scope}[xscale=-1]
						\draw[thick, black!60!green] (-3,0) to[out=0,in=185] (-1,0.1) to[out=-5,in=250] (1,1.5) to[out=-70,in=180] (3,0);
					\end{scope}
				\end{scope}
			\end{scope}
		\end{scope}
		\draw[ultra thick, \myblue] (-8,1) -- (8,3);
		\draw[ultra thick, orange] (-8,3) -- (8,1);
		\draw[burgundy, thick] (-8,0.8) to[out=5,in=180] (0,1.5) to[out=0,in=175] (8,0.8);
		\draw[black!60!green, thick] (-8,3.2) to[out=-5,in=180] (0,2.5) to[out=0,in=185] (8,3.2);
		\node[below, orange] at (-8,3) {$\kappa_2$};
		\node[above, \myblue] at (-8,1) {$\kappa_1$};
		\node[above, black!60!green] at (0,2.5) {$p_+$};
		\node[below, burgundy] at (0,1.5) {$p_-$};
	\end{tikzpicture}
	\caption{Parallel transport evolution of the levels, parameters and wave functions with time }\label{fig:two_delta}
\end{figure}
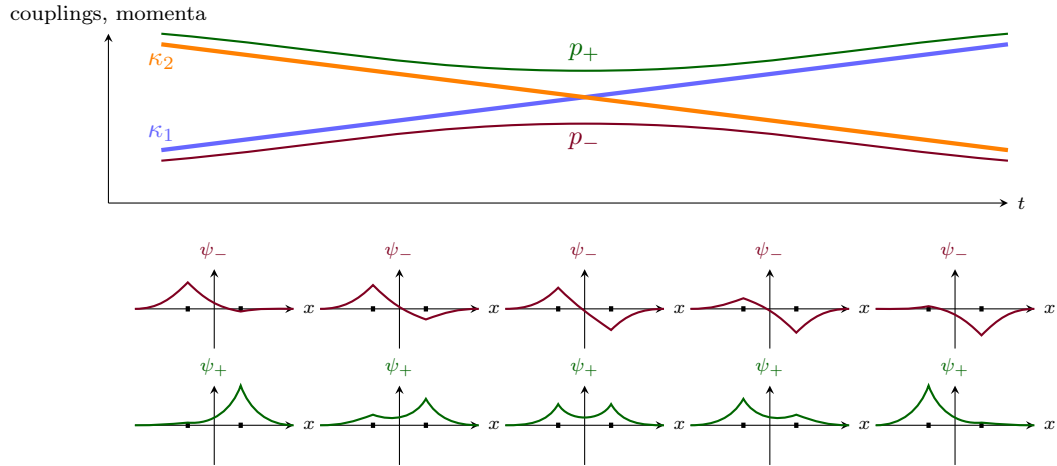

There are two solutions for the transfer problem \eqref{transfer} with appropriate boundary conditions, valid even if $\Delta\kappa=O(e^{-\kappa s})$:
\begin{equation}
	p_-=\frac{\kappa}{2}-\frac{1}{4}\sqrt{\Delta\kappa^2+4\kappa^2e^{-2\kappa s}},\quad p_+=\frac{\kappa}{2}+\frac{1}{4}\sqrt{\Delta\kappa^2+4\kappa^2e^{-2\kappa s}}\,.
\end{equation}
Clearly, these are two functions $p_-<\frac{\kappa}{2}$, $p_+>\frac{\kappa}{2}$ staying smooth even when the gap between $\kappa_1$ and $\kappa_2$ closes.
Let us track the variations of parameters, energy eigen values and wave functions as values of $\kappa_1$ and $\kappa_2$ are permuted depicted in Fig.\ref{fig:two_delta}.
Whereas energy levels do not intersect it is clear from Fig.\ref{fig:two_delta} that the particle localized at one well hops into the other well.
So we could treat the evolution matrix due to intersection of trajectories $\kappa_1(t)$ and $\kappa_2(t)$ in this case as a simple R-matrix permuting filled and empty states in the wells:
\begin{equation}\label{R-mat-delta}
	R_{12}={\bf v}_1^+{\bf v}_2^-+{\bf v}_1^-{\bf v}_2^+\,.
\end{equation}
Since levels never collide actually we are still able to use the adiabatic evolution and the respective flat Berry connection leading to the Yang-Baxter equation satisfied by our R-matrix \eqref{R-mat-delta}:
\begin{equation}\label{YBdelta}
	\begin{array}{c}
		\begin{tikzpicture}
			\node (A) at (0,0) {$\begin{array}{c}\begin{tikzpicture}
						\draw[\myblue,thick] (0,0.5) -- (2,0.5);
						\draw[burgundy,thick] (0,0) to[out=60,in=180] (2,1);
						\draw[black!60!green,thick] (0,1) to[out=0,in=120] (2,0);
						\node[left,burgundy] at (0,0) {$\scriptstyle \kappa_1$};
						\node[left,\myblue] at (0,0.5) {$\scriptstyle \kappa_2$};
						\node[left,black!60!green] at (0,1) {$\scriptstyle \kappa_3$};
						\draw[-stealth] (-0.7,-0.5) -- (2.5,-0.5);
						\draw[-stealth] (-0.7,-0.5) -- (-0.7,1.3);
						\node[right] at (2.5,-0.5) {$\scriptstyle t$};
						\node[above] at (-0.7,1.3) {$\scriptstyle \kappa$};
				\end{tikzpicture}\end{array}$};
			\node (B) at (5,0) {$\begin{array}{c}\begin{tikzpicture}
						\draw[\myblue,thick] (0,0.5) -- (2,0.5);
						\draw[burgundy,thick] (0,0) to[out=0,in=240] (2,1);
						\draw[black!60!green,thick] (0,1) to[out=300,in=180] (2,0);
						\node[left,burgundy] at (0,0) {$\scriptstyle \kappa_1$};
						\node[left,\myblue] at (0,0.5) {$\scriptstyle \kappa_2$};
						\node[left,black!60!green] at (0,1) {$\scriptstyle \kappa_3$};
						\draw[-stealth] (-0.7,-0.5) -- (2.5,-0.5);
						\draw[-stealth] (-0.7,-0.5) -- (-0.7,1.3);
						\node[right] at (2.5,-0.5) {$\scriptstyle t$};
						\node[above] at (-0.7,1.3) {$\scriptstyle \kappa$};
				\end{tikzpicture}\end{array}$};
			\path (A) edge[stealth-stealth] node[above] {\scriptsize homotopy} (B);
		\end{tikzpicture}
	\end{array},\,R_{23}R_{13}R_{12}=R_{12}R_{13}R_{23}\,.
\end{equation}

Here few comments are in order.

First, on the evolution diagrams analogous to \eqref{YBdelta} the energy levels do not intersect:
\begin{equation}
	\begin{array}{c}
		\begin{tikzpicture}
			\node (A) at (0,0) {$\begin{array}{c}\begin{tikzpicture}
						\draw[dashed,thin] (0,0.5) -- (2,0.5);
						\draw[dashed,thin] (0,0) to[out=60,in=180] (2,1);
						\draw[dashed,thin] (0,1) to[out=0,in=120] (2,0);
						\draw[thick] (0,0) to[out=60,in=180] (0.8,0.4) -- (1.2,0.4) to[out=0,in=120] (2,0);
						\draw[thick] (0,0.5) to[out=0,in=180] (1,0.7) to[out=0,in=180] (2,0.5);
						\draw[thick] (0,1) --  (2,1);
						\draw[-stealth] (-0.7,-0.5) -- (2.5,-0.5);
						\draw[-stealth] (-0.7,-0.5) -- (-0.7,1.3);
						\node[right] at (2.5,-0.5) {$\scriptstyle t$};
						\node[above] at (-0.7,1.3) {$\scriptstyle p$};
				\end{tikzpicture}\end{array}$};
			\node (B) at (5,0) {$\begin{array}{c}\begin{tikzpicture}
						\begin{scope}[shift={(0,1)}]
						\begin{scope}[yscale=-1]
						\draw[dashed,thin] (0,0.5) -- (2,0.5);
						\draw[dashed,thin] (0,0) to[out=60,in=180] (2,1);
						\draw[dashed,thin] (0,1) to[out=0,in=120] (2,0);
						\draw[thick] (0,0) to[out=60,in=180] (0.8,0.4) -- (1.2,0.4) to[out=0,in=120] (2,0);
						\draw[thick] (0,0.5) to[out=0,in=180] (1,0.7) to[out=0,in=180] (2,0.5);
						\draw[thick] (0,1) --  (2,1);
						\end{scope}
						\end{scope}
						\draw[-stealth] (-0.7,-0.5) -- (2.5,-0.5);
						\draw[-stealth] (-0.7,-0.5) -- (-0.7,1.3);
						\node[right] at (2.5,-0.5) {$\scriptstyle t$};
						\node[above] at (-0.7,1.3) {$\scriptstyle p$};
				\end{tikzpicture}\end{array}$};
			\path (A) edge[stealth-stealth] node[above] {\scriptsize homotopy} (B);
		\end{tikzpicture}
	\end{array}
\end{equation}

Second, we see a transition in Fig.\ref{fig:two_delta} goes via an anti-symmetric function for lower $p_-$ and via a symmetric for greater $p_+$.
There is no contradiction with the number of wave function zeroes and the energy level arrangement since the energy eigen value $\epsilon=-p^2$, so that actually $\epsilon_->\epsilon_+$.

Third, since the transition for lower $p_-$ goes via an anti-symmetric function it might seem that this sign phase should be taken into account in the R-matrix expression.
So that it would be more correct to use the following R-matrix:
\begin{equation}
	R_{12}\overset{?}{=}{\bf v}_1^+{\bf v}_2^--{\bf v}_1^-{\bf v}_2^+\,.
\end{equation}
This is true for the two-well case, yet not completely correct for generic $n$.
To estimate the sign of the wave function peak value at position $s_k$ one could use the following relation (let us assume that all the wells are located equidistantly $s_k=s_0k$):
\begin{equation}
	\psi(s_k)=c_1^+e^{s_0k}\left(\prod\lm_{i=1}^{k-1}\left(1-\frac{\kappa_i}{2p}\right)+O(e^{-p s_0})\right)\,.
\end{equation}
So the sign would depend on how many wells are between nodes $i$ and $j$ whose R-matrix $R_{ij}$ we would like to calculate.
It is simpler to just cancel all the signs by choosing a normalization by the phase ${\rm sgn} \,\psi(s_k)$.
So that eventually we will acquire signless R-matrix \eqref{R-mat-delta}.

Being a nice illustration for the Yang-Baxter equation construction as a homotopy relation the example with multiple wells \emph{does not} illustrate the relation between R-matrix structure and the tunneling algebra, unfortunately.
The reason is the following.
As we see from Fig.\ref{fig:two_delta} the transition passes through a wave function that is symmetric or anti-symmetric.
It appears in a non-perturbative way as a result of summation over all the tunneling (instanton) jump amplitudes \cite{Vainshtein:1981wh}, so we can not recover expansion \eqref{instantons} in a naive way.
We will be able to overcome this issue in supersymmetric models where selection rules allow one to perform the expected expansion explicitly.

\section{Dynamical SUSY}\label{sec:dynSUSY}

Unfortunately, in modern studies of theoretical and mathematical physics researchers implement advanced mathematical techniques from algebraic geometry and category theory, what makes the physical content of phenomena in discussion unavailable for the broader audience.
Here we attempt to simplify the consideration as much as possible.
And following these basic directions we should have discarded supersymmetry as an excessive concept.
Seemingly, all the objectives of the supersymmetry implication might have been achieved via discarding higher loops in the perturbation theory etc.
Yet, unfortunately, we were unable to discard supersymmetry completely in this note since in addition to quantum fluctuation suppression supersymmetry delivered specific selection rules for tunneling amplitudes.
These selection rules affect Ward identities for amplitudes and, eventually, tunneling algebra relations directly.

On the other hand, supersymmetry causes certain discomfort for the consideration in question due to a ban on all traditional tunneling amplitudes.
The reason is the following.
If calculated via an instanton transition in the path integral one should integrate over instanton moduli that include the temporal transition modulus at least.
Supersymmetry maps bosonic moduli into fermionic zero modes that do not contribute into the instanton action, and the path integral over them is zero due to canonical integration rules for Grassmann variables.

For this reason in the ordinary supersymmetric quantum mechanics instanton tunneling amplitudes contribute only to supercharge matrix elements \cite{Witten:1982im}, whereas if the target space is a complex manifold, and the complex structure is respected by the theory \cite[Sec. 10.3.4]{Hori:2003ic}, any instanton amplitude is zero.
To overcome this issue we would like to propose to consider non-stationary systems with explicitly time-dependent Hamiltonians.
As we will see in what follows this trick allows one to break supersymmetry enough to allow temporal ``catalytic'' instanton amplitude contributions and not to spoil supersymmetry localization advantages at the same time.

\subsection{Flat projected Berry connection}

\subsubsection{Projected Berry connection}
The basic model we are planning to exploit in this section is the simplest $\CN=2$ supersymmetric quantum mechanics (SQM) \cite{Witten:1982im,Hori:2003ic,Gaiotto:2015aoa}.
This model implements the basic localization paradigm when properties of specific preserving supersymmetry states could be identified exactly.
This paradigm is divided in two steps.
First, one notes that the ground sates are in one-to-one correspondence with cohomologies of supercharges $Q$ and $Q^{\dagger}$.
Second, one notes that cohomologies remain intact with respect to conjugation of $Q$ with other operators on one hand, and the values of parameters entering $Q$ vary.
So one could modify parameters in such a way that quasi-classical methods become applicable.

We would like to introduce time-dependence for our Hamiltonians in a such way that parameters entering potential, mass and so on vary with time.
In the models we consider this modification is achievable via a conjugation of the constant supercharges with time-dependent operators $\CO(t)$.
In addition to this conjugation one might need to rescale operators afterwards with a numerical factor $\lambda(t)$.
So we introduce the time dependence in the following way:
\begin{equation}
    Q(t)=e^{\lambda(t)}e^{-\CO(t)}Qe^{\CO(t)},\quad Q^{\dagger}(t)=e^{-\lambda(t)}e^{\CO(t)}Q^{\dagger}e^{-\CO(t)}\,.
\end{equation}
A similar conjugation was used in \cite{Witten:1982im} to argue independence of the ground Hilbert subspace of some parameters.
In SQM the Hamiltonian may be derived as an anti-commutator of supercharges $H_0=\left\{Q,Q^{\dagger}\right\}$.
We do the same for time-dependent supercharges, however the new time-dependent Hamiltonian $H_0(t):=\left\{Q(t),Q^{\dagger}(t)\right\}$ \emph{does not} preserve any of supercharges.

Furthermore we assume evolving wave functions $\Psi(t)$ satisfy the following Schr\"odinger equation:
\begin{equation}
    \p_t\Psi(t)=\xi \, H(t)\,\Psi(t)\,.
\end{equation}
Here we introduced a parameter $\xi$ to capture both Minkowskian ($\xi=-\I$) and Euclidean ($\xi=-1$) evolution patterns.

One might modify Hamiltonian $H_0(t)$ in a universal way so that either of supercharges, $Q(t)$ or $Q^{\dagger}(t)$, is preserved up to scale variance.
These modifications read:
\begin{equation}\label{Ham_mod}
	\begin{aligned}
	    & \mbox{Variant (a): } H(t)=\left\{Q(t),Q^{\dagger}(t)\right\}-\xi^{-1}\p_t \CO(t)\quad\Longrightarrow\quad\left[\p_t-\xi H(t),Q(t)\right]=\p_t\lambda(t)\,Q(t)\,,\\
	    & \mbox{Variant (b): } H(t)=\left\{Q(t),Q^{\dagger}(t)\right\}+\xi^{-1}\p_t \CO(t)\quad\Longrightarrow\quad\left[\p_t-\xi H(t),Q^{\dagger}(t)\right]=-\p_t\lambda(t)\,Q^{\dagger}(t)\,,\\
	\end{aligned}
\end{equation}
Note that this modification could \emph{break} in general the property of Hamiltonians to be self-adjoint.

Let us choose variant (a) in \eqref{Ham_mod} and integrate the resulting equation for the supercharge evolution:
\begin{equation}
    \left(e^{-\lambda(t)} Q(t)\right)\times{\rm Pexp}\left(\xi\int\lm_{0}^{t}H(t')\,dt'\right)={\rm Pexp}\left(\xi\int\lm_{0}^{t}H(t')\,dt'\right)\times\left(e^{-\lambda(0)} Q(0)\right)\,.
\end{equation}

This relation implies that any state belonging to ${\rm Ker}\,Q(0)$ or ${\rm Im}\,Q(0)$ at the initial moment remains in ${\rm Ker}\,Q(t)$ or ${\rm Im}\,Q(t)$ respectively during the whole evolution process.
The same is true for cohomologies of $Q(t)$.
If evolution is temporary, and the supercharges at the beginning and the end of times approach constant values, we could identify the supercharge cohomologies with stationary ground states of the Hamiltonians at the beginning and at the end of evolution.
Therefore despite our evolution is non-unitary, the ground states (also referred to as BPS states) are mapped onto ground states without any mixture of excited states.
In examples we are planning to consider the ground state is highly degenerate, so the evolution operator \emph{projected} to the ground states is a matrix in general.
We propose to call this matrix a \emph{projected} Berry phase in analogy with the canonical Berry phase.

In conclusion let us describe the projected Berry connection and the resulting evolution of the ground states in more explicit terms.
Let us choose an orthogonal basis $\psi_n(t)$ among harmonic forms annihilated by both $Q(t)$ and $Q^{\dagger}(t)$.
Then a state $\Psi(t)$ evolving with Hamiltonian $H(t)$ and belonging to ${\rm Ker}\,Q(t)$ can be decomposed according to the Hodge decomposition theorem in the following way:
\begin{equation}\label{cohomology}
	\Psi(t)=\sum\lm_n c_n(t)\psi_n(t)+Q(t)\chi(t)\,.
\end{equation}
This state is annihilated by $\p_t-\xi H(t)$.
Massaging this constraint a little we derive the following equation:
\begin{equation}
    \sum\lm_n\left(\p_t c_n(t)\psi_n(t)+c_n(t)\p_t\psi_n(t)+c_n(t)\p_t\CO(t)\psi_n(t)\right)+Q(t)\left[\p_{t}\lambda-\xi H(t)+\p_t\right]\chi(t)=0\,.
\end{equation}
Multiplying this equation from the left by the orthogonal basis vectors we derive the following effective projected Berry (Gauss-Manin) connection on the harmonic basis:
\begin{tcolorbox}
\begin{equation}\label{pBerry}
    \p_t c_n(t)+\sum\lm_k B_{nk}(t)\; c_k(t)=0,\quad B_{nk}(t):=\langle \psi_n(t)|\p_t+\p_t\CO(t)|\psi_k(t)\rangle\,.
\end{equation}
\end{tcolorbox}

\subsubsection{Flatness}\label{sec:flatness}
Let us demonstrate that the resulting projected Berry connection we constructed is \emph{flat}.

\begin{figure}[ht!]
    \centering
    \begin{tikzpicture}
	\draw[thick, postaction={decorate}, decoration={markings, mark= at position 0.6 with {\arrow{stealth}}}] (0,0) -- (2,0) node[pos=0.5,below] {$\scriptstyle 1$};
	\draw[thick, postaction={decorate}, decoration={markings, mark= at position 0.6 with {\arrow{stealth}}}] (0,0) -- (0,2) node[pos=0.5,left] {$\scriptstyle 3$};
	\draw[thick, postaction={decorate}, decoration={markings, mark= at position 0.6 with {\arrow{stealth}}}] (2,0) -- (2,2) node[pos=0.5,right] {$\scriptstyle 2$};
	\draw[thick, postaction={decorate}, decoration={markings, mark= at position 0.6 with {\arrow{stealth}}}] (0,2) -- (2,2) node[pos=0.5,above] {$\scriptstyle 4$};
	\draw[fill=black] (0,0) circle (0.07) (2,0) circle (0.07) (0,2) circle (0.07) (2,2) circle (0.07);
	\node[below left] at (0,0) {$\scriptstyle H_0$};
	\node[below right] at (2,0) {$\scriptstyle H_1$};
	\node[above left] at (0,2) {$\scriptstyle H_2$};
	\draw[-stealth] (-0.7,-0.7) -- (2.5,-0.7);
	\draw[-stealth] (-0.7,-0.7) -- (-0.7,2.5);
	\node[right] at (2.5,-0.7) {$\scriptstyle \Delta\CO_1,\Delta t_1$};
	\node[above] at (-0.7,2.5) {$\scriptstyle \Delta \CO_2,\Delta t_2$};
	\draw[thick,burgundy, -stealth] (0.2,0.4) to[out=90,in=180] node[pos=0.5,above left] {$\scriptstyle II$} (1.6,1.8);
	\draw[thick,burgundy, -stealth] (0.4,0.2) to[out=0,in=270] node[pos=0.5,below right] {$\scriptstyle I$} (1.8,1.6);
    \end{tikzpicture}
    \caption{Two homotopic paths in the parameter space}\label{fig:two_paths}
\end{figure}
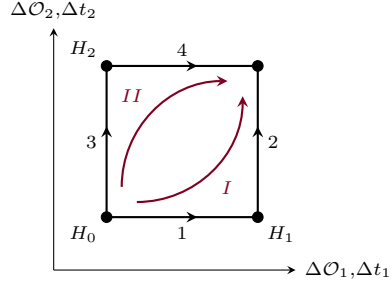

Consider a variation of parameters along two paths $I$ and $II$ forming a boundary of a small plaquette in the parameter space (see Fig.\ref{fig:two_paths}).
Along the horizontal (vertical) line time changes as $\Delta t_1$ ($\Delta t_2$) and the twisting operator changes as $\Delta\CO_1$($\Delta \CO_2$).
The initial Hamiltonian changes accordingly.
Let us denote the supercharges of the initial state as simply $Q$ and $Q^{\dagger}$.
Then we have for two paths the following evolution operators:
\begin{equation}
    \begin{aligned}
	&U_I=U_2U_1=(1+\xi H_1\Delta t_2)(1+\xi H_0\Delta t_1)=\\
	&=\left(1+\xi\left\{Q-\left[\Delta \CO_1,Q\right],Q^{\dagger}+\left[\Delta \CO_1,Q^{\dagger}\right]\right\}\Delta t_2-\Delta \CO_2\right)\left(1+\xi\left\{Q,Q^{\dagger}\right\}\Delta t_1+\Delta \CO_1\right)\,,\\
	& U_{II}=U_4 U_3=(1+\xi H_2\Delta t_1)(1+\xi H_0\Delta t_2)=U_I\left(1\leftrightarrow 2\right)\,,
    \end{aligned}
\end{equation}
where the latter line is equivalent to the former line with permuted indices 1 and 2.
The difference of these paths reads:
\begin{equation}\label{UminusU}
    U_I-U_{II}=2\xi \,Q\left(\left[Q^{\dagger},\Delta\CO_2\Delta t_1-\Delta\CO_1\Delta t_2\right]\right)+2\xi\left(\left[Q^{\dagger},\Delta\CO_2\Delta t_1-\Delta\CO_1\Delta t_2\right]\right)Q\,.
\end{equation}
We would like to learn how this difference of paths affects cohomology \eqref{cohomology}.
The second term of \eqref{UminusU} annihilates all ${\rm Ker}\, Q$, whereas the first term maps the whole wave function to ${\rm Im}\, Q$.
So we conclude that on cohomology of $Q$ parallel transports $U_I$ and $U_{II}$ are equivalent for locally homotopic paths.
This implies that the projected Berry connection \eqref{pBerry} is indeed flat.


\subsection{Parametric oscillators}

\subsubsection{Oscillator on a line}
Probably, the most conventional example of the above construction is a parametric oscillator with time-dependent frequency.
Canonical parametric oscillator is known to be a complicated problem in general.
For us this problem is spectacular since the path integral calculation of the one-loop fluctuation determinant could be reformulated naturally as a quantum mechanical problem of fluctuation evolving with the parametric oscillator Hamiltonian.
Examples when the one-loop determinant is calculated explicitly in a quantum mechanical problem are quite rare \cite{Hasanov:2024alg}, what emphasizes the complexity of this problem.
Fortunately, for the supersymmetric formulation of this problem is much simpler and is performed by revealing a correspondence between non-zero bosonic and fermionic fluctuation modes, that cancel each other in the final expression.
This observation  might be reformulated for our scenario: evolution of the ground state in the supersymmetric case is simple to describe.
Indeed the ground state in this case is non-degenerate, and the projected Berry connection \eqref{pBerry} is a simple $\IC$-connection that could be integrated easily.

Supercharges in this case read (here we choose Euclidean evolution with Euclidean time $\tau$ implying further applications to instantons):\footnote{In this note we assume for all the fermionic operators the following commutation relations and the vacuum definition:
\begin{equation}
    \left\{\psi_i,\psi_j^{\dagger}\right\}=\delta_{ij},\quad \psi_i|0\rangle=0\,.
\end{equation}
}
\begin{equation}
    Q(\tau)=\psi^{\dagger}\left(\p_x+\omega(\tau)x\right)=e^{-\frac{\omega(\tau)\, x^2}{2}}\,\psi^{\dagger}\p_x\,e^{\frac{\omega(\tau)\, x^2}{2}},\quad Q^{\dagger}(\tau)=\psi\left(-\p_x+\omega(\tau)x\right)=-e^{\frac{\omega(\tau)\, x^2}{2}}\,\psi\p_x\,e^{-\frac{\omega(\tau)\, x^2}{2}}\,.
\end{equation}
We choose a Hamiltonian that preserves $Q(\tau)$:
\begin{equation}
    H(\tau)=-\frac{1}{2}\p_x^2+\frac{1}{2}\left(\omega(\tau)^2+\omega'(\tau)\right)x^2+\frac{1}{2}\omega(\tau)\left[\psi^{\dagger},\psi\right]\,.
\end{equation}

We assume that the sign of frequency $\omega(\tau)$ does not change during evolution.
Depending on this sign there is a unique unit vector of the ground state:
\begin{equation}
    \begin{aligned}
    &\frac{1}{2}\left\{Q(\tau),Q^{\dagger}(\tau)\right\}\Psi(\tau)=\left(-\frac{1}{2}\p_x^2+\frac{1}{2}\omega(\tau)^2x^2+\frac{1}{2}\omega(\tau)\left[\psi^{\dagger},\psi\right]\right)\Psi(\tau)=0\,,\\
    &\Psi(\tau)=\left(\frac{|\omega(\tau)|}{\pi}\right)^{\frac{1}{4}}e^{-\frac{|\omega(\tau)| x^2}{2}}|0\rangle,\mbox{ if }\omega(\tau)>0; \quad \Psi(\tau)=\left(\frac{|\omega(\tau)|}{\pi}\right)^{\frac{1}{4}}e^{-\frac{|\omega(\tau)| x^2}{2}}\psi^{\dagger}|0\rangle,\mbox{ if }\omega(\tau)<0\,.
    \end{aligned}
\end{equation}
Let us stress here again that wave function $\Psi(\tau)$ is \emph{not} a solution to the Euclidean Schr\"odinger equation: apparently, it is not annihilated by operator $\p_{\tau}+H(\tau)$.
$\Psi(\tau)$ becomes a solution in the far past and the far future, when $\omega(\tau)$ stops varying.
Rather the general solution is represented in the form $c(\tau)\Psi(\tau)+Q(\tau)\chi(\tau)$, where $c(\tau)$ is a mere numerical factor, and the $Q(\tau)$-exact part is of no interest for us.
This factor $c(\tau)$ is our projected Berry phase, and it satisfies the following simple projected $\IC$-Berry connection equation:
\begin{equation}
    \p_{\tau}c(\tau)+\frac{\omega'(\tau)}{4|\omega(\tau)|}c(\tau)=0\,.
\end{equation}
Integrating this equation one arrives to the following transition amplitude as the $Q(\tau)$-exact component is projected out:
\begin{equation}
    \langle\Psi(\tau_2)|\,{\rm Pexp}\left(-\int\lm_{\tau_1}^{\tau_2}H(\tau')d\tau'\right)|\Psi(\tau_1)\rangle=\left(\frac{\omega(\tau_2)}{\omega(\tau_1)}\right)^{-\frac{1}{4}{\rm sgn}\,\omega(\tau)}\,.
\end{equation}

If function $\omega(\tau)$ does change its sign during evolution, this is a separate case, where our consideration is not completely correct.
Indeed in the point $\omega(\tau_0)=0$ our ground state wave functions leave the $L^2(\IR)$-space.
However our proposal works for Euclidean times $\tau>\tau_0+\epsilon$ and $\tau<\tau_0-\epsilon$ for some small $\epsilon$, and we see that the wave functions on the left and the right ends of interval $\left[\tau_0-\epsilon,\tau_0+\epsilon\right]$ have different fermion numbers $F=\psi^{\dagger}\psi$.
On the other hand even the time-dependent Hamiltonian commutes with $F$.
So the amplitude in this case is zero.
This is not surprising as this behavior of function $\omega(\tau)$ corresponds to frequency of the zero mode of an instanton (see e.g. \cite[Sec. 10.5.2]{Hori:2003ic}), and as we stated earlier such amplitudes are banned by supersymmetry unless other operators are inserted.

\subsubsection{Oscillator in a complex plane}
A similar discussion was conducted in \cite[Sec.5]{Dedushenko:2021mds} in relations to stable envelopes.

As the next to the simplest example let us consider here a parametric oscillator in a complex plane parameterized by complex coordinate $\phi=x_1+\I x_2$.
In this case parameter space is spanned by a triplet of frequencies we make time dependent $\vec{\omega}(\tau)=(\omega_1(\tau),\omega_2(\tau),\omega_3(\tau))$.
It is useful to combine two frequencies into complex frequency $\omega_z(\tau):=\omega_1(\tau)+\I \omega_2(\tau)$, $\omega_{\bar z}(\tau):=\omega_1(\tau)-\I \omega_2(\tau)$.
The supercharges take the following form:
\begin{equation}\label{supch}
    Q(\tau)=\psi_1^{\dagger}\left(\p_{\phi}-\omega_3(\tau)\bar\phi\right)+\psi_2\omega_z(\tau) \bar\phi,\quad Q^{\dagger}(\tau)=\psi_1\left(-\p_{\bar\phi}-\omega_3(\tau)\phi\right)+\psi^{\dagger}_2\omega_{\bar z}(\tau) \phi\,.
\end{equation}
The commutator of the supercharges reads:
\begin{equation}
    \begin{aligned}
    &\left\{Q(\tau),Q^{\dagger}(\tau)\right\}=H_0(\tau)-\omega_3(\tau)G\,,\\
    &H_0(\tau)=-\p_{\bar\phi}\p_{\phi}+\left|\vec{\omega}(\tau)\right|^2|\phi|^2+\frac{\omega_z(\tau)}{2}\left[\psi_2,\psi_1\right]+\frac{\omega_{\bar z}(\tau)}{2}\left[\psi_1^{\dagger},\psi_2^{\dagger}\right]-\frac{\omega_3(\tau)}{2}\left(\left[\psi_1^{\dagger},\psi_1\right]+\left[\psi_2^{\dagger},\psi_2\right]\right)\,,\\
    & G=\phi\p_{\phi}-\bar\phi\p_{\bar\phi}+\left[\psi_1^{\dagger},\psi_1\right]-\left[\psi_2^{\dagger},\psi_2\right]\,,
    \end{aligned}
\end{equation}
where $G$ is an operator of electric charge.
All the operators are electrically neutral, and in what follows we will work with the sole zero charge subsector of the Hilbert space where $G\Psi=0$.

We modify the Hamiltonian so that it preserves supercharge $Q(t)$ up to a scale:
\begin{equation}
    H(\tau)=H_0(\tau)-{\omega}_3'(\tau)|\phi|^2+\frac{\omega_z'(\tau)}{4\omega_z(\tau)}\left(\left[\psi_1^{\dagger},\psi_1\right]+\left[\psi_2^{\dagger},\psi_2\right]\right)\,.
\end{equation}
Eventually the supercharge satisfies the following equation:
\begin{equation}
    \left[\p_\tau+H(\tau),Q(\tau)\right]=\frac{\omega_z'(\tau)}{2\omega_z(\tau)}Q(\tau)\,.
\end{equation}
A normalized wave function annihilated by both supercharges $Q(\tau)$ and $Q^{\dagger}(\tau)$ and electric charge operator $G$ reads:
\begin{equation}\label{psi1}
    \Psi(\tau)=\frac{e^{-\left|\vec{\omega}(\tau)\right|\cdot |\phi|^2}}{\sqrt{\pi\left(\omega_3(\tau)+|\vec{\omega}(\tau)|\right)}}\left(\omega_z(\tau)-\left(\omega_3(\tau)+|\vec{\omega}(\tau)|\right)\psi_1^{\dagger}\psi_2^{\dagger}\right)|0\rangle\,.
\end{equation}
Few remarks about behavior of this function near $\omega_z(\tau)\sim 0$ are in order.
First, one might notice easily that the wave function becomes manifestly singular in the limit $\omega_z(\tau)\to 0$ when $\omega_3(\tau)<0$ belongs to the lower hemisphere.
This is not completely true.
It is a legitimate action to multiply an element of the Hilbert space by a pure phase.
By choosing a specific phase and massaging the expression a little we would arrive to the following result:
\begin{equation}\label{psi2}
    \left(\frac{\omega_{\bar z}(\tau)}{\omega_z(\tau)}\right)^{\frac{1}{2}}\Psi(\tau)=\frac{e^{-\left|\vec{\omega}(\tau)\right|\cdot |\phi|^2}}{\sqrt{\pi\left(-\omega_3(\tau)+|\vec{\omega}(\tau)|\right)}}\left(\left(-\omega_3(\tau)+|\vec{\omega}(\tau)|\right)-\omega_{\bar z}(\tau)\psi_1^{\dagger}\psi_2^{\dagger}\right)|0\rangle\,.
\end{equation}
The latter expression is manifestly smooth in the lower hemisphere when $\omega_3(\tau)<0$ and has a manifest singularity when $\omega_3(\tau)>0$.

Second, supercharges \eqref{supch} in the limit $\omega_z(\tau)\to 0$ become proportional to creation and annihilation operators $b_-=\p_{\phi}-\omega\bar\phi$, $b_-^{\dagger}=-\p_{\bar\phi}-\omega\phi$ for a charged particle in a plane subjected to the magnetic field perpendicular to the plane.
There is a second set for the creation/annihilation operators $b_+=\p_{\phi}+\omega\bar\phi$, $b_+^{\dagger}=-\p_{\bar\phi}+\omega\phi$ commuting with $b_-$, $b_-^{\dagger}$, however the Hamiltonian depends only on one set $H\sim b_-^{\dagger}b_-$.
This produces an infinitely degenerate ground state $\left(b_+^{\dagger}\right)^{\ell}\Psi$, where label $\ell$ corresponds to the Landau level.
The situation looks like at the vertical pole in the 3d space spanned by $\vec{\omega}$ new solution branches open, what makes the existence of smooth limits for \eqref{psi1} and \eqref{psi2} questionable.
This issue is resolved in the following way.
Landau levels do contribute to the ground level Hilbert space \cite[Sec.2.1]{Galakhov:2020upa}.
However the Landau level shift operator has electric charge $\left[G,b_+^{\dagger}\right]=-b_+^{\dagger}$.
Therefore the Landau levels other than $\ell=0$ do not contribute to the zero charge subsector we started this consideration with.

Again to describe the projected Berry connection in this case we represent the solution to the Euclidean Schr\"odinger equation in the form $c(\tau)\Psi(\tau)+Q(\tau)\chi(\tau)$.
Then evolution of coefficient $c(\tau)$ is described by the following connection equation:
\begin{equation}
    \p_{\tau}c(\tau)+c(\tau)\times \p_{\tau}\log\left(\frac{\omega_z(\tau)}{\omega_3(\tau)+\left|\vec{\omega}(\tau)\right|}\right)^{\frac{1}{2}}=0\,.
\end{equation}

For the evolution amplitude we derive:
\begin{equation}\label{amplchir}
    \mathsf{T}_{1\to 2}=\langle\Psi(\tau_2)|\,{\rm Pexp}\left(-\int\lm_{\tau_1}^{\tau_2}H(\tau')d\tau'\right)|\Psi_1(\tau)\rangle=\sqrt{\frac{\omega_3(\tau_2)+\left|\vec{\omega}(\tau_2)\right|}{\omega_z(\tau_2)}\cdot\frac{\omega_z(\tau_1)}{\omega_3(\tau_1)+\left|\vec{\omega}(\tau_1)\right|}}\,.
\end{equation}

Finally, we would like to mention that the projected Berry connection \eqref{pBerry} computed without contribution $\p_t\CO(t)$ corresponds to the Dirac monopole connection \cite[App.A]{Galakhov:2023aev} with a Dirac string oriented along the axis where $\Psi(\tau)$ has a singularity in the $\vec{\omega}$ space.

\section{Quantum algebras}\label{sec:QA}

\subsection{Landau-Ginzburg instantons and \texorpdfstring{$tt^*$}{}-connection}\label{sec:tt*}

In this section we will review an application of the structure discussed before to the case of quantum algebras $U_q(\fg)$.
This setting implies that $U_q(\fg)$ appears as the tunneling algebra, and evolution operators correspond to associated R-matrices, that could be represented in the form of the universal Khoroshkin-Tolstoy R-matrix \cite{khoroshkin1992uniqueness}.

This story could be naturally embedded in a broader context of supersymmetric quantum mechanics on a complex manifold.
This system could be obtained by a dimensional reduction of 2d $\CN=(2,2)$ supersymmetric Landau-Ginzburg model \cite{Hori:2003ic}, the resulting model has $\CN=4$ supercharges: $Q_+$, $Q_-$, $\bar Q_+$ and  $\bar Q_-$.
We denote the complex manifold coordinates as $\phi_i$, $i=1,\ldots,n$.
For simplicity we assume it is flat.
The potential in this case is defined be a holomorphic function $W(\phi_i)$.
Holomorphic coordinates on the parameter space we denote as $z_a$ and think of them as just parameters of the function $W(\phi_i|z_a)$.

However the dynamical Hamiltonian preserves only one of four supercharges.
First we have to choose a pair of supercharges that will represent our usual $Q$ and $Q^{\dagger}$ as a combination parameterized by a pure phase $\zeta$ \cite{Gaiotto:2015aoa}.
Various choices of $\zeta$ could be mimicked by rescaling of the superpotential $W\to\zeta W$.
Then we construct the dynamical Hamiltonian according the scheme defined in \eqref{Ham_mod}.

So here we use the following dynamical supercharges:
\begin{equation}
\begin{aligned}
	& Q(t)=\sum\lm_{i=1}^n\psi_i^{\dagger}\Big(\p_{\phi_i}-\p_{\phi_i}W(\phi_i|z_a(t))\Big)+\sum\lm_{i=1}^n\chi_i^{\dagger}\Big(\p_{\bar\phi_i}-\overline{\p_{\phi_i}W(\phi_i|z_a(t))}\Big)\,,\\
	& Q^{\dagger}(t)=\sum\lm_{i=1}^n\psi_i\Big(-\p_{\bar\phi_i}-\overline{\p_{\phi_i}W(\phi_i|z_a(t))}\Big)+\sum\lm_{i=1}^n\chi_i\Big(-\p_{\phi_i}-\p_{\phi_i}W(\phi_i|z_a(t))\Big)\,.
\end{aligned}
\end{equation}

The respective time-dependent Hamiltonian preserving $Q(t)$ (such that $Q(t)$ commutes with $\p_t-\xi H(t)$) reads:
\begin{equation}\label{LGHam}
	H(t)=\sum\lm_{i=1}^n\left(-\p_{\phi_i}\p_{\bar\phi_i}+\left|\p_{i}W\right|^2\right)-\sum\lm_{i,j=1}^n\left(\p^2_{ij}W\psi_i^{\dagger}\chi_j+\overline{\p^2_{ij}W}\psi_i\chi_j^{\dagger}\right)+\xi^{-1}\p_t{\rm Re}\,W\,.
\end{equation}

So we derive the following projected Berry connection \eqref{pBerry} in this case:
\begin{equation}\label{ttstar}
	B=\dot z^a(t)\left(A_a+C_a\right)+\dot{\bar z}^a(t) \left(\bar A_a+\bar C_a\right)\,,
\end{equation}
where
\begin{equation}
	(A_a)_{ij}=\langle \psi_i|\p_{z^a}|\psi_j\rangle,\quad (C_a)_{ij}=\langle \psi_i|\p_{z^a}W|\psi_j\rangle\,.
\end{equation}

This connection is known as the $tt^*$-connection \cite{Cecotti:1989gv, Cecotti:1991me, Cecotti:2014wea}, and there is a well-known theorem proving it being flat.

\bigskip

We are interested in tunneling instanton contributions.
Instantons interpolate between classical vacua delivering minima to the potential in  \eqref{LGHam}  -- i.e. critical points of superpotential $W$.
Suppose $W$ has $k$ critical points.
We will denote these critical points with the help of the bold face font $\bphi_i^{(a)}$, $a=1,\ldots,k$.

Simple manipulations allow one to extract the instanton equation from the quadratic Euclidean action for bosons:
\begin{equation}
	S=\int d\tau\sum\lm_i\left(\left|\p_{\tau}\phi_i\right|^2+\left|\p_{\phi_i}W\right|^2\right)=\int d\tau\sum\lm_i\left|\p_{\tau}\phi_i-\nu\overline{\p_{\phi_i}W}\right|^2+\Delta {\rm Re}\left(\nu^{-1}W\right)\,,
\end{equation}
where $\nu$ is some phase, it depends on the choice of vacua between which an instanton interpolates.
So the instanton equation reads in this case:
\begin{equation}\label{LGinstanton}
	\p_{\tau}\phi_i-\nu\overline{\p_{\phi_i}W}=0\,.
\end{equation}
Phase $\nu$ could be defined from the following constraint:
\begin{equation}
	\Delta \left(\nu^{-1}W\right)=\int d\tau\,\p_{\tau}(\nu^{-1}W)=\int d\tau\,\sum\lm_i\left|\p_{\phi_i}W\right|^2>0\,.
\end{equation}
So for the instanton interpolating from the $a^{\rm th}$ to the $b^{\rm th}$ vacuum we have:
\begin{equation}
	\nu_{ab}=\frac{\bar W_b-\bar W_a}{|W_b-W_a|}\,,
\end{equation}
where $W_{a,b}:=W\left(\bphi_i^{(a,b)}\right)$.

Each instanton has {\bf\color{burgundy} two} fermion zero modes annihilated by respective Dirac operators of two chiralities in the instanton background:
\begin{equation}
\begin{aligned}
	&\p_{\tau}\phi-\nu^{-1}\bar W'=0\;\Rightarrow\; \begin{array}{l}
		\p_{\tau}\delta\phi-\nu^{-1}\bar W''\delta\bar\phi=0\\
		\p_{\tau}\delta\bar\phi-\nu W''\delta\phi=0
	\end{array}\;\Rightarrow\\
	&\Rightarrow \left(\p_\tau-\left(\begin{array}{cc}
		0 & \bar{W}''\\
		W'' & 0 \\
	\end{array}\right)\right)\left(\begin{array}{c}
		\I\nu^{-\frac{1}{2}}\delta\phi\\
		-\I\nu^{\frac{1}{2}}\delta\bar\phi
	\end{array}\right)\xi_1=0,\; \left(\p_\tau-\left(\begin{array}{cc}
	0 & \bar{W}''\\
	W'' & 0 \\
	\end{array}\right)\right)\left(\begin{array}{c}
		\nu^{-\frac{1}{2}}\delta\phi\\
		\nu^{\frac{1}{2}}\delta\bar\phi
	\end{array}\right)\xi_2=0\,.
\end{aligned}
\end{equation}
Fermionic modes $\xi_{1,2}$ do not contribute to the quadratic action in the instanton background, rather they contribute to the path integral $\int d\xi_1d\xi_2=0$.
So there are {\bf \color{burgundy} no} ordinary non-zero tunneling amplitudes \cite[Sec.10.4.4]{Hori:2003ic}.

Nevertheless, one might obtain a non-trivial contribution from instantons in the term containing the derivative with respect to parameters.
Indeed, considering a variation with respect to parameters brings down from the exponentiated action in the path integral a term proportional to a fermion square:
\begin{equation}
\begin{aligned}
	&\langle \psi_k|\p_z|\psi_m\rangle\sim\int D\phi D\psi e^{-S_E}\int d\tau\left(\p_z W''\psi_+\psi_-+\p_{\bar z}\bar W''\bar\psi_-\bar\psi_+\right)=\\
	&=\frac{1}{\sqrt{\int d\tau\,|\p_{\tau}\phi|^2}}\int d\tau'\, d\xi\, d\bar\xi\int d\tau\,{\rm Im}\left[\Delta W''\p_{\tau}\phi^2\right]\xi\bar\xi\,e^{-|\Delta W|}\times\frac{{\rm det}'\,\CD}{\left({\rm det}'\,\CD^{\dagger}\CD\right)^{\frac{1}{2}}}\neq 0\,.
\end{aligned}
\end{equation}

Further let us consider instanton contributions to the projected Berry connection.

Suppose there are only two classical vacua.
Then equations \eqref{pBerry} descend to the following system:
\begin{equation}
	\begin{array}{l}
		\p_t c_1(t)=-\p_t{\rm Re}\left( W_1\right)c_1(t)+s_{21}c_2(t)e^{-|\Delta W_{12}|},\\
		\p_t c_2(t)=-\p_t{\rm Re}\left( W_2\right)c_2(t)+s_{12}c_1(t)e^{-|\Delta W_{12}|},
	\end{array}
\end{equation}
It is convenient to apply a change of variables $c_i(t)=e^{-{\rm Re}\,W_i}\tilde c_i(t)$.
In new terms the equations read:
\begin{equation}
	\begin{array}{l}
		\p_t \tilde c_1(t)=s_{21}e^{{\rm Re}\left( \Delta W_{12}\right)-|\Delta W_{12}|}\tilde c_2(t),\\
		\p_t \tilde c_2(t)=s_{12}e^{{\rm Re}\left(\Delta W_{21}\right)-|\Delta W_{12}|}\tilde c_1(t),\\
	\end{array}
\end{equation}

Suppose critical values $W_{1,2}$ are arranged in the plane $\zeta W$ in the following way:
\begin{equation}\label{varphi}
	\begin{array}{c}
		\begin{tikzpicture}
			\draw (-0.5,-0.5) -- (0.5,0.5);
			\draw[dashed] (-0.7,0) -- (0.7,0);
			\draw ([shift=(0:0.4)]0,0) arc (0:45:0.4);
			\node at (0.6,0.2) {$\scriptstyle \varphi$};
			\draw[fill=black] (-0.5,-0.5) circle (0.05) (0.5,0.5) circle (0.05);
			\node[right] at (0.5,0.5) {\tiny 1};
			\node[left] at (-0.5,-0.5) {\tiny 2};
		\end{tikzpicture}
	\end{array},\quad \begin{array}{l}
		{\rm Re}\,\Delta W_{12}=|\Delta W_{12}|\cos\varphi\\
		\\
		{\rm Re}\, \Delta W_{21}=-|\Delta W_{12}|\cos\varphi\\
	\end{array},\quad \begin{array}{l}
		e^{{\rm Re}\, \Delta W_{12}-|\Delta W_{12}|}\sim\frac{\delta(\varphi)}{\sqrt{|\Delta W|}}\,,\\
		e^{{\rm Re}\,\Delta W_{21}-|\Delta W_{12}|}\sim e^{-2|\Delta W_{12}|} \sim 0\,.
	\end{array}
\end{equation}
Therefore the evolution is described by an \emph{upper-triangular Stokes matrix}:
\begin{tcolorbox}
	\begin{equation}\label{Stokes}
		\left(\begin{array}{c}
			\tilde c_1(+\infty)\\ \tilde c_2(+\infty)
		\end{array}\right)=\left(\begin{array}{cc}
			1 & S_{21}\\
			0 & 1\\
		\end{array}\right)\left(\begin{array}{c}
			\tilde c_1(-\infty)\\ \tilde c_2(-\infty)
		\end{array}\right)\,.
	\end{equation}
\end{tcolorbox}

Yet the major difference with the tunneling in the non-supersymmetric case depicted in Fig.\ref{fig:two_delta} is that in the supersymmetric case transition is \emph{directed}.
So the tunneling can go only one-way from vacuum 2 to vacuum 1, so that $\Delta W\in \IR_{>0}$.

\subsection{Picard-Lefschetz monodromy in complex systems}

Consider an $N$-dimensional integral:
\begin{equation}\label{integral}
	\Psi=\int\lm_{\CL} \prod\lm_{i=1}^{N}d\phi_i\,e^{W(\phi_i)}\,.
\end{equation}
Saddle points $\bphi_i^{(\alpha)}$, $\alpha=1,\ldots,K$ are solutions to a system of algebraic equations $\p_{\phi_i}W=0$, $\forall i$.

A Lefschetz thimble $\CL_{\alpha}$ is defined as a \emph{union} of all trajectories $\phi_i(s)$ satisfying the following equations and boundary conditions:
\begin{equation}\label{Lefschetz_thi}
	\p_s \phi_i(s)=-\overline{\p_{\phi_i}W\left(\phi_k(s)\right)},\quad\lim\lm_{s\to-\infty}\phi_i(s)=\bphi_i^{(\alpha)}\,.
\end{equation}
By construction it has the following properties:
\begin{itemize}
	\item There are as many $\CL_{\alpha}$ as saddle points $\alpha=1,\ldots,K$.
	\item Each $\CL_{\alpha}$ is a Lagrangian submanifold of \emph{real} dimension $N$ with respect to the following symplectic form $\omega=\sum\lm_{i=1}^N d\phi_i\wedge d\bar\phi_i$.
	\item Function $W$ maps the whole set of points $\CL_{\alpha}$ on a straight ray:
	\begin{equation}
		W(\CL_{\alpha})=\left\{W(\bphi_{\alpha})- t|\,t\in[0,+\infty)\right\}.
	\end{equation}
	\item Lefschetz thimbles form a basis of contours corresponding to Borel resummed expansions around saddle points.
\end{itemize}

We should remark that for a canonical choice of conjugated coordinates, say, ${\rm Re}\,\phi_i$ and ${\rm Im}\,\phi_i$ equation \eqref{Lefschetz_thi} describes Hamiltonian evolution with ``time'' $s$ and Hamiltonian ${\rm Im}\,W$.

Stokes $(\alpha,\beta)$-``surface'' has real co-dimension 1 in the parameter space and is defined again by a constraint:
\begin{equation}
	W(\bphi_{\alpha})-W(\bphi_{\beta})\in \IR_{\leq 0}\,.
\end{equation}
On this surface geometry of the Lefschetz thimbles might turn out to be ill-defined as some of the trajectories solving  \eqref{Lefschetz_thi} solve the instanton equation \eqref{LGinstanton} for $\nu=1$ as well.
If a thimble contains such an instanton trajectory its topology is broken.
On two sides of the $(\alpha,\beta)$-Stokes surface to critical point $\alpha$ one assigns two \emph{different} homotopy classes of Lefschetz thimbles, yet these classes are related by a linear \emph{Picard-Lefschetz transform}.
Now if one decomposes an integration over some cycle $\CC$ over Lefschetz thimbles:
\begin{equation}
	\int\lm_{\CC}\prod\lm_{i}d\phi_i\,e^{-W}=\sum\lm_{k}\tilde c_k \int\lm_{\CL_k}\prod\lm_{i}d\phi_i\,e^{-W}\,,
\end{equation}
as the basis of Lefschetz thimbles jumps across the Stokes surface, one is forced to transform linear coefficients accordingly:
\begin{equation}
	\left(\begin{array}{c}
		\tilde c_{\alpha}'\\ \tilde c_{\beta}'
	\end{array}\right)=\left(\begin{array}{cc}
		1 & S_{\beta\alpha}\\
		0 & 1\\
	\end{array}\right)\left(\begin{array}{c}
		\tilde c_{\alpha}\\ \tilde c_{\beta}
	\end{array}\right)\,.
\end{equation}
This transform describes a discontinuity in the asymptotic of the integrals and is analogous to \eqref{Stokes}.
Generic transport of the basis of Lefschetz thimbles and the respective transformation of the coefficients we will call \emph{Picard-Lefschetz transport}.

So we see an $(\alpha,\beta)$-Stokes surface is equivalent to a constraint $\varphi=0$ in \eqref{varphi}.
So we treat it as a geometric locus in the parameter space where an instanton interpolating form vacuum $\bphi_i^{(\alpha)}$ to vacuum $\bphi_i^{(\beta)}$ contributes to the tunneling amplitude the most.
Eventually we could treat Stokes surfaces as sole delta-type supports for non-zero off-diagonal contributions in the evolution matrix.
Note that the anti-instanton contribution is suppressed as well -- the Stokes matrix is upper-triangular.

In general, to arrive to the same conclusion we could have searched for an asymptotic behavior of the flat $tt^*$-sections \cite{Cecotti:1989gv,Cecotti:1991me}.
The Stokes coefficients $S_{ij}$ emerging in \eqref{Stokes} have also a meaning of Maslov indices \cite{arnol1967d}, and it is easier to calculate them explicitly in the framework of asymptotic solutions to $tt^*$-sections \cite[App.A]{Galakhov:2016cji}.

\subsection{A model for quantum algebras}

Quantum algebras $U_{q}(\fg)$ could be embedded in the context discussed so far in the following way.
There is a well-known connection between tensor products of $U_q(\fg)$ representations \cite{Knizhnik:1984nr,drinfeld1991quasi} and conformal blocks for $\fg$ WZWN models  \cite{Dotsenko:1984nm,1990IJMPA...5.2495G,schechtman1991arrangements,Dunin-Barkowski:2012cuv}.
The latter have a natural integral representations \cite{etingof1998lectures} in the form of \eqref{integral}.

Suppose algebra $\fg$ has simple roots $\alpha_i$.
We consider $k$ punctures ``colored'' with the highest weight $\mu_a$ representations, $a=1,\ldots,k$.
The total weight of the conformal block is $\mu=-\sum\lm_a\mu_a+\sum\lm_i n_i \alpha_i$.
There are $n_i$ complex fields $\phi_{i,u}$ corresponding to root $\alpha_i$.
The integrand (superpotential) to \eqref{integral} reads in this case \cite{etingof1998lectures}:
\begin{equation}\label{Uqsup}
	\mathsf{W}=\sum\lm_{a,b=1}^{k}\langle\mu_a,\mu_b\rangle\log(z_a-z_b)-\sum\lm_{i=1}^{r}\sum\lm_{u=1}^{n_i}\sum\lm_{a=1}^{k}\langle\alpha_i,\mu_a\rangle\log\left(\phi_{i,u}-z_a\right)+\frac{1}{2}\sum\lm_{i,j=1}^r\sum\lm_{u=1}^{n_i}\sum\lm_{v=1}^{n_j}\langle\alpha_i,\alpha_j\rangle\log\left(\phi_{i,u}-\phi_{j,v}\right)\,.
\end{equation}
The actual superpotential $W$ is related to $\mathsf{W}$ by a proportionality relation: $W=\mathsf{W}/\kappa$, where $\kappa$ is related to a complexified coupling constant of the WZWN model \cite{Knizhnik:1984nr}.
It is related to a parameter $q$ ``quantizing'' ordinary Lie algebra $\fg$:
\begin{equation}
	q=e^{\frac{\pi\I}{\kappa}}\,.
\end{equation}

The respective projected Berry ($tt^*$-)connection for such a model takes a form of the Knizhink-Zamolodchikov connection \cite{Knizhnik:1984nr}.
The basic evolution operator corresponds to a clockwise or counterclockwise braiding of the punctures $z_a$:
\begin{equation}
	R_{a,a+1}=\begin{array}{c}
		\begin{tikzpicture}
			\draw (-2.1,-0.3) -- (-2.5,-0.7) -- (1.6,-0.7) -- (2,-0.3) -- cycle;
			\draw[thick] (-0.5,-0.5) to[out=90,in=210] (0,0) to[out=30,in=270] (0.5,0.5) (0.5,-0.5) to[out=90,in=330] (0.1,-0.07) (-0.1,0.07) to[out=150,in=270] (-0.5,0.5);
			\draw[thick] (-1.5,-0.5) -- (-1.5,0.5) (-2,-0.5) -- (-2,0.5) (1.5,-0.5) -- (1.5,0.5);
			\draw[fill=black] (-2,-0.5) circle (0.07) (-1.5,-0.5) circle (0.07) (-0.5,-0.5) circle (0.07) (0.5,-0.5) circle (0.07) (1.5,-0.5) circle (0.07)
			(-2,0.5) circle (0.07) (-1.5,0.5) circle (0.07) (-0.5,0.5) circle (0.07) (0.5,0.5) circle (0.07) (1.5,0.5) circle (0.07)
			(-1.25,-0.5) circle (0.03) (-1,-0.5) circle (0.03) (-0.75,-0.5) circle (0.03)
			(1.25,-0.5) circle (0.03) (1,-0.5) circle (0.03) (0.75,-0.5) circle (0.03)
			(-1.25,0.5) circle (0.03) (-1,0.5) circle (0.03) (-0.75,0.5) circle (0.03)
			(1.25,0.5) circle (0.03) (1,0.5) circle (0.03) (0.75,0.5) circle (0.03);
			\begin{scope}[shift={(0,1)}]
				\draw (-2.1,-0.3) -- (-2.5,-0.7) -- (1.6,-0.7) -- (2,-0.3) -- cycle;
			\end{scope}
			\node[right] at (2,-0.5) {$\IC$};
			\node[right] at (2,0.5) {$\IC$};
			\node[below] at (-2,-0.7) {$1$};
			\node[below] at (-1.5,-0.7) {$2$};
			\node[below] at (-0.5,-0.7) {$a$};
			\node[below] at (0.5,-0.7) {$a+1$};
			\node[below] at (1.5,-0.7) {$k$};
			\node[above] at (-2,0.7) {$1$};
			\node[above] at (-1.5,0.7) {$2$};
			\node[above] at (-0.5,0.7) {$a$};
			\node[above] at (0.5,0.7) {$a+1$};
			\node[above] at (1.5,0.7) {$k$};
		\end{tikzpicture}
	\end{array},\;
	R_{a,a+1}^{-1}=\begin{array}{c}
		\begin{tikzpicture}
			\draw (-2.1,-0.3) -- (-2.5,-0.7) -- (1.6,-0.7) -- (2,-0.3) -- cycle;
			\begin{scope}[xscale=-1]
				\draw[thick] (-0.5,-0.5) to[out=90,in=210] (0,0) to[out=30,in=270] (0.5,0.5) (0.5,-0.5) to[out=90,in=330] (0.1,-0.07) (-0.1,0.07) to[out=150,in=270] (-0.5,0.5);
			\end{scope}
			\draw[thick] (-1.5,-0.5) -- (-1.5,0.5) (-2,-0.5) -- (-2,0.5) (1.5,-0.5) -- (1.5,0.5);
			\draw[fill=black] (-2,-0.5) circle (0.07) (-1.5,-0.5) circle (0.07) (-0.5,-0.5) circle (0.07) (0.5,-0.5) circle (0.07) (1.5,-0.5) circle (0.07)
			(-2,0.5) circle (0.07) (-1.5,0.5) circle (0.07) (-0.5,0.5) circle (0.07) (0.5,0.5) circle (0.07) (1.5,0.5) circle (0.07)
			(-1.25,-0.5) circle (0.03) (-1,-0.5) circle (0.03) (-0.75,-0.5) circle (0.03)
			(1.25,-0.5) circle (0.03) (1,-0.5) circle (0.03) (0.75,-0.5) circle (0.03)
			(-1.25,0.5) circle (0.03) (-1,0.5) circle (0.03) (-0.75,0.5) circle (0.03)
			(1.25,0.5) circle (0.03) (1,0.5) circle (0.03) (0.75,0.5) circle (0.03);
			\begin{scope}[shift={(0,1)}]
				\draw (-2.1,-0.3) -- (-2.5,-0.7) -- (1.6,-0.7) -- (2,-0.3) -- cycle;
			\end{scope}
			\node[right] at (2,-0.5) {$\IC$};
			\node[right] at (2,0.5) {$\IC$};
			\node[below] at (-2,-0.7) {$1$};
			\node[below] at (-1.5,-0.7) {$2$};
			\node[below] at (-0.5,-0.7) {$a$};
			\node[below] at (0.5,-0.7) {$a+1$};
			\node[below] at (1.5,-0.7) {$N$};
			\node[above] at (-2,0.7) {$1$};
			\node[above] at (-1.5,0.7) {$2$};
			\node[above] at (-0.5,0.7) {$a$};
			\node[above] at (0.5,0.7) {$a+1$};
			\node[above] at (1.5,0.7) {$k$};
		\end{tikzpicture}
	\end{array}\,.
\end{equation}
and represents an $U_q(\fg)$ R-matrix.
The flatness of the projected Berry connection guarantees that the resulting R-matrix solves the Yang-Baxter equations, and therefore is a representation of the braid group on $z_a$.

The universal Khoroshkin-Tolstoy R-matrix \cite{cmp/1104248397} acts in the space of a representation tensor power, where the factors are vectors corresponding to the punctures.
The universal R-matrix has the following form of a product over roots:
\begin{equation}
	R\sim\prod\lm_{\alpha}\sum\lm_{n=0}^{\infty}\frac{\left(q-q^{-1}\right)^n}{[n]_q!}f_{\alpha}^n\otimes e_{\alpha}^n\,,
\end{equation}
where $e_{\alpha}$ and $f_{\alpha}$ are generators of $U_q(\fg)$ corresponding to $\alpha$ and $-\alpha$ respectively.

Interactions between particles in the respective Hamiltonian are obtained from derivatives of $W$, so an interaction potential contribution is proportional to the inverse distance between particle positions in the complex plane.
We expect that the particles gather around the punctures in forms of clouds or droplets (that become the usual Wigner droplets in matrix models when the number of particles is huge \cite{Morozov:2010cq,Mironov:2010pi,Mironov:2011jn}).
If we move punctures far apart, the mutual interaction between particles from different clouds becomes negligible.
It is expected that the state of each individual cloud around puncture $z_a$ containing $m_{\alpha}$ particles of type $\alpha$ corresponds to a vector of weight $-\mu_a+\sum\lm_{\alpha}\alpha m_{\alpha}$ of the $U_q(\fg)$ representation of the highest weight $\mu_a$.
This is a straightforward incarnation of the model of multiple potential wells we considered in Sec.\ref{sec:TAC}.

In this model we expect that the first off-diagonal terms $\mathsf{T}\sim(q-q^{-1})\, f_{\alpha}\otimes e_{\alpha}$ in the R-matrix.
Operator $f_{\alpha}$ lowers the vector weight and, therefore, the number $m_{\alpha}$ of particles in the corresponding cloud.
Operator $e_{\alpha}$ implements the inverse process.
So together $f_{\alpha}$ and $e_{\alpha}$ correspond to tunneling algebra operators ${\bf v}^-$ and ${\bf v}^+$, and we conclude:
\begin{tcolorbox}
	\centering
	In this case the tunneling algebra is isomorphic to quantum algebra $U_q(\fg)$.
\end{tcolorbox}

To conclude this section let us demonstrate in detail how this prescription works in the simplest case of $U_q(\fs\fl_2)$ when the majority of punctures is colored by the fundamental representation $\Box$.

It is useful to put two punctures with non-integral weights in two far singular points $0$ and $\infty$ and to assume there are no particles hovering around those points.
Alternatively one could add a linear term to \eqref{Uqsup}:
\begin{equation}
	W=\frac{1}{\kappa}\left(\sum\lm_{a\neq b}\log(z_a-z_b)-\sum\lm_i\sum\lm_a\log(z_a-\phi_i)+\frac{1}{2}\sum\lm_{i\neq j}\log(\phi_i-\phi_j)+c\sum\lm_i\phi_i\right)\,.
\end{equation}
There are two solutions \cite[App.A]{Galakhov:2017pod} for a cloud of particles corresponding to spin-up and spin down vectors:
\begin{itemize}
	\item Spin-down $|\downarrow\rangle$ at puncture $z_a$: puncture $z_a$ is empty, there are no particles hovering around it.
	\item Spin-up $|\uparrow\rangle$ at puncture $z_a$: there is a particle at location $\bphi_i=z_a+c^{-1}+O(c^{-2})$.
\end{itemize}

The mutual interaction between particles contributes the same term $(\phi_i-\phi_j)^{-1}$ in the Lefschetz thimble equation \eqref{Lefschetz_thi}, that could be neglected if punctures $z_a$ are widely separated.
So as a manifold we could approximate the Lefschetz thimble in this case as a Cartesian product $\ldots\times\CL_{i-1}\times\CL_i\times \CL_{i+1}\times \ldots$ of individual Lefschetz thimbles for fields $\phi_i$ constructed for the respective choices of vacua corresponding to $|\uparrow\rangle$.
Such an individual Lefschetz thimble is a solution to \eqref{Lefschetz_thi} for the single field $\phi$.\footnote{In this case the form of the Lefschetz thimble may be predicted exactly.
Equation \eqref{Lefschetz_thi} describes Hamiltonian evolution with Hamiltonian ${\rm Im}\,W$ in the 2d phase space with coordinates $({\rm Re}\,\phi,{\rm Im}\,\phi)$.
This equation is integrable.
}
Being depicted in the complex $\phi$-plane it is represented by two arcs encircling the corresponding puncture and flowing towards $0$ or $\infty$ depending on the complex phase of $c$.
For example:
\begin{equation}
	\begin{array}{c}
		\begin{tikzpicture}
			\draw[fill=white] (-1.5,0) circle (0.07) (-0.5,0) circle (0.07) (0.5,0) circle (0.07) (1.5,0) circle (0.07);
			\draw (-1.8,0) to[out=90,in=180] (-1.5,0.3) to[out=0,in=90] (-1.2,0) (-1.8,0) to[out=270,in=180] (0,-1) (-1.2,0) to[out=270,in=180] (0,-1);
			\begin{scope}[xscale=-1]
				\draw (-1.8,0) to[out=90,in=180] (-1.5,0.3) to[out=0,in=90] (-1.2,0) (-1.8,0) to[out=270,in=180] (0,-1) (-1.2,0) to[out=270,in=180] (0,-1);
				\draw (-0.8,0) to[out=90,in=180] (-0.5,0.3) to[out=0,in=90] (-0.2,0) (-0.8,0) to[out=270,in=135] (0,-1) (-0.2,0) to[out=270,in=135] (0,-1);
			\end{scope}
			\draw[fill=burgundy] (-1.5,0.3) circle (0.04) (0.5,0.3) circle (0.04) (1.5,0.3) circle (0.04);
			\node[below] at (0,-1) {$\scriptstyle \infty$};
			\node[below] at (-1.5,-0.08) {$\scriptstyle z_1$};
			\node[below] at (-0.5,-0.08) {$\scriptstyle z_2$};
			\node[below] at (0.5,-0.08) {$\scriptstyle z_3$};
			\node[below] at (1.5,-0.08) {$\scriptstyle z_4$};
			\node at (-1.5,0.6) {$\scriptstyle |\uparrow\rangle$};
			\node at (-0.5,0.6) {$\scriptstyle |\downarrow\rangle$};
			\node at (0.5,0.6) {$\scriptstyle |\uparrow\rangle$};
			\node at (1.5,0.6) {$\scriptstyle |\uparrow\rangle$};
			\node at (-1,0.6) {$\scriptstyle\otimes$};
			\node at (0,0.6) {$\scriptstyle\otimes$};
			\node at (1,0.6) {$\scriptstyle\otimes$};
		\end{tikzpicture}
	\end{array}\,.
\end{equation}

\begin{figure}[ht!]
	\centering
	\begin{tikzpicture}
		\node[draw,rounded corners=2,rotate=-90] (A) at (0,0) {$\begin{array}{c}
			\begin{tikzpicture}[rounded corners=0]
				\draw[fill=white] (-1.5,0) circle (0.07) (-0.5,-0.3) circle (0.07) (0.5,0.3) circle (0.07) (1.5,0) circle (0.07);
				\draw (0,-1) to[out=150,in=270] (-0.8,-0.3) to[out=90,in=180] (-0.5,0) to[out=0,in=90] (-0.2,-0.3) to[out=270,in=150] (0,-1);
				\draw (0,-1) to[out=60,in=270] (0.2,0.3) to[out=90,in=180] (0.5,0.6) to[out=0,in=90] (0.8,0.3) to[out=270,in=60] (0,-1);
			\end{tikzpicture}
			\end{array}$};
		\node[draw,rounded corners=2,rotate=-90] (B) at (2.5,0) {$\begin{array}{c}
				\begin{tikzpicture}[rounded corners=0]
					\draw[fill=white] (-1.5,0) circle (0.07) (-0.15,-0.3) circle (0.07) (0.15,0.3) circle (0.07) (1.5,0) circle (0.07);
					\draw (0,-1) to[out=120,in=270] (-0.45,-0.3) to[out=90,in=180] (-0.15,0) to[out=0,in=90] (0.15,-0.3) to[out=270,in=120] (0,-1);
					\draw (0.15,0.6) to[out=0,in=90] (0.45,0.3) to[out=270,in=80] (0,-1);
					\draw[burgundy, thick, postaction={decorate}, decoration={markings, mark= at position 0.8 with {\arrow{stealth}}}] (0.15,0.6) to[out=180,in=90] (-0.15,0.3) -- (-0.15,0);
				\end{tikzpicture}
			\end{array}$};
		\node[draw,rounded corners=2,rotate=-90] (C) at (5,0) {$\begin{array}{c}
				\begin{tikzpicture}[rounded corners=0]
					\draw[fill=white] (-1.5,0) circle (0.07) (0,-0.3) circle (0.07) (0,0.3) circle (0.07) (1.5,0) circle (0.07);
					\draw (0,-1) to[out=120,in=270] (-0.3,-0.3) to[out=90,in=180] (0,0) to[out=0,in=90] (0.3,-0.3) to[out=270,in=60] (0,-1);
					\draw (0,-1) to[out=120,in=270] (-0.4,0.1) to[out=90,in=180] (0,0.6) to[out=0,in=90] (0.4,0.1) to[out=270,in=60] (0,-1);
				\end{tikzpicture}
			\end{array}$};
		\node[draw,rounded corners=2,rotate=-90] (D) at (7.5,0) {$\begin{array}{c}
				\begin{tikzpicture}[rounded corners=0, xscale=-1]
					\draw[fill=white] (-1.5,0) circle (0.07) (-0.15,-0.3) circle (0.07) (0.15,0.3) circle (0.07) (1.5,0) circle (0.07);
					\draw (0,-1) to[out=120,in=270] (-0.45,-0.3) to[out=90,in=180] (-0.15,0) to[out=0,in=90] (0.15,-0.3) to[out=270,in=120] (0,-1);
					\draw (0.15,0.6) to[out=0,in=90] (0.45,0.3) to[out=270,in=80] (0,-1);
					\draw[{black!60!green}, thick, postaction={decorate}, decoration={markings, mark= at position 0.8 with {\arrow{stealth}}}] (0.15,0.6) to[out=180,in=90] (-0.15,0.3) -- (-0.15,0);
				\end{tikzpicture}
			\end{array}$};
		\node[draw,rounded corners=2,rotate=-90] (E) at (10,0) {$\begin{array}{c}
				\begin{tikzpicture}[rounded corners=0, xscale=-1]
					\draw[fill=white] (-1.5,0) circle (0.07) (-0.5,-0.3) circle (0.07) (0.5,0.3) circle (0.07) (1.5,0) circle (0.07);
					\draw (0,-1) to[out=150,in=270] (-0.8,-0.3) to[out=90,in=180] (-0.5,0) to[out=0,in=90] (-0.2,-0.3) to[out=270,in=150] (0,-1);
					\draw (0,-1) to[out=60,in=270] (0.2,0.3) to[out=90,in=180] (0.5,0.6) to[out=0,in=90] (0.8,0.3) to[out=270,in=60] (0,-1);
				\end{tikzpicture}
			\end{array}$};
		\draw[-stealth] (-1.5,-2) -- (11.5,-2);
		\node[right] at (11.5,-2) {$t$};
		\draw[ultra thick] (-1,2.5) -- (11,2.5);
		\draw[ultra thick] (-1,3) to[out=0,in=180] (11,3.5);
		\draw[line width=2mm, white] (-1,3.5) to[out=0,in=180] (11,3);
		\draw[ultra thick] (-1,3.5) to[out=0,in=180] (11,3);
		\draw[ultra thick] (-1,4) -- (11,4);
		\node[left] at (-1,4) {$\scriptstyle z_1$};
		\node[left] at (-1,3.5) {$\scriptstyle z_2$};
		\node[left] at (-1,3) {$\scriptstyle z_3$};
		\node[left] at (-1,2.5) {$\scriptstyle z_4$};
		\node[right] at (11,4) {$\scriptstyle z_1$};
		\node[right] at (11,3.5) {$\scriptstyle z_3$};
		\node[right] at (11,3) {$\scriptstyle z_2$};
		\node[right] at (11,2.5) {$\scriptstyle z_4$};
		\draw[dashed, thick] (A.south west) to[out=90,in=270] (0,2.5) (A.north west) to[out=90,in=270] (0,2.5) (0,2.5) -- (0,4);
		\draw[dashed, thick] (B.south west) to[out=90,in=270] (2.5,2.5) (B.north west) to[out=90,in=270] (2.5,2.5) (2.5,2.5) -- (2.5,4);
		\draw[dashed, thick] (C.south west) to[out=90,in=270] (5,2.5) (C.north west) to[out=90,in=270] (5,2.5) (5,2.5) -- (5,4);
		\draw[dashed, thick] (D.south west) to[out=90,in=270] (7.5,2.5) (D.north west) to[out=90,in=270] (7.5,2.5) (7.5,2.5) -- (7.5,4);
		\draw[dashed, thick] (E.south west) to[out=90,in=270] (10,2.5) (E.north west) to[out=90,in=270] (10,2.5) (10,2.5) -- (10,4);
		\node[burgundy] at (2.5,-1) {inst. I};
		\node[black!60!green] at (7.5,-1) {inst. II};
	\end{tikzpicture}
	\caption{$U_q(\fs\fl_2)$ R-matrix as an evolution process in the parameter space $z_i(t)$}\label{fig:UqRmat}
\end{figure}
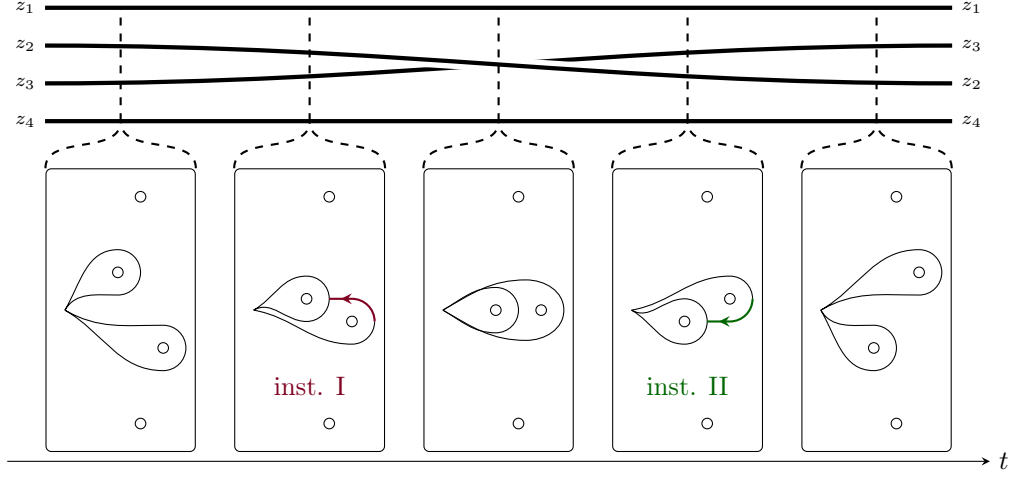

Finally we re-construct the R-matrix in this case as an evolution process (see Fig.\ref{fig:UqRmat}) when two neighboring punctures are intertwined.
Clearly, the first off-diagonal contribution is due to two emerging instanton contributions following \eqref{Stokes}:
\begin{equation}
	R=q^{\frac{1}{4}h\otimes h}\left(1+({\color{burgundy}\underbrace{q}_{\rm inst.\,I}}-{\color{black!60!green}\underbrace{q^{-1}}_{\rm inst.II}})\underbrace{f}_{{\bf v}^-}\otimes \underbrace{e}_{{\bf v}^+}\right)q^{\frac{1}{4}h\otimes h}\,.
\end{equation}
A similar analysis of R-matrices as Picard-Lefschetz monodromies was performed in \cite{Gaiotto:2011nm}.

\section{Yangians}\label{sec:Yangians}

In this section we will work with a family of models known as gauged linear sigma models with a quiver target space.
This model might be derived as a dimensional reduction of 4d $\CN=1$ SYM-Higgs theory \cite{Wess:1992cp} to $\CN=4$ SQM \cite{Denef:2002ru,Ohta:2014ria,Galakhov:2020vyb,Raeymaekers:2024ics,Bykov:2024wft}.
This model in general has 4 supercharges $Q_1$, $Q_2$, $Q_1^{\dagger}$, $Q_2^{\dagger}$, dynamical SUSY preserves only one combination as in Sec.\ref{sec:dynSUSY}, and without loss of generality we choose it to be given by $Q_1+Q_2$.
For the notations and conventions we refer the reader to \cite{Galakhov:2024bzs}.
For this section we solely choose specific notations for the fields acquired by dimensional reduction from the 4d gauge field $A_{\mu=0,1,2,3}$.
We choose the gauge $A_0=0$ and define a real and a complex scalars $\Sigma:=A_3$, $\Phi:=A_1+\I A_2$.

\subsection{Meridian instantons}\label{sec:meridian}

We start with a revision of a particle on a $\IC\IP^1$-target space.
One might think of this model as the simplest non-trivial framed quiver target space with one gauge and one framing node connected by a pair of oriented edges:
\begin{equation}
	\begin{array}{c}
		\begin{tikzpicture}
			\draw[thick, postaction={decorate}, decoration={markings, mark= at position 0.6 with {\arrow{stealth}}}] (0,0) --  (3,0) node[pos=0.5, above] {$\left(\begin{array}{c}
					q_1\\ q_2
				\end{array}\right)$};
			\draw[fill=burgundy] (3,0) circle (0.1);
			\draw[fill=\myblue] (-0.1,-0.1) -- (-0.1,0.1) -- (0.1,0.1) -- (0.1,-0.1) -- cycle;
			\node[right, draw=black] at (3.2,0) {$U(1): \; \Sigma=(\sigma), \; \Phi=(\phi)$};
			\node[left, draw=black] at (-0.2,0) {$U(2)_{\rm flav}: \; \Sigma_{\rm fxd}=\left(\begin{array}{cc}u_1 & 0\\ 0& u_2\end{array}\right), \;\Phi_{\rm fxd}=\left(\begin{array}{cc}\mu_1 & 0 \\ 0 & \mu_2\end{array}\right)$};
		\end{tikzpicture}
	\end{array}\,.
\end{equation}
The projective coordinate on the target space of $\IC\IP^1$ is given by a ratio $(q_1:q_2)$.
From two complex fields $q_1$ and $q_2$ one real degree of freedom is eliminated by the real D-term constraint forcing the particle to stay confined on $S^3$.
Whereas the complex phase is fixed by the gauge invariance requirement with respect to the dynamical $U(1)$ quiver node.
So the resulting target space is indeed $\IC\IP^1\cong S^3/S^1$.

For this system we exploit standard supercharges  obtained by a dimensional reduction of $\CN=1$ 4d supercharges in the SYM theory:
\begin{equation}\label{CP1Qs}
	\begin{aligned}
	    & Q=\lambda\left(\p_\sigma+\left[r-\sum\lm_{i=1,2} |q_i|^2\right]\right)+\xi\p_{\phi}+\sum\lm_{i=1,2}\psi_i\left(\p_{q_i}-\left[\sigma-u_i\right]\bar q_i\right)+\sum\lm_{i=1,2}\chi_i\left(\bar\phi-\bar\mu_i\right)\bar q_i\,,\\
		& Q^{\dagger}=\bar\lambda\left(-\p_\sigma+\left[r-\sum\lm_{i=1,2} |q_i|^2\right]\right)-\bar\xi\p_{\bar\phi}+\sum\lm_{i=1,2}\bar\psi_{i}\left(-\p_{\bar q_i}-\left[\sigma-u_i\right]q_i\right)+\sum\lm_{i=1,2}\bar\chi_i\left(\phi-\mu_i\right) q_i\,,
	\end{aligned}
\end{equation}

The physical Hilbert space is cut by the zero electric charge constraint $G\Psi_{\rm phys}=0$, where the electric charge operator is given by the following expression:
\begin{equation}
	G=\sum\lm_{i=1,2}\left(\left[\bar q_i\p_{\bar q_i}-q_i\p_{q_i}\right]+\frac{1}{2}\left\{\bar\psi_i,\psi_i\right\}+\frac{1}{2}\left\{\bar\chi_i,\chi_i\right\}\right)\,.
\end{equation}

\begin{figure}[ht!]
	\begin{center}
		\begin{tikzpicture}[scale=0.4]
			\draw[thick, \myblue, postaction={decorate}, decoration={markings, mark= at position 0.8 with {\arrow{stealth}}}] (0.,4.33013) to[out=8.42736,in=158.419] (2.42064,4.12038) to[out=-21.5814,in=118.338] (4.30271,2.5427) to[out=-61.6623,in=90.9911] (4.75294,0.90817) to[out=-89.0089,in=57.1496] (4.01502,-1.66792) to[out=-122.85,in=26.8642] (1.75054,-3.74166) to[out=-153.136,in=10.0243] (0.,-4.33013);
			\draw[thick, \myblue, postaction={decorate}, decoration={markings, mark= at position 0.8 with {\arrow{stealth}}}] (0.,4.33013) to[out=66.3814,in=212.086] (0.38269,4.92625) to[out=32.0865,in=107.377] (0.82141,4.58619) to[out=-72.6226,in=92.6613] (1.02115,3.21344) to[out=-87.3387,in=87.4513] (1.02115,1.59067) to[out=-92.5487,in=78.8817] (0.58432,-2.20686) to[out=-101.118,in=67.5296] (0.,-4.33013);
			\draw[thick, \myblue, postaction={decorate}, decoration={markings, mark= at position 0.8 with {\arrow{stealth}}}] (0.,4.33013) to[out=156.641,in=12.1067] (-1.79006,4.60041) to[out=-167.893,in=64.8587] (-3.36209,3.3573) to[out=-115.141,in=89.223] (-3.71389,1.80801) to[out=-90.777,in=108.258] (-3.41012,-0.18446) to[out=-71.742,in=134.402] (-1.6869,-3.09872) to[out=-45.5982,in=154.896] (0.,-4.33013);
			\draw[thick, \myblue, postaction={decorate}, decoration={markings, mark= at position 0.8 with {\arrow{stealth}}}] (0.,4.33013) to[out=-169.976,in=35.7717] (-2.42064,3.33387) to[out=-144.228,in=65.3504] (-4.30271,1.14466) to[out=-114.65,in=89.0144] (-4.75294,-0.63615) to[out=-90.9856,in=118.338] (-4.36418,-2.42871) to[out=-61.6623,in=150.444] (-3.03113,-3.82886) to[out=-29.556,in=188.427] (0.,-4.33013);
			\draw[thick, \myblue, postaction={decorate}, decoration={markings, mark= at position 0.8 with {\arrow{stealth}}}] (0.,4.33013) to[out=-112.47,in=78.8817] (-0.55702,2.34575) to[out=-101.118,in=86.3994] (-0.98868,-0.9876) to[out=-93.6006,in=95.8183] (-0.98868,-3.66377) to[out=-84.1817,in=114.198] (-0.77978,-4.69786) to[out=-65.8024,in=197.314] (-0.47195,-4.96835) to[out=17.3138,in=246.381] (0.,-4.33013);
			\draw[thick, \myblue, postaction={decorate}, decoration={markings, mark= at position 0.8 with {\arrow{stealth}}}] (0.,4.33013) to[out=-25.1035,in=136.988] (1.36785,3.41024) to[out=-43.012,in=118.858] (2.78721,1.60876) to[out=-61.1422,in=92.2759] (3.70839,-1.39763) to[out=-87.7241,in=55.435] (3.19828,-3.64409) to[out=-124.565,in=12.1067] (1.89146,-4.57866) to[out=-167.893,in=336.641] (0.,-4.33013);
			\draw[dashed] (4.75528,0.77254) to[out=125.756,in=341.232] (2.93893,2.02254) to[out=161.232,in=10.2046] (-1.54508,2.37764) to[out=-169.795,in=65.9128] (-4.84292,0.62172) to[out=-114.087,in=135.541] (-4.52414,-1.06445) to[out=-44.4594,in=165.707] (-2.40877,-2.19077) to[out=-14.2928,in=190.205] (1.54508,-2.37764) to[out=10.2046,in=209.559] (3.64484,-1.71137) to[out=29.5593,in=252.423] (4.91144,-0.46845) to[out=72.4234,in=300.159] (4.75528,0.77254);
			\draw[ultra thick] (0,0) circle (5);
			\draw[fill=burgundy] (0., 4.33013) circle (0.15) (0., -4.33013) circle (0.15);
			\node[above] at (0,5) {$\scriptstyle\mathscr{N}$};
			\node[below] at (0,-5) {$\scriptstyle\mathscr{S}$};
		\end{tikzpicture}
		\caption{$\IC\IP^1$ embedding as the Riemann sphere} \label{fig:sphere}
	\end{center}
\end{figure}
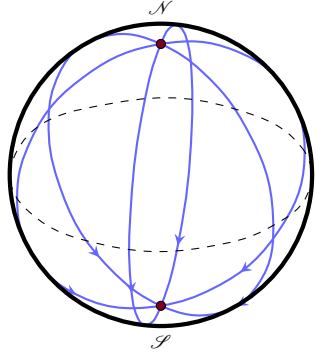

Supercharges \eqref{CP1Qs} could be treated as supercharges in SQM with a Morse height function:
\begin{equation}\label{height}
	h=r\sigma-\sum\lm_{i=1,2}(\sigma-u_i)|q_i|^2\,.
\end{equation}
Enforcing the particle to stay confined on $S^3$ $|q_1|^2+|q_2|^2=r$ one would arrive to the height function (see \cite[Sec.4.4]{Galakhov:2023aev}) $h=(u_1-u_2)(|q_1|^2-|q_2|^2)$ that corresponds to the third coordinate of the canonical $\IC\IP^1$ Riemann sphere embedding in $\IR^3$ (see Fig.\ref{fig:sphere}).
Two classical vacua are localized near the north and the south poles.
A natural instanton corresponds to a flow between poles along a meridian induced by the height function gradient.
Alternatively we could have massaged the original instanton equations to the form of a soliton equation in a cosine potential \cite[Sec.4.3]{Galakhov:2020upa}.

This instanton has two apparent moduli: as usual we could have moved the instanton core position in Euclidean time and rotated the instanton flow trajectory around the sphere.

As in the case of the LG model (see Sec.\ref{sec:tt*}) the presence of two zero modes annihilates instanton amplitude path integrals.
Moreover the instanton trajectory does not minimize the part of the action containing equivariant parameters $\mu_i$.
To deal with these issues simultaneously rather than computing  an instanton trajectory and studying fluctuation around it we would manage to calculate explicit wave functions of a particle localized near either pole beyond instanton approximation.
An overlap of these wave functions is saturated by the instanton amplitude if we were calculating by the quasi-classical approximation method.

Let us switch to coordinates on the Riemann sphere:
\begin{equation}
	\begin{aligned}
		q_1=e^{\I\theta}\rho\frac{z}{\sqrt{1+|z|^2}},\quad q_2=e^{\I\theta}\rho\frac{1}{\sqrt{1+|z|^2}},\quad \bar q_1=e^{-\I\theta}\rho\frac{\bar z}{\sqrt{1+|z|^2}},\quad \bar q_2=e^{-\I\theta}\rho\frac{1}{\sqrt{1+|z|^2}}
	\end{aligned}
\end{equation}
Respective derivatives read:
\begin{equation}
	\begin{aligned}
		& \p_{q_1}=\frac{e^{-\I \theta}\bar z}{2\sqrt{1+|z|^2}}\p_{\rho}+\frac{e^{-\I\theta}\sqrt{1+|z|^2}}{\rho}\p_{z}\,,\\
		& \p_{q_2}=\frac{e^{-\I \theta}}{2\sqrt{1+|z|^2}}\p_{\rho}-\frac{e^{-\I\theta}\sqrt{1+|z|^2}}{2\rho}\I\p_{\theta}-\frac{e^{-\I\theta}\sqrt{1+|z|^2}}{\rho}z\p_{z}\,,\\
		& \p_{\bar q_1}=\frac{e^{\I \theta}z}{2\sqrt{1+|z|^2}}\p_{\rho}+\frac{e^{\I\theta}\sqrt{1+|z|^2}}{\rho}\p_{\bar z}\,,\\
		& \p_{\bar q_2}=\frac{e^{\I \theta}}{2\sqrt{1+|z|^2}}\p_{\rho}+\frac{e^{\I\theta}\sqrt{1+|z|^2}}{2\rho}\I\p_{\theta}-\frac{e^{\I\theta}\sqrt{1+|z|^2}}{\rho}\bar z\p_{\bar z}\,.
	\end{aligned}
\end{equation}
For the charge operator we have:
\begin{equation}
	G=\I\p_{\theta}+\sum\lm_i\left(\bar\psi_i\psi_i+\bar\chi_i\chi_i\right)-2\,.
\end{equation}
For the volume form we derive:
\begin{equation}\label{norm}
	dq_1d\bar q_1 dq_2d\bar q_2=\frac{2\I\rho^3}{\left(1+|z|^2\right)^2}d\rho\, d\theta\, dz\, d\bar z\,.
\end{equation}

We rewrite the supercharges in the following form ($\rho\to \sqrt{r}+\rho$, $\sigma\to \sigma+\frac{u_1+u_2}{2}$, $\phi\to \phi+\frac{\mu_1+\mu_2}{2}$, $u:=u_1-u_2$, $\mu=\mu_1-\mu_2$):
\begin{equation}
	\begin{aligned}
		& Q=\lambda\left(\p_{\sigma}-2A\sqrt{r}\rho\right)+\left(\xi\p_{\phi}+A\sqrt{r}e^{-\I\theta}\frac{\chi_1\bar z+\chi_2}{\sqrt{1+|z|^2}}\bar \phi\right)+e^{-\I\theta}\frac{\bar z \psi_1+\psi_2}{\sqrt{1+|z|^2}}\left(\frac{\p_{\rho}}{2}-A\sqrt{r}\sigma\right)+\\
		& +e^{-\I\theta}\frac{\psi_1-z \psi_2}{\sqrt{1+|z|^2}}\frac{1+|z|^2}{\sqrt{r}}\left(\p_z+\frac{ u r\bar z}{\left(1+|z|^2\right)^2}\right)-e^{-\I\theta}\psi_2\frac{\sqrt{1+|z|^2}}{2\sqrt{r}}\I\p_{\theta}-\frac{\bar\mu \sqrt{r}\bar z}{1+|z|^2}e^{-\I\theta}\frac{\chi_1-z\chi_2}{\sqrt{1+|z|^2}}\,,\\
		& Q^{\dagger}=\bar\lambda\left(-\p_{\sigma}-2A\sqrt{r}\rho\right)+\left(-\bar\xi\p_{\bar\phi}+A\sqrt{r}e^{\I\theta}\frac{\bar \chi_1 z+\bar\chi_2}{\sqrt{1+|z|^2}}\phi\right)+e^{\I\theta}\frac{z \bar\psi_1+\bar\psi_2}{\sqrt{1+|z|^2}}\left(\frac{\p_{\rho}}{2}-A\sqrt{r}\sigma\right)+\\
		& +e^{\I\theta}\frac{\bar\psi_1-\bar z\bar \psi_2}{\sqrt{1+|z|^2}}\frac{1+|z|^2}{\sqrt{r}}\left(-\p_{\bar z}+\frac{ u r z}{\left(1+|z|^2\right)^2}\right)-e^{\I\theta}\bar\psi_2\frac{\sqrt{1+|z|^2}}{2\sqrt{r}}\I\p_{\theta}-\frac{\mu \sqrt{r}z}{1+|z|^2}e^{\I\theta}\frac{\bar \chi_1-\bar z\bar\chi_2}{\sqrt{1+|z|^2}}\,.
	\end{aligned}
\end{equation}
As well we have introduced parameter $A$ to perform the necessary localization procedures.

In the limit $A\to\infty$ the first lines of supercharges contribute only, this allows us to fix the first order wave function approximation depending on coordinates $\rho$, $\sigma$ and $\phi$ but having $z$ as an external parameter:
\begin{equation}
	\Psi_0=e^{-2A\sqrt{r}\rho^2-A\frac{\sqrt{r}}{2}\sigma^2-A\sqrt{r}|\phi|^2+2\I\theta}\left(\bar\xi+e^{-\I\theta}\frac{\bar z \bar\chi_1+\bar\chi_2}{\sqrt{1+|z|^2}}\right)\left(\bar\lambda+e^{-\I\theta}\frac{\bar z \bar\psi_1+\bar\psi_2}{\sqrt{1+|z|^2}}\right)|0\rangle\,.
\end{equation}
Let us note that $G\Psi_0=0$.

We look for the second order approximation to this function in the following form:
\begin{equation}\label{second-order}
	\Psi=\left(|\mu|f\left(|z|^2\right)\frac{\bar\psi_1-\bar z\bar\psi_2}{\sqrt{1+|z|^2}}+\bar\mu g\left(|z|^2\right)\frac{\bar\chi_1-\bar z\bar\chi_2}{\sqrt{1+|z|^2}}\right)\Psi_0\,.
\end{equation}
By constructing matrix elements $\langle \Psi_0|\nu Q|\Psi\rangle$, $\langle \Psi_0|\nu Q^{\dagger}|\Psi\rangle$, where $\nu$ are 4 monomials of fermions entering \eqref{second-order} we arrive to the following equations for functions $f$ and $g$:
\begin{equation}
	\begin{aligned}
		f'(x)+\frac{r\,u\, f(x)-r\,|\mu|\, g(x)}{(1+x)^2}=0,\quad g'(x)+\frac{-r\,|\mu|\,f(x)-r\,u\,g(x)}{(1+x)^2}=0\,.
	\end{aligned}
\end{equation}
These equations have precisely two solutions localized near the north $\mathscr{N}$ at $x=0$ and the south $\mathscr{S}$ at $x=\infty$ poles respectively:
\begin{equation}
	\left(\begin{array}{c}
		f(x)\\ g(x)
	\end{array}\right)_{\mathscr{N}}=\left(\begin{array}{c}
		-\left(u+\sqrt{|\mu|^2+u^2}\right)\\ |\mu|
	\end{array}\right)e^{\frac{r}{2}\sqrt{|\mu|^2+u^2}\frac{1-x}{1+x}},\quad
	\left(\begin{array}{c}
		f(x)\\ g(x)
	\end{array}\right)_{\mathscr{S}}=\left(\begin{array}{c}
		-\left(u-\sqrt{|\mu|^2+u^2}\right)\\ |\mu|
	\end{array}\right)e^{\frac{r}{2}\sqrt{|\mu|^2+u^2}\frac{x-1}{1+x}}\,.
\end{equation}

As effective wave functions describing two vacuum states at the north and the south poles we use the following expressions normalized with respect to effective norm $(1+|z|^2)^{-2}d|z|^2$ (cf. \eqref{norm}):
\begin{equation}
    \begin{aligned}
	&\Psi_{\mathscr{N}}=\left(-\mu,\sqrt{u^2+|\mu|^2}-u\right)\sqrt{\frac{r}{2\left(\sqrt{u^2+|\mu|^2}-u\right)\sinh\left(r\sqrt{u^2+|\mu|^2}\right)}}\exp\left(\frac{r}{2}\sqrt{u^2+|\mu|^2}\frac{1-|z|^2}{1+|z|^2}\right)\,,\\
	&\Psi_{\mathscr{S}}=\left(\mu,\sqrt{u^2+|\mu|^2}+u\right)\sqrt{\frac{r}{2\left(\sqrt{u^2+|\mu|^2}+u\right)\sinh\left(r\sqrt{u^2+|\mu|^2}\right)}}\exp\left(\frac{r}{2}\sqrt{u^2+|\mu|^2}\frac{|z|^2-1}{|z|^2+1}\right)\,,
    \end{aligned}
\end{equation}
For the scalar product and the matrix element of the derivative operator we have respectively:
\begin{tcolorbox}
\begin{equation}\label{amplitudesCP1}
    \begin{aligned}
	&\langle \Psi_{\mathscr{S}}|\Psi_{\mathscr{N}}\rangle=0\,,\\
	&\langle \Psi_{\mathscr{S}}|\p_u|\Psi_{\mathscr{N}}\rangle=-\frac{r|\mu|}{2\sqrt{u^2+|\mu|^2}\sinh\left(r\sqrt{u^2+|\mu|^2}\right)}\sim{\rm sgn}\, u\,\frac{r|\mu|}{|u|}e^{-r|u|},\quad |u|\gg |\mu|\,.
    \end{aligned}
\end{equation}
\end{tcolorbox}

Let us first remark that we predicted result $\langle \Psi_{\mathscr{S}}|\Psi_{\mathscr{N}}\rangle=0$ from the instanton zero mode counting.
Yet inserting a differential operator with respect to the parameters produces a non-zero matrix element.
This implies that the instanton amplitude with this insertion will be \emph{non-zero} as well.

Now we would like to compare an expression for the non-trivial amplitude in \eqref{amplitudesCP1} with instanton counting.
To do so, first, let us restore parts of the instanton trajectories.
As we mentioned earlier in Sec.\ref{sec:tt*} instantons saturate matrix elements of $Q$, whereas anti-instantons saturate $Q^{\dagger}$.
To re-construct instanton equations we consider operators in front of the fermions in \eqref{CP1Qs}, perform a process inverse to the usual quantization procedure ($\dot q\leftrightarrow -\I \p_q$) and, finally, apply the Wick rotation ($t\to-\I \tau$, $\p_t\to \I\p_{\tau}$):
\begin{equation}\label{CP1inst}
    \p_{q_i}-[\sigma-u]\bar q_i\quad\longrightarrow\quad -\dot q_i(\tau)-[\sigma(\tau)-u_i]q_i(\tau)=0\,.
\end{equation}
Similar manipulations are in order for field $\sigma$.
The resulting equations have an instantonic solution localized in Euclidean time $\tau$ (see \cite[Sec.4.3]{Galakhov:2020upa}).

The projective coordinate on $\IC\IP^1$ is a ratio of chiral fields $z=q_1/q_2$.
Without loss of generality we would like to consider an instanton that flows from the north pole to the south pole:
\begin{equation}
    \begin{aligned}
	&\mathscr{N}:\quad z=0,\quad (q_1,q_2)=(0,r),\quad \sigma=u_2\,,\\
	&\mathscr{S}:\quad z=\infty, \quad (q_1,q_2)=(r,0),\quad \sigma=u_1\,.
    \end{aligned}
\end{equation}
For such an instanton trajectory $\dot q_1/q_1>0$, $\dot q_2/q_2<0$.
Then using \eqref{CP1inst} we conclude that along the instanton trajectory:
\begin{equation}
    u_2<\sigma(\tau)<u_1,\quad u=u_1-u_2>0\,.
\end{equation}

In the vicinity of the north pole we could distinguish a local coordinate as field $q_1=r z$, the other field $q_2$ acquires an expectation value.
And the opposite expansion is valid near the south pole $q_2=r/z$.
In these terms the wave functions are approximated by wave functions of free chirals \eqref{psi1} for respective choices of parameters:
\begin{equation}
\begin{aligned}
    &\Psi_{\mathscr{N}}\sim\frac{\left(-\mu,\sqrt{u^2+|\mu|^2}-u\right)}{\sqrt{\sqrt{u^2+|\mu|^2}-u}}e^{-\sqrt{u^2+|\mu|^2}|q_1|^2},\quad (\omega_z,\omega_3)=(-\mu,-u)\,,\\
    &\Psi_{\mathscr{S}}\sim\frac{\left(\mu,\sqrt{u^2+|\mu|^2}+u\right)}{\sqrt{\sqrt{u^2+|\mu|^2}+u}}e^{-\sqrt{u^2+|\mu|^2}|q_2|^2},\quad (\omega_z,\omega_3)=(\mu,u)\,.
\end{aligned}
\end{equation}
This observation allows us to treat coordinate $z$ on $\IC\IP^1$ as a free chiral field that experiences a transition $(-\mu,-u)\to (\mu,u)$ in its frequencies $(\omega_z,\omega_3)$.

Let us consider further the limit $u\gg|\mu|$ for the wave functions:
\begin{equation}
    \Psi_{\mathscr{N}}\sim\underline{\left(1-\frac{\bar \mu}{u}\psi_1^{\dagger}\psi_2^{\dagger}\right)}\times \sqrt{u}e^{-u|q_1|^2}|0\rangle,\quad \Psi_{\mathscr{S}}\sim\underline{\left(1+\frac{\mu}{u}\psi_2\psi_1\right)}\times \sqrt{u}e^{-u|q_2|^2}\psi_1^{\dagger}\psi_2^{\dagger}|0\rangle\,.
\end{equation}
The underlined operators are close to a unit operator in the zeroes order of $|\mu|/u$, so initially we could ignore them.
It becomes clear that even if the scalar parts of the wave functions overlap due to an (anti)instanton, these states have different fermion numbers.
These fermions $\psi_1^{\dagger}\psi_2^{\dagger}$ are exactly those two zero fermion modes confined to the instanton annihilating the respective matrix element.
The derivative with respect to $u$ produces exactly a two-fermion term of order $|\mu|/u$ that can not be ignored.

If we had started with the action deformations instead, we would find the following term in the action:
\begin{equation}
    S\ni \int dt\; u(t)(\bar\psi_1(t)\bar\psi_2(t)+\psi_2(t)\psi_1(t))\,,
\end{equation}
whose variation with respect to $u$ brings down a quadratic fermion operator similarly to the case of the Landau-Ginzburg model discussed in Sec.\ref{sec:tt*}.
Comparing with amplitude \eqref{CP1inst} we conclude that the integral over fermion zero modes is ``responsible'' exactly for the one-loop determinant coefficient in front of the action suppression:
\begin{equation}
    \int\lm_{\mbox{\tiny zero-modes}} D\psi (\bar\psi_1\bar\psi_2+\psi_2\psi_1)\sim r\frac{|\mu|}{u}=:\Delta\,.
\end{equation}

Let us consider further the induced projected Berry connection for this system \eqref{pBerry}:
\begin{equation}
\begin{aligned}
    &\p_tc_{\mathscr{N}}+\Delta_{\mathscr{NS}}e^{-r|u|}c_{\mathscr{S}}+\p_t\left(r u_2+\frac{1}{2}\log\frac{-\mu}{\sqrt{u^2+|\mu|^2}-u}\right)\times c_{\mathscr{N}}=0\,,\\
    &\p_tc_{\mathscr{S}}+\Delta_{\mathscr{SN}}e^{-r|u|}c_{\mathscr{N}}+\p_t\left(r u_1+\frac{1}{2}\log\frac{\mu}{\sqrt{u^2+|\mu|^2}+u}\right)\times c_{\mathscr{S}}=0\,,
\end{aligned}
\end{equation}
where the logarithms are contributions of the Berry connection of the free chiral field \eqref{amplchir}.
Let us redefine variables as:
\begin{equation}
    c_{\mathscr{N}}=e^{-r u_2}\sqrt{\frac{\sqrt{u^2+|\mu|^2}-u}{-\mu}}\tilde c_{\mathscr{N}},\quad c_{\mathscr{S}}=e^{-r u_1}\sqrt{\frac{\sqrt{u^2+|\mu|^2}+u}{\mu}}c_{\mathscr{S}}\,.
\end{equation}
Then we find for new variables (we separate a contribution of parallel transport \eqref{amplchir} from the other contributions):
\begin{equation}
\begin{aligned}
    &\p_t\tilde c_{\mathscr{N}}=-\Delta_{\mathscr{NS}}\times\mathsf{T}_{(\mu,u)\to(-\mu,-u)}\times e^{-r|u|-ru}\times \tilde c_{\mathscr{S}}\,,\\
    &\p_t\tilde c_{\mathscr{S}}=-\Delta_{\mathscr{SN}}\times\mathsf{T}_{(-\mu,-u)\to(\mu,u)}\times e^{-r|u|+ru}\times \tilde c_{\mathscr{N}}\,.
\end{aligned}
\end{equation}
Clearly when $u>0$ the anti-instanton flowing from $\mathscr{S}$ to $\mathscr{N}$ is suppressed by $e^{-2r u}$ in the r.h.s. of the first line, whereas the instanton flowing from $\mathscr{N}$ to $\mathscr{S}$ is not suppressed for $u>0$ as we predicted.

Eventually, we conclude that the Stokes matrix acquires in this case the following form:
\begin{tcolorbox}
\begin{equation}\label{CP1Stokes}
    \left.\left(\begin{array}{c}\tilde c_{\mathscr{N}}\\ \tilde c_{\mathscr{S}}\end{array}\right)\right|_{t=+\infty}=\left(\begin{array}{cc} 1 & 0\\ \alpha \mathsf{T}_{\mathscr{N}\to \mathscr{S}} & 1\end{array}\right)\times\left.\left(\begin{array}{c}\tilde c_{\mathscr{N}}\\ \tilde c_{\mathscr{S}}\end{array}\right)\right|_{t=-\infty}\,,
\end{equation}
\end{tcolorbox}
where $\alpha$ is an inessential factor.

\subsection{Equivariant integrals calculated by instantons}

In general we expect the Stokes coefficients to be given by transition amplitudes $\mathsf{T}$ as in \eqref{CP1Stokes}.
Similarly to the quantum algebra consideration here we would like to make our model ``coarse-grained'', to strip off various factors related to the continuous basis rotations etc.
For this purpose it is rather convenient to assign to transition amplitudes $\mathsf{T}$ certain geometric meaning in terms of the target space for the theory in question -- a quiver variety.

We start with the free chiral wave function \eqref{psi1} at $\omega_3=0$.
Its geometric meaning is that it corresponds to a Thom representative of the equivariant Euler homology class \cite{Cordes:1994fc} on $\IC$ parameterized by $\phi$ with equivariant action of weight $\omega_z$.
To observe this we substitute fermions by forms in the usual way:
\begin{equation}
    {\bf e}(\phi):=\left(\omega_z-|\omega_z|\psi_1^{\dagger}\psi_2^{\dagger}\right)e^{-|\omega_z||\phi^2|}|0\rangle=\left(\omega_z+|\omega_z|\,d\phi\wedge d\bar\phi\right)e^{-|\omega_z||\phi|^2}\,.
\end{equation}
Here the Euler class is normalized canonically, so that the following primitive equivariant integrals localized following the Atiyah-Bott formula read:
\begin{equation}\label{equiints}
    \int\lm_{\IC_{\omega_z}}{\bf e}=1,\quad \int\lm_{\IC_{\omega_z}}1=\frac{1}{\omega_z},\quad \int\lm_{\rm pt}{\bf e}=\omega_z,\quad \int\lm_{\rm pt}1=1\,.
\end{equation}
and this normalization \emph{differs} from the canonical ``physical'' normalization in \eqref{psi1}.
To preserve the geometric meaning of wave functions it is natural to use the geometric normalization condition in what follows and to redefine the transit amplitude \eqref{amplchir} accordingly.
For physically normalized states $\Psi_1$ and $\Psi_2$ transit amplitude \eqref{amplchir} is just a Berry phase induced by evolution:
\begin{equation}
    {\rm Pexp}\left(-\int\lm_{\tau_1}^{\tau_2}H(\tau')\,d\tau\right)|\Psi_1\rangle=\mathsf{T}_{1\to 2}^{\rm (old)}\;|\Psi_2\rangle\,.
\end{equation}
It is natural to define the transit amplitude for the geometric states $\Phi_1=g_1\Psi_1$, $\Phi_2=g_2\Psi_2$ in a similar fashion as a phase:
\begin{equation}
    {\rm Pexp}\left(-\int\lm_{\tau_1}^{\tau_2}H(\tau')\,d\tau\right)|\Phi_1\rangle=\mathsf{T}_{1\to 2}^{\rm (new)}\;|\Phi_2\rangle\,.
\end{equation}
Then, clearly,
\begin{equation}
    \mathsf{T}_{1\to 2}^{(\rm new)}=\frac{g_1}{g_2}\times\mathsf{T}_{1\to 2}^{(\rm old)}\,.
\end{equation}

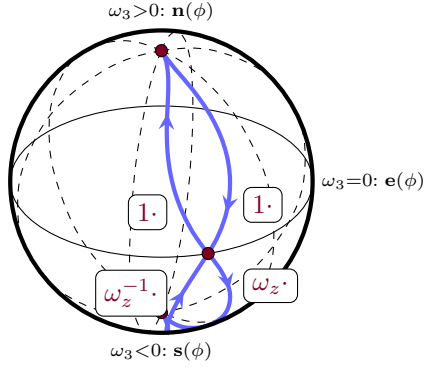
\begin{figure}[ht!]
    \centering
    $\begin{array}{c}
	\begin{tikzpicture}[scale=0.4]
	    \draw[dashed] (0.,4.33013) to[out=8.42736,in=158.419] (2.42064,4.12038) to[out=-21.5814,in=118.338] (4.30271,2.5427) to[out=-61.6623,in=90.9911] (4.75294,0.90817) to[out=-89.0089,in=57.1496] (4.01502,-1.66792) to[out=-122.85,in=26.8642] (1.75054,-3.74166) to[out=-153.136,in=10.0243] (0.,-4.33013);
	    \draw[dashed] (0.,4.33013) to[out=66.3814,in=212.086] (0.38269,4.92625) to[out=32.0865,in=107.377] (0.82141,4.58619) to[out=-72.6226,in=92.6613] (1.02115,3.21344) to[out=-87.3387,in=87.4513] (1.02115,1.59067) to[out=-92.5487,in=78.8817] (0.58432,-2.20686) to[out=-101.118,in=67.5296] (0.,-4.33013);
	    \draw[dashed] (0.,4.33013) to[out=156.641,in=12.1067] (-1.79006,4.60041) to[out=-167.893,in=64.8587] (-3.36209,3.3573) to[out=-115.141,in=89.223] (-3.71389,1.80801) to[out=-90.777,in=108.258] (-3.41012,-0.18446) to[out=-71.742,in=134.402] (-1.6869,-3.09872) to[out=-45.5982,in=154.896] (0.,-4.33013);
	    \draw[dashed] (0.,4.33013) to[out=-169.976,in=35.7717] (-2.42064,3.33387) to[out=-144.228,in=65.3504] (-4.30271,1.14466) to[out=-114.65,in=89.0144] (-4.75294,-0.63615) to[out=-90.9856,in=118.338] (-4.36418,-2.42871) to[out=-61.6623,in=150.444] (-3.03113,-3.82886) to[out=-29.556,in=188.427] (0.,-4.33013);
	    \draw[dashed] (0.,4.33013) to[out=-112.47,in=78.8817] (-0.55702,2.34575) to[out=-101.118,in=86.3994] (-0.98868,-0.9876) to[out=-93.6006,in=95.8183] (-0.98868,-3.66377) to[out=-84.1817,in=114.198] (-0.77978,-4.69786) to[out=-65.8024,in=197.314] (-0.47195,-4.96835) to[out=17.3138,in=246.381] (0.,-4.33013);
	    \draw[dashed] (0.,4.33013) to[out=-25.1035,in=136.988] (1.36785,3.41024) to[out=-43.012,in=118.858] (2.78721,1.60876) to[out=-61.1422,in=92.2759] (3.70839,-1.39763) to[out=-87.7241,in=55.435] (3.19828,-3.64409) to[out=-124.565,in=12.1067] (1.89146,-4.57866) to[out=-167.893,in=336.641] (0.,-4.33013);
	    \begin{scope}[yscale=0.5]
	    \draw (0,0) circle (5);
	    \end{scope}
	    \draw[ultra thick, \myblue, postaction={decorate}, decoration={markings, mark= at position 0.8 with {\arrow{stealth}}}] (0.,4.33013) to[out=-56.4429,in=126.094] (1.19919,2.82234) to[out=-53.9062,in=115.19] (1.83008,1.79074) to[out=-64.8099,in=107.636] (2.06238,1.21516) to[out=-72.3637,in=99.0162] (2.21369,0.60506) to[out=-80.9838,in=89.7018] (2.26905,-0.03232) to[out=-90.2982,in=80.119] (2.21745,-0.68665) to[out=-99.881,in=70.6651] (2.0526,-1.34445) to[out=-109.335,in=60.5435] (1.73158,-2.06814) to[out=-119.457,in=57.3305] (1.54508,-2.37764);
	    \draw[ultra thick, \myblue, postaction={decorate}, decoration={markings, mark= at position 0.3 with {\arrowreversed{stealth}}}] (0.,4.33013) to[out=-58.2765,in=97.2861] (0.19718,3.74287) to[out=-82.7139,in=90.342] (0.22669,3.29485) to[out=-89.658,in=91.4492] (0.18056,1.44754) to[out=-88.5508,in=95.9675] (0.21613,0.83688) to[out=-84.0325,in=101.554] (0.30742,0.21498) to[out=-78.4457,in=107.812] (0.46536,-0.40639) to[out=-72.188,in=114.366] (0.69467,-1.01405) to[out=-65.6338,in=124.061] (1.16509,-1.8682) to[out=-55.939,in=129.36] (1.54508,-2.37764);
	    \draw[ultra thick, \myblue, postaction={decorate}, decoration={markings, mark= at position 0.3 with {\arrow{stealth}}}](1.54508,-2.37764) to[out=-50.5573,in=113.598] (2.12786,-3.25206) to[out=-66.4023,in=83.4188] (2.26488,-3.92332) to[out=-96.5812,in=28.6551] (1.76144,-4.65665) to[out=-151.345,in=1.70086] (1.1116,-4.83765) to[out=-178.299,in=332.647] (0.42507,-4.70043) to[out=152.647,in=304.319] (0.,-4.33013);
	    \draw[ultra thick, \myblue, postaction={decorate}, decoration={markings, mark= at position 0.4 with {\arrowreversed{stealth}}}] (1.54508,-2.37764) to[out=-123.25,in=56.9412] (1.03467,-3.1523) to[out=-123.059,in=58.7358] (0.72831,-3.6348) to[out=-121.264,in=63.08] (0.44175,-4.13917) to[out=-116.92,in=68.4937] (0.29253,-4.46173) to[out=-111.506,in=76.1404] (0.20892,-4.71034) to[out=-103.86,in=112.364] (0.18683,-4.97717) to[out=-67.6361,in=284.082] (0.2097,-4.7846) to[out=104.082,in=295.283] (0.13361,-4.57005) to[out=115.283,in=0] (0.,-4.33013);
	    \draw[ultra thick] (0,0) circle (5);
	    \draw[fill=burgundy] (0., 4.33013) circle (0.2) (0., -4.33013) circle (0.2) (1.54508, -2.37764) circle (0.2);
	    \node[above] at (0,5) {$\scriptstyle\omega_3>0:\;{\bf n}({\phi})$};
	    \node[right] at (5,0) {$\scriptstyle\omega_3=0:\;{\bf e}({\phi})$};
	    \node[below] at (0,-5) {$\scriptstyle\omega_3<0:\; {\bf s}({\phi})$};
	    \node[right=0.2, fill=white, draw=black, rounded corners=3] at (2.2031, -0.768991) {$\color{burgundy} 1\cdot$};
	    \node[left=0.2, fill=white, draw=black, rounded corners=3] at (0.694666, -1.01405) {$\color{burgundy} 1\cdot$};
	    \node[right=0.2, fill=white, draw=black, rounded corners=3] at (2.2031, -3.44516) {$\color{burgundy} \omega_z\cdot$};
	    \node[left=0.2, fill=white, draw=black, rounded corners=3] at (0.694666, -3.69022) {$\color{burgundy} \omega_z^{-1}\cdot$};
    \end{tikzpicture}
    \end{array}$
    \caption{Parallel transport of free chirals}\label{fig:transports}
\end{figure}

Another relevant regime we will need to interpret geometrically is $|\omega_3|\gg |\omega_z|$.
There are two options $\omega_3<0$ and $\omega_3>0$.
We call respective wave-functions as ${\bf s}(\phi)$ (``south'') and ${\bf n}(\phi)$ (``north'').
For these functions a dependence on $\omega_z$, is negligible and we propose the following canonical normalization rules:
\begin{equation}
    {\bf n}(\phi)=e^{-|\omega_3||\phi|^2}|\omega_3|\psi_1^{\dagger}\psi_2^{\dagger}|0\rangle,\quad {\bf s}(\phi)=e^{-|\omega_3||\phi|^2}|0\rangle\,.
\end{equation}
Then we calculate the following transition amplitudes (see Fig.\ref{fig:transports}):
\begin{equation}\label{elemampls}
    \begin{aligned}
	&A)\quad\mathsf{T}_{{\bf e}\to{\bf n}}=\lim\lm_{\omega_3\to+\infty}\frac{\sqrt{|\omega_z|}}{\sqrt{\omega_3}}\sqrt{\frac{\omega_3+|\vec\omega|}{\omega_z}\frac{\omega_z}{|\omega_z|}}=1\,,\\
	&B)\quad\mathsf{T}_{{\bf n}\to{\bf e}}=\mathsf{T}_{{\bf e}\to{\bf n}}^{-1}=1\,,\\
	&C)\quad\mathsf{T}_{{\bf e}\to{\bf s}}=\lim\lm_{\omega_3\to-\infty}\frac{\sqrt{|\omega_z|}}{\sqrt{|\vec \omega|+\omega_3}/\omega_z}\sqrt{\frac{\omega_3+|\vec\omega|}{\omega_z}\frac{\omega_z}{|\omega_z|}}=\omega_z\,,\\
	&D)\quad\mathsf{T}_{{\bf s}\to{\bf e}}=\mathsf{T}_{{\bf e}\to{\bf s}}^{-1}=\omega_z^{-1}\,.
    \end{aligned}
\end{equation}
These transition amplitudes have a clear meaning of elementary equivariant integrals \eqref{equiints}.

Following \cite{Galakhov:2020upa} we assume that instantons perform the following type of geometric integration.
An instanton interpolates between points of target spaces we call $X$ at $\tau=-\infty$ and $Y$ at $\tau=+\infty$.
Matching endpoints of instanton trajectories carves a subspace (kernel) $\CK\subset X\times Y$.
One might imagine an instanton as a tiny string connecting D-branes of $X$ and $Y$.
Topological string theory \cite{Aganagic:2013jpa,Hori:2003ic,Dijkgraaf:2004te,Aspinwall:2009isa} endows $X$ and $Y$ with the structure of coherent sheave categories, where a natural transform from sheaves on $X$ to sheaves on $Y$ is a Fourier-Mukai transform with some kernel.
Since our ``string'' exists in $\CK$, it is natural to consider as a kernel for the Fourier-Mukai transform the structure sheaf of $\CK$.
For cohomologies the Fourier-Mukai transform descends to a simple pullback-pushforward transform $H^*(X)\to H^*(Y)$ or an equivariant integral.\footnote{To the reader a mystifying notion of the Fourier-Mukai transform might seem overcomplicated and unsuitable for this simple quantum mechanics discussion.
	However we should stress that in this case it is quite simple and easily applicable, more or less it descends to the ordinary Fourier transform with a delta-function kernel, yet supported on some submanifold.
	Due to localization \cite{Pestun:2016zxk} all the integration loci shrink to neighborhoods of fixed points where integrals are approximated by Gaussian integrals over target spaces -- linear spaces $\ldots\oplus\IC_{w_1}\oplus\IC_{w_2}\oplus\IC_{w_3}\oplus\ldots$ with an equivariant action, where integrals are calculated by simply applying \eqref{equiints} iteratively.}
So we propose the following geometric meaning for amplitudes:
\begin{equation}\label{geommeanampl}
    \mathsf{T}_{X\to Y}\overset{?}{=}\int\lm_{\CK\in X\times Y}{\bf e}(X)\,.
\end{equation}

Let us check our proposal in the case of elementary amplitudes.
To do so we should first restore the instanton trajectories for these degrees of freedom.
Let us remind that we agreed upon the canonical form of the supercharge containing operator $\p_q-\omega_3 \bar q$ that leads to an instanton equation $-\dot q-\omega_3 q=0$.
As before we assume that along the instanton field $\sigma$ flows as $u_2<\sigma<u_1$, and $u=u_1-u_2>0$.
We assume that functions $\omega_3$ for each degree of freedom in \eqref{elemampls} is linear in $\sigma$.
This allows us to restore the dependence of $\omega_3$ for all types of transitions in \eqref{elemampls} and solve respective instanton equations ($\Lambda\gg 1$):
\begingroup
\renewcommand*{\arraystretch}{1.3}
\begin{equation}
    \begin{array}{c|c|c|c|c}
	\mbox{Case} & \omega_3(u_2) & \omega_3(u_1) & \omega_3(\sigma) & \mbox{Inst. trj.}\\
	\hline
	A & 0 & u_1-u_2>0 & \sigma-u_2\geq 0 & A(+\infty)=e^{-\Lambda}A(-\infty)\\
	\hline
	B & u_1-u_2>0 & 0 & u_1-\sigma \geq 0 & B(+\infty)=e^{-\Lambda} B(-\infty)\\
	\hline
	C & 0 & u_2-u_1<0 &u_2-\sigma\leq 0 & C(+\infty)=e^{\Lambda}C(-\infty)\\
	\hline
	D & u_2-u_1< 0 & 0 & \sigma-u_1\leq 0 & D(+\infty)=e^{\Lambda}D(-\infty)\\
    \end{array}
\end{equation}
\endgroup
From this table we learn which field is heavy $|\omega_3|\neq 0$, it does not contribute to the Euler class and the variety as well.
The respective variety $X$ or $Y$ is just a point.
Also we note that if fields are related as $\Phi'=e^{\Lambda}\Phi$ this implies that field $\Phi=0$ on the instanton trajectory since $e^{\Lambda}$ is a large multiplier, whereas $\Phi'$ is not constrained, there is a family of trajectories with arbitrary value $\Phi'$ implying that the instanton string endpoint could travel across the whole $\Phi'$ complex plane.
This allows us to construct $\CK$ as a variety inside $X\times Y$ satisfying respective boundary conditions.

In the cases of all fields in \eqref{elemampls} we confirm \eqref{geommeanampl} by applying elementary integrals \eqref{equiints}:
\begingroup
\renewcommand*{\arraystretch}{1.3}
\begin{equation}\label{ABCD}
    \begin{array}{c|ccccc|c}
	\mbox{Case} & \CK & \subset & X & \times & Y & \int_{\CK}{\bf e}(X)\\
	\hline
	A & \IC_A\times\{{\rm pt}\} & & \IC_A & \times & \{{\rm pt}\} & \omega_z/\omega_z=1\\
	\hline
	B & \{{\rm pt}\}\times \{{\rm pt}\} & & \{{\rm pt}\} & \times & \IC_B & 1\\
	\hline
	C & \{{\rm pt}\}\times \{{\rm pt}\} & & \IC_C & \times & \{{\rm pt}\} & \omega_z\\
	\hline
	D & \{{\rm pt}\}\times \IC_D & & \{{\rm pt}\} & \times & \IC_D & \omega_z^{-1}
    \end{array}
\end{equation}
\endgroup

We conclude this subsection with a revision of the geometric tunneling amplitude physical meaning.
We think of (algebraic) varieties $X$ and $Y$ as neighborhoods (tangent spaces) of fixed points (quasi-classical vacua).
These tangent spaces are spanned by chiral degrees of freedom $\phi_i$ with $(\omega_z,\omega_3)=(w_i,0)$.
Respective wave functions contain pieces corresponding to respective Euler classes:
\begin{equation}
    \Psi(X)=\prod\lm_{a\in X}{\bf e}(\phi_a|w_a)\times \Psi(\mbox{other d.o.f.}),\quad \Psi(Y)=\prod\lm_{b\in Y}{\bf e}(\phi_b|w_b)\times \Psi(\mbox{other d.o.f.})\,.
\end{equation}
Points in $X$ and $Y$ connected by instanton (gradient flow) trajectories is called a kernel $\CK\in X\times Y$.
This constraint of connectedness imposes $\CK$ as algebraic (even linear since we are interested in the tangent spaces only) relations on coordinates of $X$ and $Y$ allowing us to decompose them in the following way:
\begin{equation}
    X=A\oplus C\oplus E, \quad Y=D\oplus B\oplus E'\,,
\end{equation}
so that $\CK$ imposes the following relations (cf. \eqref{ABCD}):
\begin{equation}
    C=0,\quad B=0,\quad E=E',\quad A\mbox{ and }D\mbox{ are unconstrained}\,.
\end{equation}
In \eqref{ABCD} we have not seen the type of fields $E$ and $E'$, we treat them as those degrees of freedom not coupled to the instanton and remaining steady along the tunneling trajectory.
In this case the very space $\CK$ is decomposed as the following:
\begin{equation}
    \CK=A\oplus D\oplus E=A\oplus D\oplus E'\,.
\end{equation}
Thus following \eqref{ABCD} we arrive to the following geometric tunneling amplitude:
\begin{tcolorbox}
\begin{equation}\label{geomtunnel}
    \mathsf{T}_{X\to Y}=\int\lm_{\CK}{\bf e}(X)=\frac{\prod\lm_{a\in X}w_a}{\prod\lm_{b\in \CK}w_b}=\frac{\prod\lm_{a\in (A\oplus C\oplus E)}w_a}{\prod\lm_{b\in (A\oplus D\oplus E)}w_b}=\frac{\prod\lm_{a\in C}w_a}{\prod\lm_{b\in D}w_b}\,.
\end{equation}
\end{tcolorbox}

\subsection{Towards quiver Yangians}

\subsubsection{Young diagrams}

\begin{figure}[ht!]
	\centering
	\begin{tikzpicture}
		\draw[thick, postaction={decorate}, decoration={markings, mark= at position 0.8 with {\arrow{stealth}}}] (0,0) to[out=30, in=90] (1,0) to[out=270,in=-30] (0,0);
		\node[right] at (1,0) {$\scriptstyle B_2$};
		\begin{scope}[rotate=120]
			\draw[thick, postaction={decorate}, decoration={markings, mark= at position 0.8 with {\arrow{stealth}}}] (0,0) to[out=30, in=90] (1,0) to[out=270,in=-30] (0,0);
			\node[above] at (1,0) {$\scriptstyle B_1$};
		\end{scope}
		\begin{scope}[rotate=-120]
			\draw[thick, postaction={decorate}, decoration={markings, mark= at position 0.8 with {\arrow{stealth}}}] (0,0) to[out=30, in=90] (1,0) to[out=270,in=-30] (0,0);
			\node[below] at (1,0) {$\scriptstyle B_3$};
		\end{scope}
		\draw [thick, postaction={decorate}, decoration={markings, mark= at position 0.6 with {\arrow{stealth}}}]  (-2,0.08) -- (0,0.08) node[pos=0.5,above] {$\scriptstyle R$};
		\draw [thick, postaction={decorate}, decoration={markings, mark= at position 0.6 with {\arrow{stealth}}}]  (0,-0.08) -- (-2,-0.08) node[pos=0.5,below] {$\scriptstyle S$};
		\draw[fill=burgundy] (0,0) circle (0.15);
		\begin{scope}[shift={(-2,0)}]
			\draw[fill=\myblue] (-0.15,-0.15) -- (-0.15,0.15) -- (0.15,0.15) -- (0.15,-0.15) -- cycle;
		\end{scope}
		\node[below] at (-2,-0.15) {$\scriptstyle k$};
		\node[above right] at (0.15,0.15) {$\scriptstyle n$};
		\node[right] at (2,0) {$W=\Tr \,B_3\left(\left[B_1,B_2\right]+RS\right)$};
	\end{tikzpicture}
	\caption{A quiver for ADHM moduli space}\label{fig:ADHMquiver}
\end{figure}
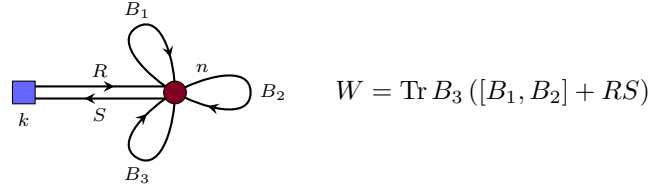

Here we do not attempt to cover the most generic cases of quiver Yangian algebras, rather we would revise the simplest case of affine Yangian $Y(\widehat{\fg\fl}_1)$, and the corresponding quiver encoding gauge-matter content of the theory is depicted in Fig.\ref{fig:ADHMquiver}.
This theory is known as an ADHM quiver description of the Hilbert scheme of points on $\IC^2$ \cite{nakajima1999lectures}.
Fixed points on a more generic family of quiver varieties corresponding to toric Calabi-Yau 3-folds \cite{Okounkov:2003sp,Ooguri:2009ijd,Yamazaki:2010fz,Li:2020rij,Bao:2024ygr} -- ground states of respective Hamiltonians -- are classified by crystals, that are reduced in the case of a quiver depicted in Fig.\ref{fig:ADHMquiver} to Young diagrams.

First, we discuss the case $k=1$.

The details of this identification could be found, for example, in \cite{Galakhov:2024bzs}.
Field $B_3$ plays the role of a Lagrange multiplier in superpotential $W$, so the motion of a particle on the quiver variety is confined to an algebraic subvariety given by $B_3=0$, $\left[B_1,B_2\right]+RS=0$.
Moreover, some of the degrees of freedom are gauge.
The gauge group may be lifted from the unitary group $U(n)$ to its complexification $GL(n,\IC)$ \cite{Galakhov:2020vyb}, and we will work here under this assumption.
Fields $B_i$, $R$ and $S$ are charged with respect to the gauge and flavor fields and in addition to the constant $GL(1,\IC)$ field of the $\Omega$-background \cite{Nekrasov:2003rj}.
We could summarize all these charges in the following table:
\begingroup
\renewcommand*{\arraystretch}{1.3}
\begin{equation}
	\begin{array}{c|c|c|c|c|c}
		& B_1 & B_2 & B_3 & R & S\\
		\hline
		GL(n,\IC)_{\rm gauge} & {\rm Ad}_{\Phi} & {\rm Ad}_{\Phi} & {\rm Ad}_{\Phi} & \Phi\cdot * & *\cdot(-\Phi) \\
		\hline
		GL(1,\IC)_{\rm flav} & 0 & 0 & 0 & -\mu & \mu \\
		\hline
		GL(1,\IC)_{\Omega} & -\epsilon_1 & -\epsilon_2 & \epsilon_1+\epsilon_2 & 0 & -\epsilon_1-\epsilon_2
	\end{array}
\end{equation}
\endgroup

Young diagrams classify all the zeroes of the potential up to $GL(n,\IC)$.
The identification goes as follows.
First, one chooses a preferred $GL(n,\IC)$ basis.
It is convenient to enumerate boxes $i=1,\ldots,n$ of Young diagram $\lambda$ in any order and to identify the $i^{\rm th}$ box with the $i^{\rm th}$ vector of the basis.
The fields in the gauge multiplet are all diagonal and are defined by coordinates $x_i$, $y_i$ of boxes.
So, for the $i^{\rm th}$ eigen value we have:
\begin{equation}
	(\Phi)_{ii}=\mu+\epsilon_1 x_i+\epsilon_2 y_i,\quad (\Sigma)_{ii}=u\,,
\end{equation}
whereas for the flavor fields we set the following parameterization: $\Sigma_{\rm flav}=(u)$, $\Phi_{\rm flav}=(\mu)$.
Matrix elements of the chiral fields are chosen in the following way.
For $B_a$, for boxes $i$ and $j$, if $(x_i,y_i)=(x_j,x_j)+e_a$, where $e_1=(1,0)$ and $e_2=(0,1)$, then element $(B_a)_{ij}$ is non-zero.
For vector $R$ the only $k^{\rm th}$ element is non-zero, where $k$ is the index of a box with coordinates $(0,0)$.
Remaining matrix elements are set to zero.

For example,
\begin{equation}
    \begin{aligned}
	&\begin{array}{c}
		\begin{tikzpicture}[scale=0.6]
			\foreach \x/\y/\z/\w in {0/0/4/0, 0/1/4/1, 0/2/3/2, 0/3/1/3, 0/0/0/3, 1/0/1/3, 2/0/2/2, 3/0/3/2, 4/0/4/1}
			{
				\draw[thick] (\x,\y) -- (\z,\w);
			}
			\foreach \x/\y in {0/0,1/0,2/0,3/0, 0/1, 1/1, 2/1, 0/2}
			{
				\node at (\x+0.5,\y+0.5) {\tiny (\x,\y)};
			}
		\end{tikzpicture}
	\end{array},\quad \begin{array}{c}
		\begin{tikzpicture}[scale=0.7]
			\foreach \x/\y in {0/0, 1/0, 2/0, 0/1, 1/1}
			{
				\draw[thick, postaction={decorate}, decoration={markings, mark= at position 0.6 with {\arrow{stealth}}}] (\x,\y) -- (\x+1,\y);
			}
			\foreach \x/\y in {0/0, 1/0, 2/0, 0/1}
			{
				\draw[thick, postaction={decorate}, decoration={markings, mark= at position 0.6 with {\arrow{stealth}}}] (\x,\y) -- (\x,\y+1);
			}
			\tikzset{box/.style ={fill=white!80!violet}}
			\foreach \x/\y/\z in {0/0/1, 1/0/2, 2/0/3, 3/0/4, 0/1/5, 1/1/6, 2/1/7, 0/2/8}
			{
				\draw[box, thick, rounded corners=2] (\x-0.25, \y-0.25) -- (\x-0.25, \y+0.25) -- (\x+0.25, \y+0.25) -- (\x+0.25, \y-0.25) -- cycle;
				\node at (\x,\y) {\tiny \z};
			}
		\end{tikzpicture}
	\end{array}\,,
	\begin{array}{l}
	\Phi=\mu\bbone_{8\times 8}+{\rm diag}\left(0, \epsilon_1,2\epsilon_1,3\epsilon_1,\epsilon_2,\epsilon_1+\epsilon_2,2\epsilon_1+\epsilon_2,2\epsilon_2\right)\,,\\
	B_1=E_{21}+E_{32}+E_{43}+E_{65}+E_{76}\,,\\
	    B_2=E_{51}+E_{62}+E_{73}+E_{85}\,,\\
	    R=(1,0,\ldots,0)^T\,,\\
	    S=(0,0,\ldots,0)\,,
	\end{array}
    \end{aligned}
\end{equation}
where $E_{ij}$ is a unit matrix with an $(i,j)$ unit entry.

\subsubsection{Multiple wells, crystal chains}

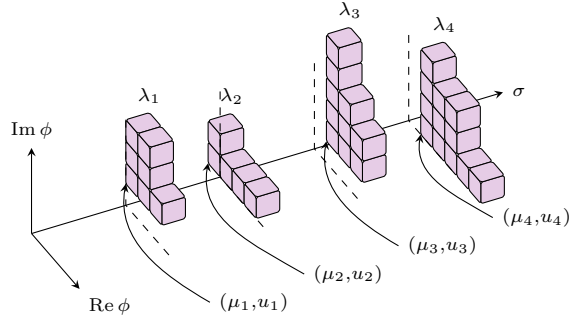
\begin{figure}[ht!]
    \centering
    \begin{tikzpicture}[scale=0.35]
    	\tikzset{box/.style ={fill=white!80!violet}}
    	\draw[-stealth] (0,0) -- (0., 3.23607);
    	\draw[-stealth] (0,0) -- (1.81596, -2.09488);
    	\draw[-stealth] (0,0) -- (17.8201, 5.33698);
    	\node[right] at (17.8201, 5.33698) {$\scriptstyle \sigma$};
    	\node[above] at (0., 3.23607) {$\scriptstyle {\rm Im}\,\phi$};
    	\node[below right] at (1.81596, -2.09488) {$\scriptstyle {\rm Re}\,\phi$};
    	\foreach \a/\b/\c/\d/\e/\f/\g/\h in {15.1641/5.64919/16.0551/5.91604/16.0551/6.72506/15.1641/6.45821, 16.0721/2.98372/16.9631/3.25057/16.9631/4.05958/16.0721/3.79273, 16.0721/3.79273/16.9631/4.05958/16.9631/4.8686/16.0721/4.60175, 16.98/1.12726/17.8711/1.39411/17.8711/2.20313/16.98/1.93628, 14.7101/3.74586/15.1641/3.22214/15.1641/4.03116/14.7101/4.55488, 14.7101/4.55488/15.1641/4.03116/15.1641/4.84018/14.7101/5.3639, 14.7101/5.3639/15.1641/4.84018/15.1641/5.64919/14.7101/6.17291, 14.7101/6.17291/15.1641/5.64919/15.1641/6.45821/14.7101/6.98193, 15.1641/3.22214/15.6181/2.69842/15.6181/3.50744/15.1641/4.03116, 15.1641/4.03116/15.6181/3.50744/15.6181/4.31646/15.1641/4.84018, 15.1641/4.84018/15.6181/4.31646/15.6181/5.12547/15.1641/5.64919, 15.6181/2.69842/16.0721/2.1747/16.0721/2.98372/15.6181/3.50744, 15.6181/3.50744/16.0721/2.98372/16.0721/3.79273/15.6181/4.31646, 15.6181/4.31646/16.0721/3.79273/16.0721/4.60175/15.6181/5.12547, 16.0721/2.1747/16.5261/1.65098/16.5261/2.46/16.0721/2.98372, 16.5261/1.65098/16.98/1.12726/16.98/1.93628/16.5261/2.46, 14.7101/6.98193/15.1641/6.45821/16.0551/6.72506/15.6011/7.24878, 15.1641/5.64919/15.6181/5.12547/16.5091/5.39232/16.0551/5.91604, 15.6181/5.12547/16.0721/4.60175/16.9631/4.8686/16.5091/5.39232, 16.0721/2.98372/16.5261/2.46/17.4171/2.72685/16.9631/3.25057, 16.5261/2.46/16.98/1.93628/17.8711/2.20313/17.4171/2.72685}
    	{
    		\draw[box, rounded corners=1] (\a,\b) -- (\c,\d) -- (\e,\f) -- (\g,\h) -- cycle;
    	}
    	\foreach \a/\b/\c/\d/\e/\f/\g/\h in {11.6001/5.39081/12.4911/5.65766/12.4911/6.46668/11.6001/6.19983, 11.6001/6.19983/12.4911/6.46668/12.4911/7.2757/11.6001/7.00885, 12.054/4.05808/12.9451/4.32493/12.9451/5.13394/12.054/4.86709, 12.508/1.91632/13.399/2.18317/13.399/2.99219/12.508/2.72534, 12.508/2.72534/13.399/2.99219/13.399/3.8012/12.508/3.53436, 11.1461/3.48748/11.6001/2.96376/11.6001/3.77278/11.1461/4.2965, 11.1461/4.2965/11.6001/3.77278/11.6001/4.5818/11.1461/5.10552, 11.1461/5.10552/11.6001/4.5818/11.6001/5.39081/11.1461/5.91453, 11.1461/5.91453/11.6001/5.39081/11.6001/6.19983/11.1461/6.72355, 11.1461/6.72355/11.6001/6.19983/11.6001/7.00885/11.1461/7.53257, 11.6001/2.96376/12.054/2.44004/12.054/3.24906/11.6001/3.77278, 11.6001/3.77278/12.054/3.24906/12.054/4.05808/11.6001/4.5818, 11.6001/4.5818/12.054/4.05808/12.054/4.86709/11.6001/5.39081, 12.054/2.44004/12.508/1.91632/12.508/2.72534/12.054/3.24906, 12.054/3.24906/12.508/2.72534/12.508/3.53436/12.054/4.05808, 11.1461/7.53257/11.6001/7.00885/12.4911/7.2757/12.0371/7.79942, 11.6001/5.39081/12.054/4.86709/12.9451/5.13394/12.4911/5.65766, 12.054/4.05808/12.508/3.53436/13.399/3.8012/12.9451/4.32493}
    	{
    		\draw[box, rounded corners=1] (\a,\b) -- (\c,\d) -- (\e,\f) -- (\g,\h) -- cycle;
    	}
    	\foreach \a/\b/\c/\d/\e/\f/\g/\h in {7.12805/2.94381/8.01906/3.21066/8.01906/4.01967/7.12805/3.75283, 8.49002/0.56363/9.38103/0.830479/9.38103/1.6395/8.49002/1.37265, 6.67406/2.65851/7.12805/2.13479/7.12805/2.94381/6.67406/3.46753, 6.67406/3.46753/7.12805/2.94381/7.12805/3.75283/6.67406/4.27655, 7.12805/2.13479/7.58204/1.61107/7.58204/2.42009/7.12805/2.94381, 7.58204/1.61107/8.03603/1.08735/8.03603/1.89637/7.58204/2.42009, 8.03603/1.08735/8.49002/0.56363/8.49002/1.37265/8.03603/1.89637, 6.67406/4.27655/7.12805/3.75283/8.01906/4.01967/7.56507/4.54339, 7.12805/2.94381/7.58204/2.42009/8.47305/2.68694/8.01906/3.21066, 7.58204/2.42009/8.03603/1.89637/8.92704/2.16322/8.47305/2.68694, 8.03603/1.89637/8.49002/1.37265/9.38103/1.6395/8.92704/2.16322}
    	{
    		\draw[box, rounded corners=1] (\a,\b) -- (\c,\d) -- (\e,\f) -- (\g,\h) -- cycle;
    	}
    	\foreach \a/\b/\c/\d/\e/\f/\g/\h in {4.47201/1.63799/5.36301/1.90484/5.36301/2.71385/4.47201/2.44701, 4.47201/2.44701/5.36301/2.71385/5.36301/3.52287/4.47201/3.25602, 4.926/0.305251/5.817/0.5721/5.817/1.38112/4.926/1.11427, 3.56403/1.87641/4.01802/1.35269/4.01802/2.16171/3.56403/2.68543, 3.56403/2.68543/4.01802/2.16171/4.01802/2.97073/3.56403/3.49445, 3.56403/3.49445/4.01802/2.97073/4.01802/3.77974/3.56403/4.30346, 4.01802/1.35269/4.47201/0.828972/4.47201/1.63799/4.01802/2.16171, 4.01802/2.16171/4.47201/1.63799/4.47201/2.44701/4.01802/2.97073, 4.01802/2.97073/4.47201/2.44701/4.47201/3.25602/4.01802/3.77974, 4.47201/0.828972/4.926/0.305251/4.926/1.11427/4.47201/1.63799, 3.56403/4.30346/4.01802/3.77974/4.90902/4.04659/4.45503/4.57031, 4.01802/3.77974/4.47201/3.25602/5.36301/3.52287/4.90902/4.04659, 4.47201/1.63799/4.926/1.11427/5.817/1.38112/5.36301/1.90484}
    	{
    		\draw[box, rounded corners=1] (\a,\b) -- (\c,\d) -- (\e,\f) -- (\g,\h) -- cycle;
    	}
    	\draw[dashed] (3.56403,4.30346) -- (3.56403,1.0674) -- (5.37999,-1.02749) (7.12805,5.37086) -- (7.12805,2.13479) -- (8.94401,0.0399094) (10.6921,6.43826) -- (10.6921,3.20219) -- (12.508,1.10731) (14.2561,7.50565) -- (14.2561,4.26958) -- (16.0721,2.1747);
    	\node[above] at (4.45503, 4.57031) {$\scriptstyle \lambda_1$};
    	\node[above] at (7.56507, 4.54339) {$\scriptstyle \lambda_2$};
    	\node[above] at (12.0371, 7.79942) {$\scriptstyle \lambda_3$};
    	\node[above] at (15.6011, 7.24878) {$\scriptstyle \lambda_4$};
    	\node[right] at (6.74196, -2.59865) {$\scriptstyle (\mu_1,u_1)$};
    	\node[right] at (10.306, -1.53125) {$\scriptstyle (\mu_2,u_2)$};
    	\node[right] at (13.87, -0.463856) {$\scriptstyle (\mu_3,u_3)$};
    	\node[right] at (17.434, 0.603539) {$\scriptstyle (\mu_4,u_4)$};
    	\draw[-stealth] (6.74196, -2.59865) to[out=150, in=260] (3.56403, 1.8764);
    	\draw[-stealth] (10.306, -1.53125) to[out=150, in=260] (6.67406, 2.65851);
    	\draw[-stealth] (13.87, -0.463856) to[out=150, in=260] (11.1461, 3.48748);
    	\draw[-stealth] (17.434, 0.603539) to[out=150, in=260] (14.7101, 3.74586);
    \end{tikzpicture}
    \caption{Crystal chain mimicking spectral tensor product $(\lambda_1)_{\mu_1}\otimes (\lambda_2)_{\mu_2}\otimes (\lambda_3)_{\mu_3}\otimes (\lambda_4)_{\mu_4}\otimes \ldots$. Ordering of tensor factors is defined by ordering of $u_1<u_2<u_3<u_4<\ldots$.}\label{fig:chain}
\end{figure}

Now we consider the case $k>1$.
So that the fields for the flavor node are given by the following fixed values:
\begin{equation}
    U(k)_{\rm flav}: \quad \Sigma_{\rm fxd}={\rm diag}\left(u_1,u_2,\ldots,u_k\right),\quad \Phi_{\rm fxd}={\rm diag}\left(\mu_1,\mu_2,\ldots,\mu_k\right)\,.
\end{equation}

In this case fixed points -- classical vacua -- are described by $k$-\emph{tuples} of Young diagrams.
One places these tuples in the 3d space (see Fig.\ref{fig:chain}).
Coordinates in this space correspond to eigenvalues $\phi$ and $\sigma$ of matrices $\Phi$ and $\Sigma$.
The $k^{\rm th}$ tuple element is located in the respective $\sigma=u_k$ plane, and we order elements in the tuple according to $u_1<u_2<\ldots<u_k$.
A shift in the $\phi$-plane of the $i^{\rm th}$ Young diagram is given by $\mu_i$.
We will identify such a tuple with an element of a spectral tensor degree of $Y(\widehat{\fg\fl}_1)$ modules, where $\mu_i$ are spectral parameters:
\begin{equation}
    |\vec{\lambda}\rangle=|\lambda_1\rangle_{\mu_1}\otimes |\lambda_2\rangle_{\mu_2}\otimes\ldots|\lambda_k\rangle_{\mu_k}\,.
\end{equation}
This 3d arrangement of boxes should not be confused in general with 3d Young diagrams \cite{Galakhov:2020vyb}, however under a certain choice of parameters this arrangement becomes a 3d Young diagram constructed from the tensor degrees of 2d diagrams \cite{FJMM, Noshita:2021dgj}.

Let us denote $n_i=|\lambda_i|$.
And assume that the enumeration of boxes runs continuously form a layer to a layer, so we enumerate boxes of $\lambda_i$ by numbers in between $n_1+\ldots+n_{i-1}+1$ and $n_1+\ldots+n_{i}$.
In this case chiral fields acquire a transparent block-diagonal structure:
\begin{equation}
    B=\left(\begin{array}{ccc}
	B(\lambda_1) &&\\
	& \ddots &\\
	&& B(\lambda_k)\\
    \end{array}\right)\,,
\end{equation}
where $B(\lambda_i)$ are expectation values of the respective fields for diagrams $\lambda_i$.
So for the vector multiplet expectation values we have:
\begin{equation}
    \Phi=\left(\begin{array}{ccc}
	\mu_1\bbone_{n_1\times n_1}+\Phi(\lambda_1) &&\\
	& \ddots &\\
	&& \mu_k\bbone_{n_k\times n_k}+\Phi(\lambda_k) \\
    \end{array}\right),\;\Sigma=\left(\begin{array}{ccc}
	u_1\bbone_{n_1\times n_1}&&\\
	& \ddots &\\
	&& u_k\bbone_{n_k\times n_k}\\
    \end{array}\right)\,.
\end{equation}

The height function for such a configuration has the following value:
\begin{equation}
    h=r\sum\lm_{i=1}^kn_k u_k\,,
\end{equation}
so that moving, say, a box between the $i^{\rm th}$ and $j^{\rm th}$ partitions would change the height function by $r(u_j-u_i)$.
In other words this action requires work, that makes partition planes $u_i$ separated by potential walls.
The case of $k>1$ for quiver SQM is the \emph{direct analog} of the $k$-well potential we considered in Sec.\ref{sec:TAC}.

\subsubsection{Raising and lowering operators}

\begin{figure}[ht!]
	\centering
	\begin{tikzpicture}[scale=0.27,yscale=-1]
		\foreach \x/\y/\z/\w in {0/0/1/0, 0/-1/1/-1, 0/0/0/-1, 1/0/1/-1, 0/-2/1/-2, 0/-1/0/-2, 1/-1/1/-2, 0/-3/1/-3, 0/-2/0/-3, 1/-2/1/-3, 0/-4/1/-4, 0/-3/0/-4, 1/-3/1/-4, 0/-5/1/-5, 0/-4/0/-5, 1/-4/1/-5, 0/-6/1/-6, 0/-5/0/-6, 1/-5/1/-6, 1/0/2/0, 1/-1/2/-1, 2/0/2/-1, 1/-2/2/-2, 2/-1/2/-2, 1/-3/2/-3, 2/-2/2/-3, 1/-4/2/-4, 2/-3/2/-4, 1/-5/2/-5, 2/-4/2/-5,  2/0/3/0, 2/-1/3/-1, 3/0/3/-1, 2/-2/3/-2, 3/-1/3/-2, 2/-3/3/-3, 3/-2/3/-3, 2/-4/3/-4, 3/-3/3/-4, 3/0/4/0, 3/-1/4/-1, 4/0/4/-1, 3/-2/4/-2, 4/-1/4/-2, 3/-3/4/-3, 4/-2/4/-3, 3/-4/4/-4, 4/-3/4/-4, 4/0/5/0, 4/-1/5/-1, 5/0/5/-1, 4/-2/5/-2, 5/-1/5/-2, 4/-3/5/-3, 5/-2/5/-3, 5/0/6/0, 5/-1/6/-1, 6/0/6/-1, 5/-2/6/-2, 6/-1/6/-2, 5/-3/6/-3, 6/-2/6/-3, 6/0/7/0, 6/-1/7/-1, 7/0/7/-1, 6/-2/7/-2, 7/-1/7/-2, 7/0/8/0, 7/-1/8/-1, 8/0/8/-1, 8/0/9/0, 8/-1/9/-1, 9/0/9/-1}
		{
			\draw (\x,\y) -- (\z,\w);
		}
		\draw[-stealth] (0,0) -- (0,-7);
		\draw[-stealth] (0,0) -- (10,0);
		\node[left] at (0,-7) {$\scriptstyle y$};
		\node[right] at (10,0) {$\scriptstyle x$};
		\node[below] at (4,1) {(a) $\lambda$};
		\begin{scope}[shift={(15,0)}]
			\foreach \x/\y/\z/\w in {0/0/1/0, 0/-1/1/-1, 0/0/0/-1, 1/0/1/-1, 0/-2/1/-2, 0/-1/0/-2, 1/-1/1/-2, 0/-3/1/-3, 0/-2/0/-3, 1/-2/1/-3, 0/-4/1/-4, 0/-3/0/-4, 1/-3/1/-4, 0/-5/1/-5, 0/-4/0/-5, 1/-4/1/-5, 0/-6/1/-6, 0/-5/0/-6, 1/-5/1/-6, 1/0/2/0, 1/-1/2/-1, 2/0/2/-1, 1/-2/2/-2, 2/-1/2/-2, 1/-3/2/-3, 2/-2/2/-3, 1/-4/2/-4, 2/-3/2/-4, 1/-5/2/-5, 2/-4/2/-5,  2/0/3/0, 2/-1/3/-1, 3/0/3/-1, 2/-2/3/-2, 3/-1/3/-2, 2/-3/3/-3, 3/-2/3/-3, 2/-4/3/-4, 3/-3/3/-4, 3/0/4/0, 3/-1/4/-1, 4/0/4/-1, 3/-2/4/-2, 4/-1/4/-2, 3/-3/4/-3, 4/-2/4/-3, 3/-4/4/-4, 4/-3/4/-4, 4/0/5/0, 4/-1/5/-1, 5/0/5/-1, 4/-2/5/-2, 5/-1/5/-2, 4/-3/5/-3, 5/-2/5/-3, 5/0/6/0, 5/-1/6/-1, 6/0/6/-1, 5/-2/6/-2, 6/-1/6/-2, 5/-3/6/-3, 6/-2/6/-3, 6/0/7/0, 6/-1/7/-1, 7/0/7/-1, 6/-2/7/-2, 7/-1/7/-2, 7/-2/6/-3, 7/0/8/0, 7/-1/8/-1, 8/0/8/-1, 8/0/9/0, 8/-1/9/-1, 9/0/9/-1}
			{
				\draw (\x,\y) -- (\z,\w);
			}
			\foreach \x/\y in {0/6, 4/3, 7/1, 9/0}
			{
				\draw[fill=burgundy] (\x,-\y) -- (\x+1,-\y) -- (\x+1,-\y-1) -- (\x,-\y-1) -- cycle;
			}
			\foreach \x/\y in {1/5, 2/4, 6/2}
			{
				\draw[fill=burgundy] (\x,-\y) -- (\x+1,-\y) -- (\x+1,-\y-1) -- (\x,-\y-1) -- cycle;
			}
			\node[below] at (4,1) {(b) ${\rm Add}(\lambda)$};
		\end{scope}
		\begin{scope}[shift={(30,0)}]
			\foreach \x/\y/\z/\w in {0/0/1/0, 0/-1/1/-1, 0/0/0/-1, 1/0/1/-1, 0/-2/1/-2, 0/-1/0/-2, 1/-1/1/-2, 0/-3/1/-3, 0/-2/0/-3, 1/-2/1/-3, 0/-4/1/-4, 0/-3/0/-4, 1/-3/1/-4, 0/-5/1/-5, 0/-4/0/-5, 1/-4/1/-5, 0/-6/1/-6, 0/-5/0/-6, 1/-5/1/-6, 1/0/2/0, 1/-1/2/-1, 2/0/2/-1, 1/-2/2/-2, 2/-1/2/-2, 1/-3/2/-3, 2/-2/2/-3, 1/-4/2/-4, 2/-3/2/-4, 1/-5/2/-5, 2/-4/2/-5, 2/0/3/0, 2/-1/3/-1, 3/0/3/-1, 2/-2/3/-2, 3/-1/3/-2, 2/-3/3/-3, 3/-2/3/-3, 2/-4/3/-4, 3/-3/3/-4, 3/0/4/0, 3/-1/4/-1, 4/0/4/-1, 3/-2/4/-2, 4/-1/4/-2, 3/-3/4/-3, 4/-2/4/-3, 3/-4/4/-4, 4/-3/4/-4, 4/0/5/0, 4/-1/5/-1, 5/0/5/-1, 4/-2/5/-2, 5/-1/5/-2, 4/-3/5/-3, 5/-2/5/-3, 5/0/6/0, 5/-1/6/-1, 6/0/6/-1, 5/-2/6/-2, 6/-1/6/-2, 5/-3/6/-3, 6/-2/6/-3, 6/0/7/0, 6/-1/7/-1, 7/0/7/-1, 6/-2/7/-2, 7/-1/7/-2, 7/0/8/0, 7/-1/8/-1, 8/0/8/-1, 8/0/9/0, 8/-1/9/-1, 9/0/9/-1}
			{
				\draw (\x,\y) -- (\z,\w);
			}
			\foreach \x/\y in {0/5, 1/4, 5/2, 6/1}
			{
				\draw[fill=\myblue] (\x,-\y) -- (\x+1,-\y) -- (\x+1,-\y-1) -- (\x,-\y-1) -- cycle;
			}
			\foreach \x/\y in {3/3, 8/0}
			{
				\draw[fill=\myblue] (\x,-\y) -- (\x+1,-\y) -- (\x+1,-\y-1) -- (\x,-\y-1) -- cycle;
			}
			\node[below] at (4,1) {(c) ${\rm Rem}(\lambda)$};
		\end{scope}
	\end{tikzpicture}
	\caption{Sets ${\rm Add}(\star)$ and ${\rm Rem}(\star)$.}
	\label{fig:AddRemYD}
\end{figure}
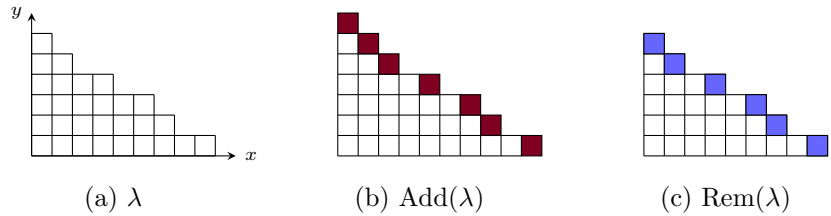

To construct raising and lowering operators following the prescription discussed in Sec.\ref{sec:TAC} we consider first just a pair of wells ($k=2$) and put in the second probe well just a single particle at maximum.
So, first, we consider tunneling processes:
\begin{equation}\label{trajectoriesYang}
	\begin{aligned}
		&\mbox{raising:} \quad |\Box\rangle_{\mu}\otimes |\lambda\rangle_{0}\longrightarrow |\varnothing\rangle_{\mu}\otimes (\lambda+\Box)_0, \quad u<0\,,\\
		&\mbox{lowering:} \quad |\lambda\rangle_{0}\otimes |\varnothing\rangle_{\mu}\longrightarrow (\lambda-\Box)_0\otimes |\Box\rangle_{\mu},\quad u>0\,.
	\end{aligned}
\end{equation}
Without loss of generality we have put the ``heavy'' well containing crystal $\lambda$ in the center of coordinates, and the probe well in generic position $(\phi,\sigma)=(\mu,u)$.
As we have discussed in Sec.\ref{sec:meridian} the tunneling is supported by instantons only flowing in the direction of increasing $\sigma$.
So for these processes to take place $u<0$ for the raising process and $u>0$ for the lowering process.
The resulting states in these two processes in the main well are again classical vacua and correspond to Young diagrams have one box more or less respectively we denoted as $\lambda+\Box$ and $\lambda-\Box$.
The sets of boxes in the lattice that could be added to or removed from diagram $\lambda$ are denoted as ${\rm Add}(\lambda)$ and ${\rm Rem}(\lambda)$ (see an example in Fig.\ref{fig:AddRemYD}).

We will describe both tunneling trajectories \eqref{trajectoriesYang} at the same moment, as it is clear one is the inverse of the other.
Let us assume $|\lambda|=n-1$, and the migrating box has index $n$.
Also assume that its nearest neighbors in $\lambda$ have indices $i$ and $j$, then we parameterize the quiver matrices in the following way:
\begin{equation}
	\begin{aligned}
	&\lambda+\Box=\begin{array}{c}
		\begin{tikzpicture}[scale=0.3, yscale=-1]
			\foreach \x/\y/\z/\w in {0/0/1/0, 0/-1/1/-1, 0/0/0/-1, 1/0/1/-1, 0/-2/1/-2, 0/-1/0/-2, 1/-1/1/-2, 0/-3/1/-3, 0/-2/0/-3, 1/-2/1/-3, 0/-4/1/-4, 0/-3/0/-4, 1/-3/1/-4, 0/-5/1/-5, 0/-4/0/-5, 1/-4/1/-5, 0/-6/1/-6, 0/-5/0/-6, 1/-5/1/-6, 1/0/2/0, 1/-1/2/-1, 2/0/2/-1, 1/-2/2/-2, 2/-1/2/-2, 1/-3/2/-3, 2/-2/2/-3, 1/-4/2/-4, 2/-3/2/-4, 1/-5/2/-5, 2/-4/2/-5,  2/0/3/0, 2/-1/3/-1, 3/0/3/-1, 2/-2/3/-2, 3/-1/3/-2, 2/-3/3/-3, 3/-2/3/-3, 2/-4/3/-4, 3/-3/3/-4, 3/0/4/0, 3/-1/4/-1, 4/0/4/-1, 3/-2/4/-2, 4/-1/4/-2, 3/-3/4/-3, 4/-2/4/-3, 3/-4/4/-4, 4/-3/4/-4, 4/0/5/0, 4/-1/5/-1, 5/0/5/-1, 4/-2/5/-2, 5/-1/5/-2, 4/-3/5/-3, 5/-2/5/-3, 5/0/6/0, 5/-1/6/-1, 6/0/6/-1, 5/-2/6/-2, 6/-1/6/-2, 5/-3/6/-3, 6/-2/6/-3, 6/0/7/0, 6/-1/7/-1, 7/0/7/-1, 6/-2/7/-2, 7/-1/7/-2, 7/0/8/0, 7/-1/8/-1, 8/0/8/-1, 8/0/9/0, 8/-1/9/-1, 9/0/9/-1}
			{
				\draw (\x,\y) -- (\z,\w);
			}
			\node at (4.5, -2.5) {$\scriptstyle i$};
			\node at (3.5, -3.5) {$\scriptstyle j$};
			\draw[fill=burgundy] (4,-3) -- (4,-4) -- (5,-4) -- (5,-3) -- cycle;
			\node[white] at (4.5,-3.5) {$\scriptstyle n$};
		\end{tikzpicture}
	\end{array}\,,\\
	&B_1=\left(\begin{array}{cc}
		B_1(\lambda)& 0\\
		q_1 e_{n,i} & 0
	\end{array}\right),\quad B_2=\left(\begin{array}{cc}
	B_2(\lambda)& 0\\
	q_1 e_{n,j} & 0
	\end{array}\right),\quad R=\left(\begin{array}{cc}
	R(\lambda)&0\\
	0 & q_2\\
	\end{array}\right)\,,\\
	&\Phi=\left(\begin{array}{cc}
	\Phi(\lambda)& 0\\
	0 & \phi\\
	\end{array}\right),\quad \Sigma=\left(\begin{array}{cc}
	0& 0\\
	0 & \sigma\\
	\end{array}\right),\quad \Phi_{\rm flav}=\left(\begin{array}{cc}
	0& 0\\
	0 & \mu\\
	\end{array}\right),\quad \Sigma_{\rm flav}=\left(\begin{array}{cc}
	0& 0\\
	0 & u\\
	\end{array}\right)\,,
	\end{aligned}
\end{equation}
where $e_{n,a}$ is a unit line with a unit at the $a^{\rm th}$ column.

This parameterization is very similar to the parameterization of the $\IC\IP^1$ case in Sec.\ref{sec:meridian}.
Similarly, we have two explicit vacuum solutions corresponding to the $\IC\IP^1$ poles:
\begin{equation}
\begin{aligned}
	&|\lambda\rangle_{0}\otimes|\Box\rangle_{\mu}:\quad q_1=0,\; q_2=1,\; \sigma=u,\; \phi=\mu\,,\\
	&|\lambda+\Box\rangle_{0}\otimes|\varnothing\rangle_{\mu}:\quad q_1=1,\; q_2=0,\; \sigma=0,\; \phi=0\,.
\end{aligned}
\end{equation}
The height function in this case is quite similar to \eqref{height} and we expect that there is a meridian instanton trajectory interpolating one way or the other depending on the sign of parameter $u$.

In general, it is rather hard to solve instanton equations explicitly.
Therefore instead we use some estimation tricks.
The chiral fields are expected to solve the following equations:
\begin{equation}\label{Quivinst}
    -\p_{\tau}B_i(\tau)-\left[\Sigma(\tau),B_i(\tau)\right]=0,\quad -\p_{\tau}R(\tau)-(\Sigma(\tau)R(\tau)-R(\tau)\Sigma_{\rm flav})=0,\quad -\p_{\tau}S(\tau)-(\Sigma_{\rm flav}S(\tau)-S(\tau)\Sigma(\tau))=0\,.
\end{equation}

Let us denote the fields at the ``north pole'' $(q_1,q_2)=(0,1)$ by ordinary variables as $B_i$, and the fields at the ``south pole'' $(q_1,q_2)=(1,0)$ by variables with a prime as $B_i'$.
Then the integral form of equations \eqref{Quivinst} states that the fields at $\tau=-\infty$ and $\tau=+\infty$ are related by $GL(n,\IC)$ transform:
\begin{equation}\label{InstQuivInt}
    G_1B_1'-B_1G_1=G_1B_2'-B_2G_1=G_1R'-R G_0=G_0S'-SG_1=0\,,
\end{equation}
where
\begin{equation}
    G_1=\left(\begin{array}{cc}
	\bbone_{(n-1)\times (n-1)} & \\
	& h_1\\
	\end{array}\right),\quad G_0=\left(\begin{array}{cc}
	    1 & 0\\
	    0 & h_0\\
    \end{array}\right)\,.
\end{equation}
Coefficient $h_i$ are given by the following expressions when $u>\sigma(\tau)>0$:
\begin{equation}
    h_1=\exp\left(-\int \sigma(\tau)\,d\tau\right)\to 0,\quad h_2=\exp\left(-\int u\,d\tau\right)\to 0\,.
\end{equation}
So for the instanton trajectory we set them to zeroes.

Now we would like to consider a small neighborhood of the above trajectory.
We denote gauge-invariant direction satisfying the F-term constraint along the moduli spaces at both vacua as $\delta B_i$, $\delta B_i'$ and so on.
These degrees of freedom span the tangent spaces on the quiver moduli spaces to the fixed points:
\begin{equation}
    X:={\rm Span}\left(\delta B_1,\delta B_2,\delta R,\delta S\right),\quad Y:={\rm Span}\left(\delta B_1',\delta B_2',\delta R',\delta S'\right)\,.
\end{equation}
For deformed trajectories we consider a deformed gauge transform $\delta G_1$.
Since at the framing node there are no dynamical degrees of freedom we set $\delta G_0=0$.
Further we decompose integral equations \eqref{InstQuivInt} up to the first order in $\delta$:
\begin{equation}
    \begin{aligned}
	&\delta G_1B_1'+G_1\delta B_1'-\delta B_1 G_1 -  B_1\delta G_1=0\,,\\
	&\delta G_1B_2'+G_1\delta B_2'-\delta B_2G_1-B_2\delta G_1=0\,,\\
	&\delta G_1R'+G_1\delta R'-\delta R G_0=0\,,\\
	&G_0\delta S'-\delta SG_1-S\delta G_1=0\,.
    \end{aligned}
\end{equation}
Excluding $\delta G_1$ from these equations (there is a \emph{unique} substitution $\delta G_1$ compatible with this system) we arrive to a collection of algebraic equations describing locus $\CK\subset X\times Y$.

Then for raising and lowering amplitudes following \eqref{geomtunnel} we derive:
\begin{equation}\label{estimateampl}
\begin{aligned}
    & \tilde{T}_{(\Box)_{\mu}\otimes (\lambda)_0\to(\varnothing)_{\mu}\otimes(\lambda+\Box)_0}=\int\lm_{\CK}{\bf e}(X)=\frac{1}{\mu^2}{\bf E}_{\lambda,\lambda+\Box}+O\left(\mu^{-3}\right)\,,\\
    & \tilde{T}_{(\lambda)_0\otimes(\varnothing)_{\mu}\to (\lambda-\Box)_0\otimes (\Box)_{\mu}}=\int\lm_{\CK}{\bf e}(Y)=\frac{1}{\epsilon_1\epsilon_2 \mu^2}{\bf F}_{\lambda,\lambda-\Box}+O\left(\mu^{-3}\right)\,.
\end{aligned}
\end{equation}
where (see Fig.\ref{fig:vert_hor_arm_leg})
\begin{equation}\label{EF}
\begin{aligned}
    &{\bf E}_{\lambda,\lambda+\Box}=\frac{1}{\epsilon_1\epsilon_2}\prod\lm_{\Box'\in\CH_{\lambda}(\Box)}\frac{-\epsilon_1(a_{\lambda}(\Box'))+\epsilon_2(l_{\lambda}(\Box')+1)}{-\epsilon_1(a_{\lambda}(\Box')+1)+\epsilon_2(l_{\lambda}(\Box')+1)}\prod\lm_{\Box'\in\CV_{\lambda}(\Box)}\frac{-\epsilon_1(a_{\lambda}(\Box')+1)+\epsilon_2(l_{\lambda}(\Box'))}{-\epsilon_1(a_{\lambda}(\Box')+1)+\epsilon_2(l_{\lambda}(\Box')+1)}\,,\\
    &{\bf F}_{\lambda+\Box,\lambda}=\prod\lm_{\Box'\in\CH_{\lambda}(\Box)}\frac{-\epsilon_1(a_{\lambda}(\Box')+2)+\epsilon_2(l_{\lambda}(\Box'))}{-\epsilon_1(a_{\lambda}(\Box')+1)+\epsilon_2(l_{\lambda}(\Box'))}\prod\lm_{\Box'\in\CV_{\lambda}(\Box)}\frac{-\epsilon_1(a_{\lambda}(\Box'))+\epsilon_2(l_{\lambda}(\Box')+2)}{-\epsilon_1(a_{\lambda}(\Box'))+\epsilon_2(l_{\lambda}(\Box')+1)}\,.\\
\end{aligned}
\end{equation}
Matrix coefficients ${\bf E}_{\lambda,\lambda+\Box}$ and ${\bf F}_{\lambda,\lambda-\Box}$ have their own geometric interpretation as integrals over respective quiver moduli spaces where locus $\CK$ is cut from $X\times Y$ by a homomorphism existence for quiver representations \cite{Galakhov:2020vyb}.

\definecolor{palette1}{rgb}{0.603922, 0.466667, 0.811765}
\definecolor{palette2}{rgb}{0.329412, 0.219608, 0.517647}
\definecolor{palette3}{rgb}{0.0156863, 0.282353, 0.333333}
\definecolor{palette4}{rgb}{0.631373, 0.211765, 0.439216}
\definecolor{palette5}{rgb}{0.92549, 0.254902, 0.462745}
\definecolor{palette6}{rgb}{1., 0.643137, 0.368627}
\definecolor{palette7}{rgb}{0.313725, 0.45098, 0.85098}
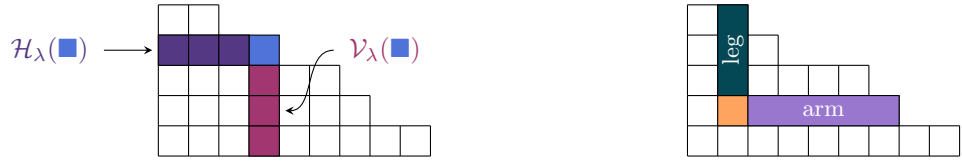
\begin{figure}[ht!]
	\begin{center}
		\begin{tikzpicture}[yscale=-1]
			\begin{scope}
				\begin{scope}[scale=0.4]
					\foreach \x/\y/\z/\w in {0/-5/1/-5, 0/-4/0/-5, 0/-4/1/-4, 0/-3/0/-4, 0/-3/1/-3, 0/-2/0/-3, 0/-2/1/-2, 0/-1/0/-2, 0/-1/1/-1, 0/0/0/-1, 0/0/1/0, 1/-5/2/-5, 1/-4/1/-5, 1/-4/2/-4, 1/-3/1/-4, 1/-3/2/-3, 1/-2/1/-3, 1/-2/2/-2, 1/-1/1/-2, 1/-1/2/-1, 1/0/1/-1, 1/0/2/0, 2/-4/2/-5, 2/-4/3/-4, 2/-3/2/-4, 2/-3/3/-3, 2/-2/2/-3, 2/-2/3/-2, 2/-1/2/-2, 2/-1/3/-1, 2/0/2/-1, 2/0/3/0, 3/-3/3/-4, 3/-3/4/-3, 3/-2/3/-3, 3/-2/4/-2, 3/-1/3/-2, 3/-1/4/-1, 3/0/3/-1, 3/0/4/0, 4/-3/5/-3, 4/-2/4/-3, 4/-2/5/-2, 4/-1/4/-2, 4/-1/5/-1, 4/0/4/-1, 4/0/5/0, 5/-3/6/-3, 5/-2/5/-3, 5/-2/6/-2, 5/-1/5/-2, 5/-1/6/-1, 5/0/5/-1, 5/0/6/0, 6/-2/6/-3, 6/-2/7/-2, 6/-1/6/-2, 6/-1/7/-1, 6/0/6/-1, 6/0/7/0, 7/-1/7/-2, 7/-1/8/-1, 7/0/7/-1, 7/0/8/0, 8/-1/9/-1, 8/0/8/-1, 8/0/9/0, 9/0/9/-1}
					{
						\draw (\x,\y) -- (\z,\w);
					}
					\draw[fill=palette7] (3,-3) -- (4,-3) -- (4,-4) -- (3,-4) -- cycle;
					\foreach \i in {0,1,2}
					{
						\draw[fill=palette2] (\i,-3) -- (\i+1,-3) -- (\i+1,-4) -- (\i,-4) -- cycle;
					}
					\node[left,palette2] at (-2,-3.5) {$\CH_{\lambda}({\color{palette7}\blacksquare})$};
					\draw[-stealth] (-1.8,-3.5) -- (-0.2,-3.5);
					\foreach \i in {0,1,2}
					{
						\draw[fill=palette4] (3,-\i) -- (3,-\i-1) -- (4,-\i-1) -- (4,-\i) -- cycle;
					}
					\node[right,palette4] at (6,-3.5) {$\CV_{\lambda}({\color{palette7}\blacksquare})$};
					\draw[-stealth] (5.8,-3.5) to[out=180,in=0] (4.2,-1.5);
				\end{scope}
			\end{scope}
			\begin{scope}[shift={(7,0)}]
				\begin{scope}[scale=0.4]
					\foreach \x/\y/\z/\w in {0/-5/1/-5, 0/-4/0/-5, 0/-4/1/-4, 0/-3/0/-4, 0/-3/1/-3, 0/-2/0/-3, 0/-2/1/-2, 0/-1/0/-2, 0/-1/1/-1, 0/0/0/-1, 0/0/1/0, 1/-5/2/-5, 1/-4/1/-5, 1/-4/2/-4, 1/-3/1/-4, 1/-3/2/-3, 1/-2/1/-3, 1/-2/2/-2, 1/-1/1/-2, 1/-1/2/-1, 1/0/1/-1, 1/0/2/0, 2/-4/2/-5, 2/-4/3/-4, 2/-3/2/-4, 2/-3/3/-3, 2/-2/2/-3, 2/-2/3/-2, 2/-1/2/-2, 2/-1/3/-1, 2/0/2/-1, 2/0/3/0, 3/-3/3/-4, 3/-3/4/-3, 3/-2/3/-3, 3/-2/4/-2, 3/-1/3/-2, 3/-1/4/-1, 3/0/3/-1, 3/0/4/0, 4/-3/5/-3, 4/-2/4/-3, 4/-2/5/-2, 4/-1/4/-2, 4/-1/5/-1, 4/0/4/-1, 4/0/5/0, 5/-3/6/-3, 5/-2/5/-3, 5/-2/6/-2, 5/-1/5/-2, 5/-1/6/-1, 5/0/5/-1, 5/0/6/0, 6/-2/6/-3, 6/-2/7/-2, 6/-1/6/-2, 6/-1/7/-1, 6/0/6/-1, 6/0/7/0, 7/-1/7/-2, 7/-1/8/-1, 7/0/7/-1, 7/0/8/0, 8/-1/9/-1, 8/0/8/-1, 8/0/9/0, 9/0/9/-1}
					{
						\draw (\x,\y) -- (\z,\w);
					}
					\draw[fill=palette6] (1,-1) -- (2,-1) -- (2,-2) -- (1,-2) -- cycle;
					\draw[fill=palette3] (1,-2) -- (2,-2) -- (2,-5) -- (1,-5) -- cycle;
					\node[rotate=90,white] at (1.5,-3.5) {\small leg};
					\draw[fill=palette1] (2,-1) -- (7,-1) -- (7,-2) -- (2,-2) -- cycle;
					\node[white] at (4.5,-1.5) {\small arm};
				\end{scope}
			\end{scope}
		\end{tikzpicture}
	\end{center}
	\caption{Vertical, horizontal sets, leg and arm functions on Young diagrams.}\label{fig:vert_hor_arm_leg}
\end{figure}

Estimate \eqref{estimateampl} has few flaws:
\begin{itemize}
    \item To observe the first flow it is easier to reduce to the simplest case $n=1$.
    The quiver target space is a bundle over $\IC\IP^1$, and this case mimics literally the case discussed in Sec.\ref{sec:meridian}.
    In this case $R=(r_1,r_2)$ and $S=(s_1,s_2)^T$.
	Fields $r_1$ and $r_2$ parameterize a projective coordinate $(r_1:r_2)$ on the $\IC\IP_1$-base.
	The F-term constraints field $S$ to satisfy equation $r_1s_1+r_2s_2$, this annihilates one of two degrees of freedom $s_1$, $s_2$, whereas the other one $(s_1,s_2)=(-r_2,r_1){\bf s}$ is dual to the tangent coordinate $(r_1:r_2)$.
	As we discussed in Sec.\ref{sec:meridian} the tangent coordinate travels from $\omega_3<0$ to $\omega_3>0$, and field $\bf s$ travels in the opposite direction, so its contribution to $\mathsf{T}$ should read $\omega_z(\omega_3=0)=\epsilon_1+\epsilon_2$.
	However as we estimated this contribution via equivariant integral it reads $\mu^{-1}$.
	To compensate this estimation error we should multiply the amplitude by $\mu(\epsilon_1+\epsilon_2)$.
    \item Each looped arrow chiral field has a matrix element $(B_i)_{n,n}$ whose charge $\omega_3=0$.
	So it remains stationary in the space $\vec{\omega}$, whereas the estimate implies it moving from $\omega_3<0$ to $\omega_3=0$.
	To compensate this contribution we should multiply the estimated amplitude by the respective weight of a looped quiver arrow.
	In our case for $B_1$ and $B_2$ the multiplier reads $\epsilon_1\epsilon_2$.
\end{itemize}
So we arrive to a relation between estimated amplitude and the actual one:
\begin{equation}
    \mathsf{T}=\mu(\epsilon_1+\epsilon_2)\times(\epsilon_1\epsilon_2) \times \tilde{T}\,.
\end{equation}

In general, we expect (cf. \eqref{sandwitch}):
\begin{tcolorbox}
\begin{equation}\label{T-matrix}
    \mathsf{T}_{(\lambda)_{\mu_1}\otimes (\lambda')_{\mu_2}\to (\lambda-\Box)_{\mu_1}\otimes (\lambda'+\Box)_{\mu_2}}=\frac{-\epsilon_1\epsilon_2(\epsilon_1+\epsilon_2)}{\mu_1-\mu_2}\times {\bf F}_{\lambda,\lambda-\Box}\times {\bf E}_{\lambda',\lambda'+\Box}\,,
\end{equation}
\end{tcolorbox}
where we neglect higher orders $O\left((\mu_1-\mu_2)^{-2}\right)$.

We define the tunneling algebra following Sec.\ref{sec:TAC} in this case as:
\begin{equation}
    \begin{aligned}
	&{\bf v}^+|\lambda\rangle=\sum\lm_{\Box\in {\rm Add}(\lambda)}{\bf E}_{\lambda,\lambda+\Box}|\lambda+\Box\rangle\,,\\
	&{\bf v}^-|\lambda\rangle=\sum\lm_{\Box\in {\rm Rem}(\lambda)}{\bf F}_{\lambda,\lambda-\Box}|\lambda-\Box\rangle\,,
    \end{aligned}
\end{equation}

\subsubsection{Affine Yangian \texorpdfstring{$Y(\widehat{\fg\fl}_1)$}{}}

Introduce canonical parameters:
\begin{equation}
    \alpha_2=-\left(\epsilon_1^2+\epsilon_1\epsilon_2+\epsilon_2^2\right),\quad \alpha_3=-\epsilon_1\epsilon_2(\epsilon_1+\epsilon_2)\,.
\end{equation}

Define Cartan generators $\psi_k$ by eigenvalue:
\begin{equation}
    \psi(z)|\lambda\rangle=\left(1+\sum\lm_{k=0}^{\infty}\frac{\psi_k}{z^{k+1}}\right)|\lambda\rangle=\frac{z+\epsilon_1+\epsilon_2}{z}\prod\lm_{\Box\in\lambda}\varphi(z-\omega_{\Box})|\lambda\rangle\,,
\end{equation}
where
\begin{equation}
\begin{aligned}
    &\omega_{\Box}=\epsilon_1 x_{\Box}+\epsilon_2 y_{\Box}\,,\\
    &\varphi(z)=\frac{\left(z+\epsilon_1\right)\left(z+\epsilon_2\right)\left(z-\epsilon_1-\epsilon_2\right)}{\left(z-\epsilon_1\right)\left(z-\epsilon_2\right)\left(z+\epsilon_1+\epsilon_2\right)}\,.
\end{aligned}
\end{equation}
So that
\begin{equation}
    \begin{aligned}
	& \psi_0|\lambda\rangle=(\epsilon_1+\epsilon_2)|\lambda\rangle\,,\\
	& \psi_1|\lambda\rangle=0\,,\\
	& \psi_2|\lambda\rangle=-2\epsilon_1\epsilon_2|\lambda|\cdot|\lambda\rangle\,,\\
	& \psi_3|\lambda\rangle=-\epsilon_1\epsilon_2(\epsilon_1+\epsilon_2)\left[6\sum\lm_{\Box\in\lambda}\omega_{\Box}-2\left(\epsilon_1+\epsilon_2\right)|\lambda|\right]|\lambda\rangle\,.\\
    \end{aligned}
\end{equation}

We define raising and lowering generators for level 0 as:
\begin{equation}
    e_0|\lambda\rangle:={\bf v}^+|\lambda\rangle,\quad f_0|\lambda\rangle:={\bf v}^-|\lambda\rangle\,.
\end{equation}
Higher order operators are defined as:
\begin{equation}
    e_{k+1}=\frac{1}{6\alpha_3}\left[\psi_3,e_k\right]-\frac{1}{3}\psi_0e_k,\quad f_{k+1}=-\frac{1}{6\alpha_3}\left[\psi_3,f_k\right]-\frac{1}{3}\psi_0f_k\,,
\end{equation}
so that:
\begin{equation}
    e_k|\lambda\rangle=\sum\lm_{\Box\in{\rm Add}(\lambda)}{\bf E}_{\lambda,\lambda+\Box}\,\omega_{\Box}^k|\lambda+\Box\rangle,\quad f_k|\lambda\rangle=\sum\lm_{\Box\in{\rm Rem}(\lambda)}{\bf F}_{\lambda,\lambda-\Box}\,\omega_{\Box}^k|\lambda-\Box\rangle\,.
\end{equation}

Generators $e_k$, $f_k$ and $\psi_k$ satisfy relations of affine Yangian $Y(\widehat{\fg\fl}_1)$:
\begin{equation}\label{Yangian}
\begin{aligned}
    & \left[\psi_m,\psi_k\right]=0\,,\\
    & \left[e_k,f_m\right]=\frac{1}{\alpha_3}\psi_{k+m}\,,\\
    & \left[e_{k+3},e_m\right]-3\left[e_{k+2},e_{m+1}\right]+3\left[e_{k+1},e_{m+2}\right]-\left[e_{k},e_{m+3}\right]+\alpha_2\left[e_{k+1},e_{m}\right]-\alpha_2\left[e_{k},e_{m+1}\right]-\alpha_3\left\{e_{k},e_{m}\right\}=0,\\
    & \left[f_{k+3},f_m\right]-3\left[f_{k+2},f_{m+1}\right]+3\left[f_{k+1},f_{m+2}\right]-\left[f_{k},f_{m+3}\right]+\alpha_2\left[f_{k+1},f_{m}\right]-\alpha_2\left[f_{k},f_{m+1}\right]+\alpha_3\left\{f_{k},f_{m}\right\}=0,\\
    & \left[\psi_{k+3},e_m\right]-3\left[\psi_{k+2},e_{m+1}\right]+3\left[\psi_{k+1},e_{m+2}\right]-\left[\psi_{k},e_{m+3}\right]+\alpha_2\left[\psi_{k+1},e_{m}\right]-\alpha_2\left[\psi_{k},e_{m+1}\right]-\alpha_3\left\{\psi_{k},e_{m}\right\}=0,\\
    & \left[\psi_{k+3},f_m\right]-3\left[\psi_{k+2},f_{m+1}\right]+3\left[\psi_{k+1},f_{m+2}\right]-\left[\psi_{k},f_{m+3}\right]+\alpha_2\left[\psi_{k+1},f_{m}\right]-\alpha_2\left[\psi_{k},f_{m+1}\right]+\alpha_3\left\{\psi_{k},f_{m}\right\}=0,\\
    & {\rm Sym}_{k_1,k_2,k_3}\left[e_{k_1},\left[e_{k_2},e_{k_3+1}\right]\right]=0\,,\\
    & {\rm Sym}_{k_1,k_2,k_3}\left[f_{k_1},\left[f_{k_2},f_{k_3+1}\right]\right]=0\,.
\end{aligned}
\end{equation}

Following \cite{Galakhov:2022uyu} we construct the R-matrix depending on the spectral parameter $a$ as:
\begin{equation}\label{R-matrix}
    R(a)=U(-a)\,R^{(0)}(a)\,U(a)^{-1}\,,
\end{equation}
where $U(a)$ is a lower-triangular matrix, and $R^{(0)}(a)$ is a trivial R-matrix consisting of a diagonal matrix times a permutation matrix.
Both R-matrices intertwine two co-products one might call the \emph{true} and a \emph{naive} ones:
\begin{equation}\label{intertwining}
    R(a_1-a_2)\Delta_{a_1,a_2}=\Delta_{a_2,a_1}R(a_1-a_2),\quad R^{(0)}(a_1-a_2)\Delta_{a_1,a_2}^{(0)}=\Delta_{a_2,a_1}^{(0)}R^{(0)}(a_1-a_2)\,.
\end{equation}
The naive co-product originates from a simple straightforward generalization of expressions \eqref{EF} when two crystals are located at the same $\sigma$-coordinate (see Fig.\ref{fig:chain}) and is not a co-product in the algebraic sense: it can not be represented as a homomorphism of an algebra to its tensor square.
Yet comparing \eqref{R-matrix} and \eqref{intertwining} it is natural to conclude that matrix $U(a)$ conjugates the true co-product into the naive one:
\begin{equation}
    U(a_1-a_2)\Delta_{a_1,a_2}=\Delta^{(0)}_{a_1,a_2}U(a_1-a_2)\,.
\end{equation}
We will not consider complete expressions for co-products, rather we will consider some first terms for $e_0$ following \cite{Galakhov:2022uyu} and \cite{Prochazka:2015deb}:
\begin{equation}
    \Delta_{a_1,a_2}(e_0)=e_0\otimes 1+1\otimes e_0,\quad \Delta^{(0)}_{a_1,a_2}(e_0)=e_0\otimes 1+\left(1+\frac{\psi_0}{a_1-a_2}+O(|a_1-a_2|^{-2})\right)\otimes e_0\,.
\end{equation}

\begin{figure}[ht!]
    \centering
    \begin{tikzpicture}
    	\draw[draw=white, fill=white!60!orange, rounded corners=2] (-1.2,-0.4) -- (1.2,-0.4) -- (1.2,-1.35) -- (-1.2,-1.35) -- cycle;
    	\begin{scope}[yscale=-1]
    	\draw[draw=white, fill=white!60!burgundy, rounded corners=2] (-1.2,-0.4) -- (1.2,-0.4) -- (1.2,-1.35) -- (-1.2,-1.35) -- cycle;
    	\end{scope}	
    	\draw[draw=white, fill=white!60!blue, rounded corners=2] (-1.2,0.2) -- (1.2,0.2) -- (1.2,-0.2) -- (-1.2,-0.2) -- cycle;
    	\draw[postaction={decorate}, decoration={markings, mark= at position 0.6 with {\arrow{stealth}}}] (-1,-1.25) -- (1,-1.25);
    	\draw[postaction={decorate}, decoration={markings, mark= at position 0.6 with {\arrow{stealth}}}] (-1,-1) -- (1,-1);
    	\draw[postaction={decorate}, decoration={markings, mark= at position 0.6 with {\arrow{stealth}}}] (-0.95,-0.75) -- (0.95,-0.75);
    	\draw[postaction={decorate}, decoration={markings, mark= at position 0.6 with {\arrow{stealth}}}] (-0.8,-0.5) -- (0.8,-0.5);
    	\begin{scope}[yscale=-1]
    	\draw[postaction={decorate}, decoration={markings, mark= at position 0.6 with {\arrow{stealth}}}] (-1,-1.25) -- (1,-1.25);
    	\draw[postaction={decorate}, decoration={markings, mark= at position 0.6 with {\arrow{stealth}}}] (-1,-1) -- (1,-1);
    	\draw[postaction={decorate}, decoration={markings, mark= at position 0.6 with {\arrow{stealth}}}] (-0.95,-0.75) -- (0.95,-0.75);
    	\draw[postaction={decorate}, decoration={markings, mark= at position 0.6 with {\arrow{stealth}}}] (-0.8,-0.5) -- (0.8,-0.5);
    	\end{scope}
    	\draw[ultra thick] (-1,-1.5) -- (-1,-1) to[out=90,in=270] (1,1) -- (1,1.5);
    	\begin{scope}[xscale=-1]
    	\draw[ultra thick] (-1,-1.5) -- (-1,-1) to[out=90,in=270] (1,1) -- (1,1.5);
    	\end{scope}
    	\draw[thick, draw=black!40!orange, rounded corners=2] (-1.2,-0.4) -- (1.2,-0.4) -- (1.2,-1.35) -- (-1.2,-1.35) -- cycle;
    	\begin{scope}[yscale=-1]
    	\draw[thick, draw=black!40!burgundy, rounded corners=2] (-1.2,-0.4) -- (1.2,-0.4) -- (1.2,-1.35) -- (-1.2,-1.35) -- cycle;
    	\end{scope}	
    	\draw[thick, draw=black!40!blue, rounded corners=2] (-1.2,0.2) -- (1.2,0.2) -- (1.2,-0.2) -- (-1.2,-0.2) -- cycle;
    	\draw[stealth-stealth] (-1.4,1.5) -- (-1.4,-2) -- (1.5,-2);
    	\node[right] at (1.5,-2) {$\scriptstyle u$};
    	\node[above left] at (-1.4,1.5) {$\scriptstyle t$};
    	\node[below] at (-1,-1.5) {$\scriptstyle u_1$};
    	\node[below] at (1,-1.5) {$\scriptstyle u_2$};
    	\node[above] at (-1,1.5) {$\scriptstyle u_2$};
    	\node[above] at (1,1.5) {$\scriptstyle u_1$};
    	\node[black!40!orange, right] at (1.5,-1) {Tunneling $1\to 2$: $U(\mu_1-\mu_2)^{-1}$};
    	\node[black!40!blue, right] at (1.5,0) {Nothing happens: $R^{(0)}(\mu_1-\mu_2)$};
    	\node[black!40!burgundy, right] at (1.5,1) {Tunneling $2\to 1$: $U(\mu_2-\mu_1)$};
    \end{tikzpicture}
    \caption{Affine Yangian R-matrix as an evolution process}\label{fig:Revolution}
\end{figure}
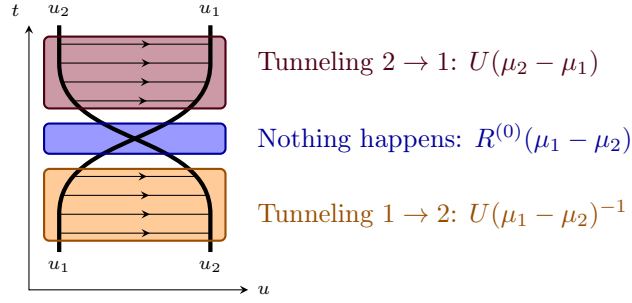

The we derive for $U(a)$ the following expansion:
\begin{equation}\label{Umatrix}
    U(a)=1\otimes 1+\frac{\alpha_3}{a}f_0\otimes e_0+O(|a|^{-2})=\left(\begin{array}{cc}
	1 & 0\\
	\mathsf{T} & 1
    \end{array}\right)\,,
\end{equation}
where $\mathsf{T}$ is the tunneling amplitude \eqref{T-matrix}.
Comparing \eqref{Umatrix} with \eqref{CP1Stokes} we conclude that matrix $U(a)$ is the evolution matrix.
This observation allows us to represent the whole R-matrix as an evolution process form moving parameters $u_1(t)$ and $u_2(t)$ form $u_1<u_2$ towards a permutation $u_2<u_1$.
While $u_1<u_2$ only tunneling processes $1\to 2$ contribute, while $u_2<u_1$ tunneling processes flow the other way around, see Fig.\ref{fig:Revolution}.
This interpretation allows one to ``predict'' immediately the Yang-Baxter equation satisfied by the R-matrix as a flatness of parallel transport (see Sec.\ref{sec:flatness}).

In this case of affine Yangian as in the case of quantum algebras the R-matrix constructed as an evolution operator is \emph{compatible} with the tunneling algebra that is represented by the affine Yangian itself.
Having calculated the R-matrices we could have tried to reconstruct the relations among the tunneling algebra generators to derive \eqref{Yangian} following \cite{Litvinov:2020zeq}.

\subsubsection{Generic quiver Yangians}

From the material presented in this section one might propose an immediate \emph{conjecture} that in general the tunneling algebra for supersymmetric quantum mechanics with a generic quiver $\fQ$ target space is the respective \emph{quiver Yangian} algebra $Y(\fQ)$ \cite{Rapcak:2018nsl,Li:2020rij,Li:2023zub}.
We should \emph{warn} the reader from making a hasty conclusion, as the quiver target space might be singular in general.
A canonical example of such singularity is the moduli space of the Hilbert scheme of points on $\IC^3$ \cite[Exercise 3.1.5]{Okounkov:2015spn} (in contrast to the case of $\IC^2$ considered in the paper).

These singularities might affect the construction quite violently: particles in the picture with multiple potential wells might fall on these singularities when leaving potential wells.
In particular, our construction the R-matrix is automatically compatible with the Yangian algebra, and simple attempts to construct the R-matrix starting from ans\"atze like \eqref{R-matrix} and \eqref{Umatrix} might fail for certain quivers choices due to singularities \cite{Galakhov:2022uyu}.
Taking into account new tunneling amplitudes between classical vacua and singularities might help to cure this situation.
However at the moment this possibility remains underinvestigated.

As another issue in this line we should mention a yet purely phenomenological observation of difficulties in attempts of quiver Yangian bosonization being present when quiver variety is singular \cite{Morozov:2022ndt} and absent when it is smooth \cite{Galakhov:2023mak}. 


\section{Conclusion}

The main suggestion of this paper is to interpret t-channel diagrams of Fig.\ref{fig:main_story}
as the generators of a new {\it tunneling algebra}.
From the quantum mechanical point of view the same diagrams can be considered
as instanton tunneling transitions between the states in the tight binding model,
which are described by external/scattering particles or branes in the QFT diagram.
Depending on the nature of the underlying particles and strings,
the tunneling algebra reproduces different representations of different algebras,
from $U(1)$ to Yangians and DIMs, and even further.
This paper is more a proposal than detailed work in this new direction.
There are still plenty details, which need to be clarified.

\bigskip

\noindent
As is quite common in the modern literature,
in this paper we relied on considerable simplifications,
provided by supersymmetry.
\begin{itemize}
\item{}
This reduced quantum corrections to one loop, and even those are made less involved.
\item{}
More important is that supersymmetry can eliminate instanton contributions
due to fermion zero modes.
We restored them by considering time-dependent configurations --
which in any case are needed to define connections and R-matrices.
However then supersymmetry helps to separate instantons from anti-instantons --
and chiral quantities are always easier to calculate and study.

\item{}
Moreover, in this paper the zero-curvature conditions and Yang-Baxter equations
were deduced from supersymmetry -- what, we suspect, is not necessary,
the origins might be much broader and not restricted to supersymmetric models.
This is one of our main {\bf open question}s.

\item{}
Supersymmetry is also behind the localization arguments, which allow to reduce
integrals over moduli spaces to particular points,
substitute strings by sticks.
This is what allows us to start building a bridge between physically observable quantities and more abstract mathematical concepts, such as integrals of Euler-class forms over moduli spaces and, furthermore, the pullback-pushforward construction and the Fourier-Mukai transform.
In other words, it is intended to provide a physical ``explanation'' of definitions and procedures usually simply postulated in mathematical literature \cite{nakajima1999lectures,Nakajima:1994nid,Kontsevich:2010px,Rapcak:2018nsl,yang2018cohomological}.
Again, it is unclear to what extent supersymmetry really matters for this
equivalence -- the algebras (Yangians and DIM) make no explicit reference to it.
This is another {\bf open question}.
\end{itemize}

\noindent
Another big issue is the study of tunneling for adiabatically changing potential
with wells.
We presented a sample example of delta-wells, which changed only depths,
not the positions and not the shapes.
Even this example appeared quite rich and instructive, but further generalizations
promise to be more interesting, both technically and conceptually.
Most important is again the understanding of the Gauss-Manin connection, Berry phases
and the zero-curvature conditions, leading to R-matrices and Yang-Baxter equations.
Supersymmetry does not look truly important for these studies.

\bigskip

We believe that the physically motivated description in terms of {\it tunneling algebras},
coming from the $t$-channel diagrams,
will be a useful alternative or complement to the widespread categorical discussions.
Most important it can make the subject available to much broader audience,
help to focus on truly relevant problems
and stimulate a faster progress in the field.

\bigskip

In conclusion let us also mention some future problems that might benefit from the discussion of the current note:
\begin{itemize}
	\item As we have seen during our discussions the computations of the tunneling amplitudes even in the supersymmetric cases start to experience certain limitations of purely analytic method applicability.
	The Landau-Ginzburg side seems to circumvent this issue at least partially by implementing numerical methods to construct pictures of Stokes lines (see e.g. \cite{Gaiotto:2012rg, Longhi:2016rjt}).
	On the sigma model side some numerical methods of analyzing parameter spaces are also developed (see e.g. \cite{Herbst:2008jq, Hori:2013ika, Brunner:2021ulc, Brunner:2024quk}), yet it seems they lack certain spectacularity  of the LG side.
	Nevertheless, it seems that the problem of calculating approximate instanton trajectories and defining respective overlaps should have a numerical solution in the framework of quantum mechanics we discussed in this note.
	It would be interesting to develop and apply these methods to compare with our analytic estimates.
	\item In general quiver varieties where a particle in the description of Yangian algebras propagates might turn out to be singular.
	A physical suggestion on a regularization procedure for these singularities was proposed in \cite{Galakhov:2020vyb}.
	It would be interesting to check if this suggestion works with the tunneling amplitudes and provides tunneling algebras in the singular case as well.
    \item The structures of the quantum algebras and Yangians is naturally extended \cite{Galakhov:2021vbo} to quantum toroidal algebras \cite{FJMM,Noshita:2021dgj,2013arXiv1302.6202N}, a.k.a. DIM algebras \cite{Ding:1996mq,Miki1,Awata:2017lqa,DIM2}, and beyond \cite{Galakhov:2023aev}.
	The structure of the underlying QFT should be usually changed for those deformations.
	Those theories might incorporate different types of instantons and defects in general.
	Surely, it would be interesting to study interaction of defects of different support dimensionality and different topological and geometrical nature, moreover effects they induce on the tunneling algebra.
    \item Evolution operators obtained by $P$-exponentiation of Berry and Gauss-Manin connections are relevant not only for R-matrices.
	Traveling across the parameter space in supersymmetric QFTs causes events of \emph{wall-crossing} in BPS spectra (see e.g. \cite{Aganagic:2009kf,Alim:2011kw} for a review).
	In practice, Stokes surfaces are considered to be marginal stability walls for BPS solitons \cite{Gaiotto:2012rg}.
	We hope that the tunneling description for these processes would provide a more rigorous approach to the study of wall-crossing effects on quiver BPS algebras \cite{Galakhov:2024foa}.
	In particular, it would be interesting to find a physical parameter  evolution analog for automorphisms in DIM algebras switching commuting families of rays \cite{Smirnov:2018drm,Crew:2020psc,Mironov:2020pcd,Mironov:2023wga}.
	\item Over recent years a topic of D-branes on toric Calabi-Yau 4-folds has acquired a rapid development \cite{Cao:2017swr,Nekrasov:2017cih,Nekrasov:2018xsb,Bonelli:2020gku,Szabo:2023ixw}.
	Nevertheless it seems the question what the BPS algebra \cite{Harvey:1995fq} for these systems should look like remain unanswered so far, despite modern literature \cite{Kimura:2023bxy,Bao:2025hfu} proposes very interesting candidates for this role.
	Since D-branes on CY4 admit an effective description \cite{Franco:2015tna} in terms of quantum mechanics we do hope that some form of the tunneling algebra would help to resolve this problem.
    \item Stable envelopes \cite{2012arXiv1211.1287M,Aganagic:2016jmx,Smirnov:2018drm,2021LMaPh.111..141O} define a certain geometric action on bases of homologies of K-theory for quiver varieties.
	In practice this action could be lifted to a natural intertwining action and allow an alternative R-matrix construction \cite{Aganagic:2016jmx} for affine Yangians and quantum toroidal (DIM) algebras.
	Geometrically one considers attracting manifolds sometimes called \emph{leaves} \cite{shenfeld2013abelianization}.
	These leaves might be thought of as an analog of Lefschetz thimbles yet constructed for the quiver gauged sigma model.
	Despite various physical models for the stable envelope construction are proposed \cite{Dedushenko:2021mds,Crew:2020psc} (among which there is an adiabatic transition between two corners of the parameter space) much work is to be done to represent the stable envelopes canonically as definite path integrals, or tunneling amplitudes.
	In particular, it would be interesting to present such a construction for models with \emph{less supersymmetry} (4 supercharges) having only a complex structure in comparison to canonical Nakajima quiver varieties having a hyperk\"ahler structure (corresponding QFTs have 8 supercharges).
	Some work in this direction from the point of view of algebraic geometry is performed in recent work \cite{Ishtiaque:2023acr}.
	\item Some of affine Yangian representations admit bosonization in terms of time variables $p_k$.
	    In these cases ground sates are enumerated by (super-, colored) Young diagrams, and the respective wave functions are orthogonal (super-)symmetric polynomials of Schur, Jack, Uglov, Macdonald type \cite{Morozov:2022ndt,Galakhov:2023mak,Galakhov:2024mbz}.
	All these types of polynomials admit a \emph{triangular} decomposition \cite{Galakhov:2024cry,Galakhov:2024zqn} over trivial symmetric functions $m_{\lambda}$: $P_{\lambda}=m_{\lambda}+\sum\lm_{\mu>\lambda} K_{\lambda\mu}m_{\mu}$, where the Young diagrams are \emph{ordered}, and coefficients $K_{\lambda\mu}$ are called Kostka numbers for Schur polynomials, or Kostka functions in general.
	Kostka functions are functions defined on (super-)Young tableaux of shape $\lambda$ and weight $\mu$, and we could reformulate the triangular property of such expansions as an absence of Young tableaux for $\lambda<\mu$.
	On the other hand, this \emph{triangular} decomposition is reminiscent of \emph{triangular} Stokes matrix transitions \eqref{Stokes} and \eqref{CP1Stokes}, and we have observed a natural order on Young diagrams in Fig.\ref{fig:chain}.
	These clues might indicate that Kostka functions are as well amplitudes of some tunneling processes, and it would be interesting to describe such a model in details if it existed.
\end{itemize}

\section*{Acknowledgments}
The work was partially funded within the state assignment of the Institute for Information Transmission Problems of RAS. 
The work of A.M. is partly supported by the grants of the Foundation for the Advancement of Theoretical Physics and Mathematics ``BASIS''.



\bibliographystyle{utphys}
\bibliography{biblio}

\end{document}